\documentclass[aps,prd,twocolumn,amsmath,amssymb,showpacs]{revtex4}

\usepackage{graphicx}
\usepackage{dcolumn}
\usepackage{bm}

\begin{document}

\title{Hadamard renormalization of the stress-energy
tensor  \\
for a quantized scalar field in a general spacetime of arbitrary
dimension}

\author{Yves D\'ecanini}
\email{decanini@univ-corse.fr} \affiliation{UMR CNRS 6134 SPE,
Equipe Physique Semi-Classique (et) de la Mati\`ere Condens\'ee, \\
Universit\'e de Corse, Facult\'e des Sciences, BP 52, 20250 Corte,
France}

\author{Antoine Folacci}
\email{folacci@univ-corse.fr} \affiliation{UMR CNRS 6134 SPE,
Equipe Physique Semi-Classique (et) de la Mati\`ere Condens\'ee, \\
Universit\'e de Corse, Facult\'e des Sciences, BP 52, 20250 Corte,
France}

\date{\today}

\begin{abstract}

We develop the Hadamard renormalization of the stress-energy tensor
for a massive scalar field theory defined on a general spacetime of
arbitrary dimension. Our formalism could be helpful in treating some
aspects of the quantum physics of extra spatial dimensions. More
precisely, for spacetime dimension up to six, we explicitly describe
the Hadamard renormalization procedure and for spacetime dimension
from seven to eleven, we provide the framework permitting the
interested reader to perform this procedure explicitly in a given
spacetime. We complete our study (i) by considering the ambiguities
of the Hadamard renormalization of the stress-energy tensor and the
corresponding ambiguities for the trace anomaly, (ii) by providing
the expressions of the gravitational counterterms involved in the
renormalization process (iii) by discussing the connections between
Hadamard renormalization and renormalization in the effective
action. All our results are expanded on standard bases for Riemann
polynomials constructed from group theoretical considerations and
thus given on irreducible forms.

\end{abstract}


\maketitle

\section{Introduction}

In semiclassical gravity, spacetime is considered from a classical
point of view, i.e. its metric $g_{\mu \nu}$ is treated classically,
while all the other fields propagating on this background (from
matter fields to the graviton field at one-loop order) are assumed
to be quantized. In the last thirty years, this approximation of
quantum gravity, usually called quantum field theory in curved
spacetime, has permitted us to obtain very interesting results
concerning more particularly i) quantum black hole physics in
connection with Hawking radiation, ii) early universe cosmology,
iii) the Casimir effect and iv) quantum violations of classical
energy conditions in connection with both the singularity theorems
of Hawking and Penrose and the existence of traversable wormholes
and time-machines... We refer to the monographs of Birrell and
Davies \cite{BirrellDavies}, Fulling \cite{Fulling89} and Wald
\cite{Wald94} as well as to references therein for various aspects
of semiclassical gravity. We also refer to a recent review by Ford
\cite{Ford2005} which is a short but rather up to date introduction
to semiclassical gravity and to its applications.  We finally refer
to Sec.~II.B of Ref.~\cite{FlanaganWald96} for a very interesting
critical account about the status and the domain of applicability of
semiclassical gravity and to
Refs.~\cite{HuVerdaguer2001,HuVerdaguerLR2004} for an extension of
semiclassical gravity, the so-called semiclassical stochastic
gravity, which also permits us to discuss and investigate its
validity.

For a quantum field in some normalized state $|\psi \rangle$, the
expectation value with respect to $|\psi \rangle$ of its
associated stress-energy-tensor operator $T_{\mu \nu}$, denoted
$\langle \psi |T_{\mu \nu} |\psi \rangle$, plays a central role in
semiclassical gravity. Indeed:

    \quad -- In curved spacetime, the particle concept is in
general very nebulous. Here, we adhere completely to the point of
view developed by Davies in Ref.~\cite{Davies1984}. It is then a
nonsense to speak about the particle content of the quantum state
$|\psi \rangle$. From the physical point of view, it is more
objectively described by a quantity such as the expectation value
$\langle \psi |T_{\mu \nu} |\psi \rangle$.

 \quad -- It is rather natural to conjecture that the classical
 metric $g_{\mu \nu}$ is coupled to the quantum field according to
 the semiclassical Einstein equations
\begin{equation}\label{SCEinsteinEq}
G_{\mu \nu}=8 \pi G \langle \psi |T_{\mu \nu} |\psi \rangle
\end{equation}
where $G_{\mu \nu}$ is the Einstein tensor $R_{\mu
\nu}-\frac{1}{2}g_{\mu \nu}R+ \Lambda g_{\mu \nu}$ (here $\Lambda$
and $G$ denote respectively the cosmological constant and the
Newton's gravitational constant) or some higher-order generalization
of this geometrical tensor. The expectation value $\langle \psi
|T_{\mu \nu} |\psi \rangle$ which acts as a source in
Eq.~(\ref{SCEinsteinEq}) then governs the back reaction of the
quantum field on the spacetime geometry.

\noindent As a consequence, in semiclassical gravity, it is
fundamental to be able to obtain an expression of the expectation
value $\langle \psi |T_{\mu \nu} |\psi \rangle$ showing in detail
the influence of the background geometry but also of the quantum
state $|\psi \rangle$. But it is well-known that this is not really
obvious \cite{BirrellDavies,Fulling89,Wald94}.

The stress-energy tensor $T_{\mu \nu}$ is an operator quadratic in
the quantum field which is, from the mathematical point of view, an
operator-valued distribution. As a consequence, the operator $T_{\mu
\nu}$ is ill-defined and the associated expectation value $\langle
\psi |T_{\mu \nu} |\psi \rangle$ is formally infinite. To deal with
such a difficulty, renormalization is required. Much work has been
done since the mid-1970s in order to renormalize the stress-energy
tensor and/or to extract from the expectation value $\langle \psi
|T_{\mu \nu} |\psi \rangle$ a finite and physically acceptable
contribution which could act as the source in the semiclassical
Einstein equations (\ref{SCEinsteinEq}) (see
Ref.~\cite{BirrellDavies} for the state of affairs of the literature
concerning this subject before 1982). Among all the methods
employed, the axiomatic approach introduced by Wald \cite{Wald77} is
certainly the most general and the most powerful. It is an extension
of the ``point-splitting method"
\cite{DeWitt75,Christensen1,Christensen2} and it has been developed
in connection with the Hadamard representation of the Green
functions by Wald \cite{Wald77,Wald78}, Adler, Lieberman and Ng
\cite{AdlerETAL1,AdlerETAL2}, Brown and Ottewill
\cite{BrownOttewill83} and Castagnino and Harari
\cite{Castagnino84}. We refer to the monographs of Fulling
\cite{Fulling89} and Wald \cite{Wald94} for rigorous presentations
of this approach which is usually called Hadamard renormalization.
It permitted us to obtain, in the most general context, the explicit
expressions of the renormalized expectation value of the
stress-energy tensor
 for the scalar field theory
\cite{BrownOttewill86,BernardFolacci86,Tadaki87} but also for some
gauge theories such as i) electromagnetism \cite{BrownOttewill86},
ii) quantum gravity at one-loop order \cite{AFO} (here the theories
described by the standard effective action as well as by the
reparametrization-invariant effective action of Vilkovsky and DeWitt
were both considered) and iii) two- and three-form field theories
\cite{Folacci91} (in this context, the Hadamard formalism allowed us
to treat carefully the phenomenon of ghosts for ghosts).

Hadamard renormalization has been exclusively considered for field
theories defined on four-dimensional curved spacetimes. (However, it
should be noted that a recent work has been achieved in a
two-dimensional framework \cite{SalehiBisabr2001} but it is
incorrect due to a wrong expression for the Hadamard representation
of the Green functions.) According to the ``recent" physical
theories such as supergravity theories, string theories and
M-theory, which were developed in order to understand gravity in a
quantum framework and to provide a unified description of all the
fundamental interactions, we should live in a spacetime with more
dimensions than the four we observe, a scenario which is a
resurgence of the old Kaluza-Klein theory \cite{Kaluza,Klein}.
Because all the previously mentioned theories are still at an early
stage of development and are far from being well understood, it is
rather difficult to make predictions by using them directly. In
fact, people studying the consequences of supergravity and string
theories in cosmology or in black hole physics often develop
analysis based on semiclassical approximations or more precisely use
the methods of quantum field theory in curved spacetime taking into
account the extra dimensions. In this context, it seems to us
crucial to extend the powerful Hadamard renormalization procedure to
be able to deal, as generally as possible, with quantum fluctuations
and with their back reaction effects. In this paper, we shall take
some steps in this direction.

It is important to note that many recent articles have already been
devoted to the role as well as to the calculation of the expectation
value of the stress-energy tensor in the presence of extra spatial
dimensions. For example:

\quad -- In the context of the Randall-Sundrum braneworld models
\cite{RandallSundrum1,RandallSundrum2} introduced in order to solve
the hierarchy problem \cite{ADD98,AADD98,ADD99}, i.e. to eliminate
the large hierarchy between the electroweak scale and the gravity
scale. The vacuum expectation value of the stress-energy tensor and
the associated vacuum energy have been called upon to stabilize the
size of the extra dimensions. There is an extensive literature on
the subject. We refer more particularly to Ref.~\cite{KnapmanToms04}
where back reaction effects are in addition considered and to
Ref.~\cite{PujolasTanaka04} where cosmological considerations in
connection with the inflationary scenario are in addition discussed
(see also
Refs~\cite{SaharianSetare03,SaharianSetare04,Setare05,Saharian05,Saharian2007}
and references therein).

\quad -- In the context of the vacuum polarization induced by
topological defects such as monopoles
\cite{demello-2002-43,demello-2006-47,demello-2006-73} or cosmic
strings (see Ref.~\cite{Saharian2007} and references therein).

\quad -- In the context of the AdS/CFT correspondence
\cite{Maldacena98,Witten98,GubserETAL98} which asserts the existence
of a duality between a theory of gravity in the $(D+1)$-dimensional
anti-de Sitter space and a conformal field theory living on its
$D$-dimensional boundary (for a review see
Ref.~\cite{AharonyMOO2000}) and which could provide a concrete
realization of the holographic principle \cite{tHooft93,Susskind95}.
A new renormalization procedure, the so-called holographic
renormalization,  has been developed. More precisely, it has been
shown that the regularized expectation value of the stress-energy
tensor corresponding to the conformal field theory living on the
boundary can be obtained from the ``regularized" action of the
gravitational field living in the bulk
\cite{HenningsonSkenderis1998,deHaroSkenderisSolodukhin2001} (see
also for a review Ref.~\cite{Skenderis2002} as well as references
therein for complements and
Refs~\cite{NojiriOdintsov1998,BalasubramanianKraus1999b,KrausLarsenSiebelink1999,Myers1999,
NojiriOdintsovOgushi2000,PapadimitriouSkenderis2004,
HollandsIshibashiMarolf2005a,HollandsIshibashiMarolf2005b,PapadimitriouSkenderis2005,
KrausLarsen2006,AstefaneseiRadu2006} for related approaches as well
as extensions). The counterterm substraction technique developed in
this context permits us to obtain the stress-energy tensor, at large
distance, for higher-dimensional black holes such as
$\mathrm{Kerr}$-$\mathrm{AdS}_5$, $\mathrm{Kerr}$-$\mathrm{AdS}_6$
and $\mathrm{Kerr}$-$\mathrm{AdS}_7$
\cite{AwadJohnson2000,AwadJohnson2001}.

\quad -- In the context of the validity of semiclassical gravity but
also of the avoidance of the singularities predicted by the
singularity theorems of Hawking and Penrose \cite{ford2003}.
Fluctuations of the stress-energy tensor induce Ricci curvature
fluctuations (see, for example, Ref.~\cite{BorgmanFord2004a}) or in
other words fluctuations of the gravitational field itself. The
existence of these fluctuations places limits on the validity of
semiclassical gravity but also could lead to important effects on
the focusing of a bundle of timelike or null geodesics. The study of
such fluctuations in the presence of compact extra spatial
dimensions has been discussed more particularly in
Ref.~\cite{BorgmanFord2004b}.

\noindent All these works have however been carried out under very
strong hypotheses: flat (or conformally flat) spacetimes with
extra-dimensions or maximally (or asymptotically maximally)
symmetric spacetimes as well as massless or conformally invariant
field theories. Of course, it is necessary, from a physical point of
view, to be able to deal with situations presenting a lower degree
of symmetry. With this aim in view, the Hadamard renormalization
procedure could be very helpful.

Finally, it should be noted that some mathematical aspects of the
Hadamard renormalization procedure for a scalar field in a general
``spacetime" of arbitrary dimension have been already considered by
Moretti in a series of recent articles
\cite{MorettiCMP1999a,MorettiCMP1999b,MorettiJMP1999,
MorettiCMP2000,MorettiCMP2003}. He has provided a rigorous proof of
the symmetry of the off-diagonal Hadamard coefficients, i.e. of the
coefficients corresponding to the short-distance divergent part of
the Hadamard representation of the Green functions for the Euclidean
and Lorentzian scalar field theories
\cite{MorettiCMP1999b,MorettiCMP2000}. He has also established a
connection between the zeta- and Hadamard- regularization procedures
in the Euclidean framework \cite{MorettiCMP1999a,MorettiJMP1999} and
he has finally discussed the possible elimination of the ambiguities
plaguing the Hadamard renormalization procedure by using microlocal
analysis in the context of the algebraic approach to quantum field
theory \cite{MorettiCMP2003}. In fact, the results we present in
this article are very different from those of Moretti. We do not
focus our attention on the mathematical aspects of Hadamard
renormalization as he did but on its practical aspects: from our
results, the interested reader should be able to obtain explicitly
the renormalized expression of the expectation value with respect to
a given state $|\psi \rangle$ of the stress-energy-tensor operator
associated with the scalar field theory if he knows (exactly or
asymptotically in a sense defined below) the Feynman propagator
corresponding to $|\psi \rangle$. With this aim in view, we have
provided in Sec.~III a step-by-step guide for the reader who simply
wishes to calculate this regularized expectation value and is not
specially interested in following the derivation of all our results.

Our article is organized as follows. In Sec.~II, we develop as
generally as possible the Hadamard renormalization of the
stress-energy tensor associated with a massive scalar field theory
defined on a general spacetime of arbitrary dimension. In Sec.~III,
we explicitly describe this procedure for arbitrary spacetimes of
dimension from 2 to 6. This is done by using recent results we
obtained in Ref.~\cite{DecaniniFolacci2005a} and which concern the
covariant Taylor series expansions of the Hadamard coefficients. For
spacetime dimension from 7 to 11, we provide the framework
permitting the interested reader to perform this regularization
procedure explicitly in a given spacetime. In Sec.~IV, we complete
our study (i) by considering the ambiguities of the Hadamard
renormalization of the stress-energy tensor and the corresponding
ambiguities for the trace anomaly, (ii) by providing the expressions
of the gravitational counterterms involved in the renormalization
process (iii) by discussing the connections between Hadamard
renormalization and renormalization in the effective action.
Finally, in Sec.~V, we briefly discuss possible extensions of our
work as well as possible applications. In a short appendix, we
provide the traces of various conserved local tensors of rank 2 and
orders 4 and 6. These results are more particularly helpful in order
to discuss the ambiguity problem for the trace anomaly considered in
Sec.~IV.

In this paper, we use units with $\hbar=c=1$ and the geometrical
conventions of Hawking and Ellis \cite{HawkingEllis} concerning the
definitions of the scalar curvature $R$, the Ricci tensor $R_{\mu
\nu}$ and the Riemann tensor $R_{\mu \nu \rho \sigma}$.  We also
extensively use the commutation of covariant derivatives in the form
\begin{eqnarray}\label{CD_NabNabTensor}
&  & T^{\rho \dots}_{\phantom{\rho} \sigma \dots ;\nu \mu} -
T^{\rho \dots}_{\phantom{\rho} \sigma \dots ;\mu \nu} = \nonumber \\
&   & \qquad  + R^{\rho}_{\phantom{\rho} \tau \mu \nu} T^{\tau
\dots}_{\phantom{\tau} \sigma \dots } + \dots -
R^{\tau}_{\phantom{\tau} \sigma \mu \nu} T^{\rho
\dots}_{\phantom{\rho} \tau \dots} - \dots
\end{eqnarray}
It is furthermore important to note that all the results we provide
in Secs.~III and IV are given on irreducible forms: indeed, by using
some of the geometrical identities displayed in our recent
unpublished report \cite{DecaniniFolacci_arXiv2008}, our results
have been systematically expanded on the standard bases constructed
from group theoretical considerations which have been proposed by
Fulling, King, Wybourne and Cummings (FKWC) in Ref.~\cite{FKWC1992}.
A reader who would like to follow or to check our calculations is
invited to have in hand these two papers and more particularly
Ref.~\cite{DecaniniFolacci_arXiv2008} which displays, in addition to
a list of useful geometrical identities, the slightly modified
version of the FKWC-bases we used in the present article.

\section{Hadamard renormalized stress-energy tensor: General considerations}

In this section, we shall describe from a general point of view the
renormalization of the stress-energy tensor associated with a
massive scalar field theory defined on a general spacetime of
arbitrary dimension $D \ge 2$. We shall assume that the scalar field
is in a normalized quantum state of Hadamard type and we shall
consider that the Wald's axiomatic approach (see
Refs.~\cite{Wald77,Wald78,Wald94}) developed in the four-dimensional
framework remains valid in the $D$-dimensional one. We shall in fact
extend various considerations previously developed in the
four-dimensional framework (see
Refs.~\cite{Wald77,AdlerETAL1,AdlerETAL2,Wald78,BrownOttewill83,Castagnino84,
BrownOttewill86,BernardFolacci86,Tadaki87,AFO,Folacci91}).

\subsection{Some aspects of the classical theory}

We begin by reviewing the classical theory of a ``free" massive
scalar field $\Phi $ propagating on a $D$-dimensional curved
spacetime $({\cal M},g_{\mu \nu})$ in order to emphasize some
results which shall play a crucial role at the quantum level. We
first recall that the associated action is given by
\begin{equation}\label{Action}
S=-\frac{1}{2}\int _{\cal M} d^D x\sqrt{-g}\left( g^{\mu
\nu}\Phi_{;\mu}\Phi_{;\nu} + m^2\Phi^2 + \xi R\Phi^2 \right)
\end{equation}
where $m$ is the mass of the scalar field and $\xi$ is a
dimensionless factor which accounts for the possible coupling
between the scalar field and the gravitational background. We
furthermore assume that spacetime has no boundary, i.e., that
$\partial M = \emptyset$. $S$ is a functional of the scalar field
$\Phi $ and of the gravitational field $g_{\mu \nu}$, i.e.
$S=S\left[\Phi, g_{\mu \nu} \right]$. The functional derivative of
$S$ with respect to $\Phi $ is given by
\begin{equation}\label{DerFunctprp}
\frac{\delta S}{\delta \Phi }=\sqrt{-g}\left( \Box -m^2 -\xi R
\right) \Phi
\end{equation}
and its extremization provides the wave (Klein-Gordon) equation
\begin{equation}\label{WEQ}
\left( \Box -m^2 -\xi R \right) \Phi =0.
\end{equation}
The functional derivative of $S$ with respect to $g_{\mu \nu}$
permits us to define the stress-energy tensor $T_{\mu \nu}$
associated with the scalar field $\Phi $ (see, for example,
Ref.~\cite{HawkingEllis}). Indeed, we have
\begin{equation}\label{STclDEF} T_{\mu \nu
}=\frac{2}{\sqrt{-g}} \frac{\delta} {\delta g^{\mu \nu
}}S\left[\Phi, g_{\mu \nu} \right]
\end{equation}
and by using that in the variation
\begin{equation}\label{varTensMet1}
g_{\mu \nu } \to g_{\mu \nu } + \delta g_{\mu \nu }
\end{equation}
of the metric tensor we have (see, for example,
Ref.\cite{ChristensenBarth83})
\begin{subequations} \label{varTensMet2}
\begin{eqnarray}
& & g^{\mu \nu } \to g^{\mu \nu } + \delta g^{\mu \nu } \label{varTensMet3a} \\
& &  \sqrt{-g}\to \sqrt{-g}+ \delta\sqrt{-g}\label{varTensMet3b}\\
& &  R \to R + \delta R  \label{varTensMet3c}
\end{eqnarray}
with
\begin{eqnarray}
& & \delta g^{\mu \nu }
= -g^{\mu \rho }g^{\nu \sigma }  \delta g_{\rho \sigma }   \label{varTensMet4a} \\
& &  \delta\sqrt{-g}=
\frac{1}{2} \sqrt{-g}g^{\mu \nu } \delta g_{\mu \nu }  \label{varTensMet4b}\\
& &  \delta R=-R^{\mu \nu}\delta g_{\mu \nu } +  \left(\delta g_{\mu
\nu }\right)^{; \mu \nu}- \left(g^{\mu \nu }\delta g_{\mu \nu
}\right)^{;\rho}_{\phantom{;\rho};\rho } \label{varTensMet4c}
\end{eqnarray}
\end{subequations}
we can explicitly find that
\begin{eqnarray}\label{STclEXP}
& & T_{\mu \nu }= (1-2\xi) \Phi_{;\mu}\Phi_{;\nu} + \left(2\xi
-\frac{1}{2}\right)g_{\mu \nu}
g^{\rho \sigma} \Phi_{;\rho}\Phi_{;\sigma}   \nonumber \\
&  & \quad -2\xi \Phi \Phi_{; \mu \nu} + 2\xi g_{\mu \nu} \Phi \Box
\Phi +\xi \left(R_{\mu \nu} - \frac{1}{2}g_{\mu \nu}R \right) \Phi^2   \nonumber \\
&  & \quad - \frac{1}{2}g_{\mu \nu}m^2 \Phi^2.
\end{eqnarray}

It is well-known that the stress-energy tensor is conserved, i.e. it
satisfies
\begin{equation}\label{STclCONSERV}
T^{\mu \nu}_{\phantom{\mu \nu};\nu}=0.
\end{equation}
This result could be obtained directly from the field equation
(\ref{WEQ}) by using the expression (\ref{STclEXP}). However, it is
more instructive from the physical point of view to derive it from
the invariance of the action (\ref{Action}) under spacetime
diffeomorphisms and therefore under the infinitesimal coordinate
transformation
\begin{equation}\label{TransInfCoord1}
x^\mu \to x^\mu + \epsilon^\mu \quad \mathrm{with}  \quad
|\epsilon^\mu| \ll 1.
\end{equation}
Indeed, under this transformation, the scalar field and the
background metric transform as
\begin{subequations} \label{TransInfCoord2}
\begin{eqnarray}
& & \Phi \to \Phi + \delta \Phi  \label{TransInfCoord3a}  \\
& & g_{\mu \nu} \to g_{\mu \nu} + \delta g_{\mu \nu}
\label{TransInfCoord3b}
\end{eqnarray}
with
\begin{eqnarray}
& & \delta \Phi = {\cal L}_{-\epsilon} \Phi =-\epsilon^{\mu}\Phi_{;\mu} \label{TransInfCoord4a}\\
& & \delta g_{\mu \nu} = {\cal L}_{-\epsilon} g_{\mu \nu}  = -
\epsilon_{\mu ;\nu} - \epsilon_{\nu ;\mu}  \label{TransInfCoord4b}
\end{eqnarray}
\end{subequations}
where ${\cal L}_{-\epsilon}$ denotes the Lie derivative with respect
to the vector $-\epsilon$. The invariance of the action
(\ref{Action}) leads to
\begin{equation}\label{ActionVAR}
\int _{\cal M} d^D x \left[\left(\frac{\delta S} {\delta \Phi}
\right)\delta \Phi  +  \left(\frac{\delta S} {\delta g_{\mu \nu }}
\right)\delta g_{\mu \nu} \right] =0
\end{equation}
which implies
\begin{equation}\label{STclCONSERV_var}
T^{\mu \nu}_{\phantom{\mu \nu};\nu}= \Phi^{;\mu} \left[\Box -m^2 -
\xi R \right]\Phi
\end{equation}
by using (\ref{TransInfCoord2}). Then, from (\ref{WEQ}) we obtain
immediately (\ref{STclCONSERV}).

It is also well-known that for
\begin{equation}\label{ICdef}
m^2=0 \quad \mathrm{and} \quad \xi= \xi_c(D)
\end{equation}
with
\begin{equation}\label{IC_xi_def}
\xi_c(D) =\frac{1}{4} \left(\frac{D-2}{D-1} \right)
\end{equation}
the stress-energy tensor is traceless, i.e. it satisfies
\begin{equation}\label{TraceSTclEXP}
T^{\mu}_{\phantom{\mu} \mu }= 0.
\end{equation}
This result could be obtained directly from the field equation
(\ref{WEQ}) by using the expression (\ref{STclEXP}). In fact, from
the physical point of view, it is more instructive to derive it by
noting that for the values of the parameters $m^2$ and $\xi$ given
by (\ref{ICdef}) the scalar field theory is conformally invariant
(see, for example, Appendix D of Ref.~\cite{Wald84}). As a
consequence, the action (\ref{Action}) is invariant under the
so-called conformal transformation
\begin{subequations} \label{TransConfGlob}
\begin{eqnarray}
& & \Phi \to {\hat \Phi} = \Omega^{(2-D)/2}  \Phi  \label{TransConfGlob_a}  \\
& & g_{\mu \nu} \to {\hat g}_{\mu \nu} = \Omega^2 g_{\mu \nu}
\label{TransConfGlob_b}
\end{eqnarray}
\end{subequations}
and therefore under the infinitesimal conformal transformation
\begin{subequations} \label{TransConfINF}
\begin{eqnarray}
& & \Phi \to {\hat \Phi} =  \Phi + \delta \Phi \label{TransConfINF_a}  \\
& & g_{\mu \nu} \to {\hat g}_{\mu \nu} =  g_{\mu \nu} + \delta
g_{\mu \nu} \label{TransConfINF_b}
\end{eqnarray}
with
\begin{eqnarray}
& & \delta \Phi = \frac{2-D}{2} \epsilon \, \Phi \label{TransConfINF_c}\\
& & \delta g_{\mu \nu} = 2 \epsilon \,  g_{\mu \nu}
\label{TransConfINF_d}
\end{eqnarray}
\end{subequations}
which corresponds to $\Omega =1+ \epsilon$ with $|\epsilon| \ll 1$.
The invariance of the action (\ref{Action}) leads to
(\ref{ActionVAR}) which now implies
\begin{equation}\label{TraceSTclEXP_var}
T^{\mu}_{\phantom{\mu} \mu }= \frac{D-2}{2} \, \Phi \left[\Box -
\xi_c(D) R \right] \Phi
\end{equation}
by using (\ref{TransConfINF}). Then, from (\ref{WEQ}) with
(\ref{ICdef}), we obtain immediately (\ref{TraceSTclEXP}).

\subsection{Hadamard quantum states and Feynman propagator}

From now on, we shall assume that the scalar field theory previously
described has been quantized and that the scalar field $\Phi$ is in
a normalized quantum state $|\psi \rangle$ of Hadamard type. The
associated Feynman propagator
\begin{equation}\label{FeynmanProp}
G^{\mathrm{F}}(x,x') = i \langle \psi | T \Phi(x) \Phi(x') |\psi
\rangle
\end{equation}
(here $T$ denotes time ordering) is, by definition, a solution of
\begin{equation}\label{WEQ_G1}
\left( \Box_x -m^2 -\xi R \right) G^{\mathrm{F}}(x,x') = -\delta^D
(x,x')
\end{equation}
with $\delta^D (x,x')= [-g(x)]^{-1/2}(x) \delta^D (x-x')$. It is
symmetric in the exchange of $x$ and $x'$ and its short-distance
behavior is of Hadamard type. Its precise form for $x'$ near $x$
depends on whether the dimension $D$ of spacetime is even or odd
(see Refs.~\cite{Hadamard,Garabedian,Friedlander} or the articles by
Moretti \cite{MorettiCMP1999a,MorettiCMP1999b,MorettiJMP1999,
MorettiCMP2000,MorettiCMP2003} as well as our recent article
\cite{DecaniniFolacci2005a} for more details). It involves the
geodetic interval $\sigma(x,x')$ and the biscalar form $\Delta
(x,x')$ of the Van Vleck-Morette determinant \cite{DeWitt65}. Here
we recall that $2\sigma(x,x')$ is a biscalar function which is
defined as the square of the geodesic distance between $x$ and $x'$
and which satisfies
\begin{equation}\label{DSrep4}
2 \sigma= \sigma^{; \mu}\sigma_{; \mu}.
\end{equation}
We have $\sigma(x,x') < 0$ if $x$ and $x'$ are timelike related,
$\sigma(x,x') = 0$ if $x$ and $x'$ are null related and
$\sigma(x,x') > 0$ if $x$ and $x'$ are spacelike related. We
furthermore recall that $\Delta (x,x')$ is given by
\begin{equation}\label{DSrep5}
\Delta (x,x')= -[-g(x)]^{-1/2} \mathrm{det} (- \sigma_{; \mu
\nu'}(x,x')) [-g(x')]^{-1/2}
\end{equation}
and satisfies the partial differential equation
\begin{subequations}\label{DSrep6}
\begin{equation}\label{DSrep6a}
\Box_x \sigma = D - 2 \Delta ^{-1/2}{\Delta ^{1/2}}_{;\mu} \sigma^{;
\mu}
\end{equation}
as well as the boundary condition
\begin{equation}\label{DSrep6b}
\lim_{x' \to x} \Delta (x,x')=1.
\end{equation}
\end{subequations}

For $D=2$, the Hadamard expansion of the Feynman propagator is given
by
\begin{eqnarray}\label{H2Rep1}
& & G^{\mathrm{F}} (x,x') = \frac{i \alpha_2}{2} \, ( V(x,x')
\ln [\sigma(x,x')+i\epsilon] \nonumber \\
& & \qquad \qquad \qquad \qquad \qquad + W(x,x') )
\end{eqnarray}
where $V(x,x')$ and $W(x,x')$ are symmetric biscalars, regular for
$x' \to x$ and which possess expansions of the form
\begin{subequations}\label{H2Rep2}
\begin{eqnarray}
& & V(x,x')= \sum_{n=0}^{+\infty} V_n(x,x')\sigma^n(x,x'), \label{H2Rep2b} \\
& & W(x,x')= \sum_{n=0}^{+\infty} W_n(x,x')\sigma^n(x,x').
\label{H2Rep2c}
\end{eqnarray}
\end{subequations}

For $D$ even with $D\not= 2$, the Hadamard expansion of the Feynman
propagator is given by
\begin{eqnarray}\label{HevRep1}
& & G^{\mathrm{F}} (x,x') = \frac{i \alpha_D}{2} \,
\left(\frac{U(x,x')}{[\sigma(x,x')+i\epsilon]^{D/2-1}} \right. \nonumber \\
& &  \left. \phantom{\frac{U}{\sigma^{d}}} + V(x,x') \ln
[\sigma(x,x')+i\epsilon] + W(x,x') \right)
\end{eqnarray}
where $U(x,x')$, $V(x,x')$ and $W(x,x')$ are symmetric biscalars,
regular for $x' \to x$ and which possess expansions of the form
\begin{subequations}\label{HevRep2}
\begin{eqnarray}
& & U(x,x')= \sum_{n=0}^{D/2-2} U_n(x,x')\sigma^n(x,x'), \label{HevRep2a} \\
& & V(x,x')= \sum_{n=0}^{+\infty} V_n(x,x')\sigma^n(x,x'), \label{HevRep2b} \\
& & W(x,x')= \sum_{n=0}^{+\infty} W_n(x,x')\sigma^n(x,x').
\label{HevRep2c}
\end{eqnarray}
\end{subequations}

For $D$ odd, the Hadamard expansion of the Feynman propagator is
given by
\begin{eqnarray}\label{HodRep1}
& & G^{\mathrm{F}} (x,x') = \frac{i \alpha_D}{2} \,
\left(\frac{U(x,x')}{[\sigma(x,x')+i\epsilon]^{D/2-1}} + W(x,x')
\right) \nonumber \\
\end{eqnarray}
where $U(x,x')$ and $W(x,x')$ are again symmetric and regular
biscalar functions which now possess expansions of the form
\begin{subequations}\label{HodRep2}
\begin{eqnarray}
& & U(x,x')= \sum_{n=0}^{+\infty} U_n(x,x')\sigma^n(x,x'), \label{Hodrep2a} \\
& & W(x,x')= \sum_{n=0}^{+\infty} W_n(x,x')\sigma^n(x,x').
\label{Hodrep2b}
\end{eqnarray}
\end{subequations}

In Eqs.~(\ref{H2Rep1}), (\ref{HevRep1}) and (\ref{HodRep1}), the
coefficient $\alpha_D$ is given by
\begin{equation}\label{Coeff_alpha}
\alpha_D= \left\{ \begin{array}{cl} 1/(2\pi) & \mathrm{for} \ D=2, \\
\Gamma(D/2-1)/(2\pi)^{D/2} & \mathrm{for} \ D\not=2,
\end{array} \right.
\end{equation}
while the factor $i\epsilon$ with $\epsilon \to 0_+$ is introduced
to give to $G^{\mathrm{F}} (x,x')$ a singularity structure that is
consistent with the definition of the Feynman propagator as a
time-ordered product (see Eq.~(\ref{FeynmanProp})).

For $D=2$, the Hadamard coefficients $V_n(x,x')$ and $W_n(x,x')$ are
symmetric and regular biscalar functions. The coefficients
$V_n(x,x')$ satisfy the recursion relations
\begin{subequations}\label{H2Rep5b}
\begin{eqnarray}\label{H2Rep5c}
& & 2(n+1)^2 V_{n+1} + 2(n+1) V_{n+1 ; \mu} \sigma ^{; \mu}
\nonumber \\
&  &  \quad  - 2(n+1) V_{n+1} \Delta ^{-1/2}{\Delta
^{1/2}}_{;\mu} \sigma^{; \mu}   \nonumber \\
&  &  \quad  + \left( \Box_x -m^2 -\xi R \right) V_n =0  \quad
\mathrm{for}~n \in \mathbb{N}
\end{eqnarray}
with the boundary condition
\begin{eqnarray}\label{H2Rep5d}
& & V_0  =- \Delta ^{1/2}.
\end{eqnarray}
\end{subequations}
The coefficients $W_n(x,x')$  satisfy the recursion relations
\begin{eqnarray}\label{H2Rep6}
& & 2(n+1)^2 W_{n+1} + 2(n+1) W_{n+1 ; \mu} \sigma ^{; \mu}
\nonumber \\
&  &  \quad  - 2(n+1) W_{n+1} \Delta ^{-1/2}{\Delta
^{1/2}}_{;\mu} \sigma^{; \mu}   \nonumber \\
& &  \quad + 4(n+1) V_{n+1} + 2 V_{n+1 ; \mu} \sigma ^{; \mu}
\nonumber \\
&  &  \quad  - 2 V_{n+1} \Delta ^{-1/2}{\Delta
^{1/2}}_{;\mu} \sigma^{; \mu}   \nonumber \\
&  &  \quad  + \left( \Box_x -m^2 -\xi R \right) W_n =0 \quad
\mathrm{for}~n \in \mathbb{N}.
\end{eqnarray}
From the recursion relations (\ref{H2Rep5c}) and (\ref{H2Rep6}), the
boundary condition (\ref{H2Rep5d}) and the relations (\ref{DSrep4})
and (\ref{DSrep6}) it is possible to prove that $G^{\mathrm{F}}
(x,x')$ given by (\ref{H2Rep1})-(\ref{H2Rep2}) solves the wave
equation (\ref{WEQ_G1}). This can be done easily by noting that we
have
\begin{equation}\label{H2Rep_EQ_V}
\left( \Box_x -m^2 -\xi R \right) V =0
\end{equation}
as a consequence of (\ref{H2Rep5b}) and
\begin{eqnarray}\label{H2Rep_EQ_W}
& & \sigma \left( \Box_x -m^2 -\xi R \right) W = \nonumber \\
& & \qquad \qquad - 2 V_{; \mu} \sigma ^{; \mu} + 2 V \Delta
^{-1/2}{\Delta ^{1/2}}_{;\mu} \sigma^{; \mu}
\end{eqnarray}
as a consequence of (\ref{H2Rep5d}) and (\ref{H2Rep6}).

For $D$ even with $D\not= 2$, the Hadamard coefficients $U_n(x,x')$,
$V_n(x,x')$ and $W_n(x,x')$ are symmetric and regular biscalar
functions. The coefficients $U_n(x,x')$ satisfy the recursion
relations
\begin{subequations}\label{HevRep5A}
\begin{eqnarray}\label{HevRep5a}
& & (n+1)(2n+4-D) U_{n+1} + (2n+4-D) U_{n+1 ; \mu} \sigma ^{; \mu}
\nonumber \\
&  &  \quad  - (2n+4-D) U_{n+1} \Delta ^{-1/2}{\Delta
^{1/2}}_{;\mu} \sigma^{; \mu}   \nonumber \\
&  &  \quad  + \left( \Box_x -m^2 -\xi R \right) U_n =0 \nonumber \\
& & \qquad \qquad\qquad\qquad \mathrm{for}~ n=0,1, \dots , D/2-3
\end{eqnarray}
with the boundary condition
\begin{equation}\label{HevRep5b}
U_0= \Delta ^{1/2}.
\end{equation}
\end{subequations}
The coefficients $V_n(x,x')$ satisfy the recursion relations
\begin{subequations}\label{HevRep5B}
\begin{eqnarray}\label{HevRep5c}
& & (n+1)(2n+D) V_{n+1} + 2(n+1) V_{n+1 ; \mu} \sigma ^{; \mu}
\nonumber \\
&  &  \quad  - 2(n+1) V_{n+1} \Delta ^{-1/2}{\Delta
^{1/2}}_{;\mu} \sigma^{; \mu}   \nonumber \\
&  &  \quad  + \left( \Box_x -m^2 -\xi R \right) V_n =0  \quad
\mathrm{for}~n \in \mathbb{N}
\end{eqnarray}
with the boundary condition
\begin{eqnarray}\label{HevRep5d}
& & (D-2) V_0 + 2 V_{0 ; \mu} \sigma ^{; \mu} - 2 V_0 \Delta
^{-1/2}{\Delta
^{1/2}}_{;\mu} \sigma^{; \mu}   \nonumber \\
&  &  \quad  + \left( \Box_x -m^2 -\xi R \right) U_{D/2-2} =0.
\end{eqnarray}
\end{subequations}
The coefficients $W_n(x,x')$  satisfy the recursion relations
\begin{eqnarray}\label{HevRep6}
& & (n+1)(2n+D) W_{n+1} + 2(n+1) W_{n+1 ; \mu} \sigma ^{; \mu}
\nonumber \\
&  &  \quad  - 2(n+1) W_{n+1} \Delta ^{-1/2}{\Delta
^{1/2}}_{;\mu} \sigma^{; \mu}   \nonumber \\
& &  \quad + (4n+2+D) V_{n+1} + 2 V_{n+1 ; \mu} \sigma ^{; \mu}
\nonumber \\
&  &  \quad  - 2 V_{n+1} \Delta ^{-1/2}{\Delta
^{1/2}}_{;\mu} \sigma^{; \mu}   \nonumber \\
&  &  \quad  + \left( \Box_x -m^2 -\xi R \right) W_n =0 \quad
\mathrm{for}~n \in \mathbb{N}.
\end{eqnarray}
From the recursion relations (\ref{HevRep5a}), (\ref{HevRep5c}) and
(\ref{HevRep6}), the boundary conditions (\ref{HevRep5b}) and
(\ref{HevRep5d}) and the relations (\ref{DSrep4}) and (\ref{DSrep6})
it is possible to prove that $G^{\mathrm{F}} (x,x')$ given by
(\ref{HevRep1})-(\ref{HevRep2}) solves the wave equation
(\ref{WEQ_G1}). This can be done easily by noting that we have
\begin{equation}\label{HevRep_EQ_V}
\left( \Box_x -m^2 -\xi R \right) V =0
\end{equation}
as a consequence of (\ref{HevRep5c}) and
\begin{eqnarray}\label{HevRep_EQ_W}
& & \sigma \left( \Box_x -m^2 -\xi R \right) W = - \left( \Box_x
-m^2 -\xi R \right) U_{D/2-2} \nonumber \\
& & \quad -(D-2) V - 2 V_{; \mu} \sigma ^{; \mu} + 2 V \Delta
^{-1/2}{\Delta ^{1/2}}_{;\mu} \sigma^{; \mu}
\end{eqnarray}
as a consequence of (\ref{HevRep5d}) and (\ref{HevRep6}).

For $D$ odd, the Hadamard coefficients $U_n(x,x')$ and $W_n(x,x')$
are symmetric and regular biscalar functions. The coefficients
$U_n(x,x')$ satisfy the recursion relations
\begin{subequations}\label{HodRep3}
\begin{eqnarray}\label{HodRep3a}
& & (n+1)(2n+4-D) U_{n+1} + (2n+4-D) U_{n+1 ; \mu} \sigma ^{; \mu}
\nonumber \\
&  &  \quad  - (2n+4-D) U_{n+1} \Delta ^{-1/2}{\Delta
^{1/2}}_{;\mu} \sigma^{; \mu}   \nonumber \\
&  &  \quad  + \left( \Box_x -m^2 -\xi R \right) U_n =0 \quad
\mathrm{for}~ n \in \mathbb{N}
\end{eqnarray}
with the boundary condition
\begin{equation}\label{HodRep3b}
U_0= \Delta ^{1/2}.
\end{equation}
\end{subequations}
The coefficients $W_n(x,x')$ satisfy the recursion relations
\begin{eqnarray}\label{HodRep4}
& & (n+1)(2n+D) W_{n+1} + 2(n+1) W_{n+1 ; \mu} \sigma ^{; \mu}
\nonumber \\
&  &  \quad  - 2(n+1) W_{n+1} \Delta ^{-1/2}{\Delta ^{1/2}}_{;\mu}
\sigma^{; \mu}
\nonumber \\
&  &  \quad  + \left( \Box_x -m^2 -\xi R \right) W_n =0 \quad
\mathrm{for}~ n \in \mathbb{N}.
\end{eqnarray}
From the recursion relations (\ref{HodRep3a}) and (\ref{HodRep4}),
the boundary conditions (\ref{HodRep3b}) and the relations
(\ref{DSrep4}) and (\ref{DSrep6}) it is possible to prove that
$G^{\mathrm{F}} (x,x')$ given by (\ref{HodRep1})-(\ref{HodRep2})
solves the wave equation (\ref{WEQ_G1}). This can be done easily
from
\begin{equation}\label{HodRep_EQ_W}
\left( \Box_x -m^2 -\xi R \right) W = 0
\end{equation}
which is a consequence of (\ref{HodRep4}).

For $D=2$, the Hadamard coefficients $V_n(x,x')$ can be formally
obtained by integrating the recursion relations (\ref{H2Rep5c})
along the unique geodesic joining $x$ to $x'$ (it is unique for $x'$
near $x$ or more generally for $x'$ in a convex normal neighborhood
of $x$). Similarly, for $D$ even with $D\not=2$, the Hadamard
coefficients $U_n(x,x')$ and $V_n(x,x')$ can be obtained by
integrating the recursion relations (\ref{HevRep5a}) and
(\ref{HevRep5c}) along the unique geodesic joining $x$ to $x'$
while, for $D$ odd, the Hadamard coefficients $U_n(x,x')$ can be
obtained by integrating the recursion relations (\ref{HodRep3a})
along the unique geodesic joining $x$ to $x'$. As a consequence, all
these Hadamard coefficients are determined uniquely and are purely
geometrical objects, i.e. they only depend on the geometry along the
geodesic joining $x$ to $x'$. By contrast, the Hadamard coefficients
$W_n(x,x')$ with $n \in \mathbb{N}$ are neither uniquely defined nor
purely geometrical. Indeed, the first coefficient of this sequence,
i.e. $W_0(x,x')$, is unrestrained by the recursion relations
(\ref{H2Rep6}) for $D=2$, (\ref{HevRep6}) for $D$ even with
$D\not=2$ and (\ref{HodRep4}) for $D$ odd. As a consequence, this is
also true for all the $W_n(x,x')$ with $n \ge 1$. This arbitrariness
is in fact very interesting and it can be used to encode the quantum
state dependence in the biscalar $W(x,x')$ by specifying the
Hadamard coefficient $W_0(x,x')$. Once it has been specified, the
recursion relations (\ref{H2Rep6}), (\ref{HevRep6}) or
(\ref{HodRep4}) uniquely determine the coefficients $W_n(x,x')$ with
$n \ge 1$ and therefore the biscalar $W(x,x')$. In other words, the
Hadamard expansions (\ref{H2Rep1})-(\ref{H2Rep2}),
(\ref{HevRep1})-(\ref{HevRep2}) and (\ref{HodRep1})-(\ref{HodRep2})
comprise a purely geometrical part, divergent for $x' \to x$ and
given by
\begin{eqnarray}\label{H2Rep1sing}
& & G^{\mathrm{F}}_{\mathrm{sing}} (x,x') = \frac{i \alpha_2}{2} \,
\left( V(x,x') \ln [\sigma(x,x')+i\epsilon] \right)
\end{eqnarray}
for $D=2$, by
\begin{eqnarray}\label{HevRep1sing}
& & G^{\mathrm{F}}_{\mathrm{sing}} (x,x') = \frac{i \alpha_D}{2} \,
\left(\frac{U(x,x')}{[\sigma(x,x')+i\epsilon]^{D/2-1}} \right. \nonumber \\
& &  \left. \phantom{\frac{U}{\sigma^{D}}} + V(x,x') \ln
[\sigma(x,x')+i\epsilon] \right)
\end{eqnarray}
for $D$ even with $D\not=2$ and by
\begin{eqnarray}\label{HodRep1sing}
& & G^{\mathrm{F}}_{\mathrm{sing}} (x,x') = \frac{i \alpha_D}{2} \,
\left(\frac{U(x,x')}{[\sigma(x,x')+i\epsilon]^{D/2-1}} \right)
\nonumber \\
\end{eqnarray}
for $D$ odd as well as a regular state-dependent part given by
\begin{eqnarray}\label{HevRepREGULAR}
& & G^{\mathrm{F}}_{\mathrm{reg}} (x,x') = \frac{i \alpha_D}{2} \,
W(x,x').
\end{eqnarray}

It should be noted that, bearing in mind practical applications, it
is very interesting to replace the Hadamard coefficients by their
covariant Taylor series expansions. Here, we shall provide some
associated results which will be helpful afterwards. As far as the
geometrical Hadamard coefficients $U_n(x,x')$ and $V_n(x,x')$ which
determine the singular part of the Feynman propagator are concerned,
they are usually obtained by looking for the solutions of the
recursion relations defining them as covariant Taylor series
expansions for $x'$ near $x$ given by
\begin{subequations}
\begin{eqnarray}
& & U_n(x,x')=u_n(x) +\sum_{p=1}^{+\infty} \frac{(-1)^p}{p!} u_{n \,
(p)}(x,x') \label{CTSExpU1} \\
& & V_n(x,x')=v_n(x) +\sum_{p=1}^{+\infty} \frac{(-1)^p}{p!} v_{n \,
(p)}(x,x') \label{CTSExpV1}
\end{eqnarray}
where the $u_{n \, (p)}(x,x')$ and $v_{n \, (p)}(x,x')$ with $p=1,2,
\dots $ are all biscalars in $x$ and $x'$ which are of the form
\begin{eqnarray}
&  & u_{n \, (p)}(x,x')=u_{n \,\, a_1 \dots a_p}(x)
\sigma^{;a_1}(x,x') \dots \sigma^{;a_p}(x,x') \nonumber \\
& & \label{CTSExpU2} \\
&  & v_{n \, (p)}(x,x')=v_{n \,\, a_1 \dots a_p}(x)
\sigma^{;a_1}(x,x') \dots \sigma^{;a_p}(x,x'). \nonumber
\\
& & \label{CTSExpV2}
\end{eqnarray}
\end{subequations}
This method, due to DeWitt \cite{DeWittBrehme,DeWitt65}, has been
used in the four-dimensional framework to construct the covariant
Taylor series expansions of $U_0(x,x')$,  $V_0(x,x')$ and
$V_1(x,x')$ (see, for example, Ref.~\cite{BrownOttewill86} and
references therein for the scalar field). In
Ref.~\cite{DecaniniFolacci2005a}, we have recently discussed the
construction of the expansions of the geometrical Hadamard
coefficients $U_n(x,x')$ and $V_n(x,x')$ of lowest orders in the
$D$-dimensional framework (with $D \ge 3$) and we intend to use
these results later. The case $D=2$ has not been explicitly treated
in Ref.~\cite{DecaniniFolacci2005a} but a comparison of Eq.~(23) of
Ref.~\cite{DecaniniFolacci2005a} with (\ref{H2Rep5b}) permits us to
express the geometrical Hadamard coefficients $V_n(x,x')$ in terms
of the mass-dependent DeWitt-coefficients ${\tilde{A}}_n(m^2;x,x')$
\cite{DecaniniFolacci2005a}. We have $V_n(x,x')= -((-1)^n/(2^n n!))
\, {\tilde{A}}_n(m^2;x,x')$ and this relation together with the
covariant Taylor series expansions of the mass-dependent
DeWitt-coefficients obtained in Ref.~\cite{DecaniniFolacci2005a}
provide the covariant Taylor series expansions of the geometrical
Hadamard coefficients $V_n(x,x')$ of lowest orders for $D=2$.

As far as the biscalar $W(x,x')$ which encodes the state-dependence
of the Feynman propagator is concerned, its covariant Taylor series
expansion is written as
\begin{subequations}
\begin{equation}
W (x,x')=w(x) +\sum_{p=1}^{+\infty} \frac{(-1)^p}{p!} w_{(p)}(x,x')
\label{CTSExpW1}
\end{equation} where the $w_{(p)}(x,x')$ with $p=1,2, \dots $ are
all biscalars in $x$ and $x'$
which are of the form
\begin{eqnarray}
&  & w_{(p)}(x,x')=w_{a_1 \dots a_p}(x)
\sigma^{;a_1}(x,x') \dots \sigma^{;a_p}(x,x'). \nonumber \\
& & \label{CTSExpW2}
\end{eqnarray}
\end{subequations}
The coefficients $w(x)$ and $w_{a_1 \dots a_p}(x)$ with $p=1,2,
\dots $ are constrained by the symmetry of $W (x,x')$ in the
exchange of $x$ and $x'$ as well as by the wave equations
(\ref{H2Rep_EQ_W}), (\ref{HevRep_EQ_W}) or (\ref{HodRep_EQ_W})
according to the values of $D$. The symmetry of $W(x,x')$ permits us
to express the odd coefficients of the covariant Taylor series
expansion of $W(x,x')$ in terms of the even ones. We have for the
odd coefficients of lowest orders (see, for example,
Refs.~\cite{BrownOttewill86,BernardFolacci86} or
Ref.~\cite{DecaniniFolacci2005a})
\begin{subequations} \label{AppConstrF_2}
\begin{eqnarray}
& & w_{a_1}=(1/2) \, w_{; a_1} \label{AppConstrF_2a} \\
& & w_{a_1 a_2 a_3}= (3/2) \, w_{(a_1 a_2; a_3)} - (1/4) \, w_{;
(a_1 a_2 a_3)}. \label{AppConstrF_2b}
\end{eqnarray}
\end{subequations}
The wave equation satisfied by $W(x,x')$ for $D$ even permits us to
write
\begin{eqnarray}\label{HevRep_EQ_W2}
& & \left(\Box_x -m^2 -\xi R \right) W \nonumber \\
& & \qquad \quad = - (D+2) V_1 - 2 V_{1 \, ; \mu} \sigma ^{; \mu} +
O \left(\sigma \right).
\end{eqnarray}
This relation is valid for $D=2$ as well as for $D$ even with
$D\not=2$. It is obtained from (\ref{H2Rep_EQ_W}) or
(\ref{HevRep_EQ_W}) by using (\ref{H2Rep2b}) and (\ref{H2Rep5d}) or
(\ref{HevRep2b}) and (\ref{HevRep5d}) as well as the following two
expansions (see, for example, Refs.~\cite{Christensen1,Christensen2}
or Ref.~\cite{DecaniniFolacci2005a})
\begin{equation}
\Delta^{1/2} = 1 + (1/12)  \, R_{a_1a_2} \sigma^{;a_1}\sigma^{;a_2}
+ O \left(\sigma^{3/2} \right) \label{App3b}
\end{equation}
and
\begin{equation}
\sigma_{; \mu \nu} = g_{\mu \nu} - (1/3)  \, R_{\mu a_1 \nu a_2}
\sigma^{;a_1}\sigma^{;a_2} + O \left(\sigma^{3/2} \right).
\label{App2a}
\end{equation}
Then, by inserting the expansion of $V_1(x,x')$ given by
(\ref{CTSExpV1}) and (\ref{CTSExpV2}) and by using (\ref{App2a}), we
have
\begin{eqnarray}\label{HevRep_EQ_W3}
& & \left( \Box_x -m^2 -\xi R \right) W \nonumber \\
& & \qquad \quad = - (D+2) v_1 + (D/2) \, v_{1 \, ; \mu} \sigma ^{;
\mu}   + O \left(\sigma \right).
\end{eqnarray}
By inserting the expansion (\ref{CTSExpW1})-(\ref{CTSExpW2}) of
$W(x,x')$ up to order $\sigma^{3/2}$ into the left-hand side of
(\ref{HevRep_EQ_W3}) and by using (\ref{AppConstrF_2}) as well as
(\ref{App2a}) we find that
\begin{subequations}
\begin{eqnarray}\label{HevRep_EQ_W4}
& & w^\rho_{\phantom{\rho} \rho} = (m^2 + \xi R)w  -(D+2) v_1  \label{HevRep_EQ_W4a}\\
& & w^\rho_{\phantom{\rho} a ;\rho} = (1/4) \, (\Box w)_{;a} +(1/2)
\, w^\rho_{\phantom{\rho} \rho ;a}
+ (1/2) R^\rho_{\phantom{\rho} a}w_{;\rho} \nonumber \\
& & \qquad \quad -(1/2) \, (m^2 + \xi R) w_{;a} + (D/2) \, v_{1 \, ;
a} \label{HevRep_EQ_W4b}
\end{eqnarray}
\end{subequations}
and by combining (\ref{HevRep_EQ_W4a}) and (\ref{HevRep_EQ_W4b}) we
establish another relation
\begin{eqnarray}\label{HevRep_EQ_W5}
& & w^\rho_{\phantom{\rho} a ;\rho} = (1/4) \, (\Box w)_{;a}
+ (1/2) R^\rho_{\phantom{\rho} a}w_{;\rho} \nonumber \\
& & \qquad \qquad +(1/2) \, \xi R_{;a}w - v_{1 \, ; a}
\end{eqnarray}
which will be helpful in the next subsection. The wave equation
(\ref{HodRep_EQ_W}) satisfied by $W(x,x')$ for $D$ odd can be worked
in the same manner. It leads to
\begin{subequations}
\begin{eqnarray}\label{HodRep_EQ_W4}
& & w^\rho_{\phantom{\rho} \rho} = (m^2 + \xi R)w  \label{HodRep_EQ_W4a}\\
& & w^\rho_{\phantom{\rho} a ;\rho} = (1/4) \, (\Box w)_{;a} +(1/2)
\, w^\rho_{\phantom{\rho} \rho ;a}
+ (1/2) R^\rho_{\phantom{\rho} a}w_{;\rho} \nonumber \\
& & \qquad \quad -(1/2) \, (m^2 + \xi R) w_{;a}
\label{HodRep_EQ_W4b}
\end{eqnarray}
\end{subequations}
and to
\begin{eqnarray}\label{HodRep_EQ_W5}
& & w^\rho_{\phantom{\rho} a ;\rho} = (1/4) \, (\Box w)_{;a}
+ (1/2) R^\rho_{\phantom{\rho} a}w_{;\rho} \nonumber \\
& & \qquad \qquad +(1/2) \, \xi R_{;a}w.
\end{eqnarray}

\subsection{Hadamard renormalization of the stress-energy tensor}

The expectation value with respect to the Hadamard quantum state
$|\psi \rangle$ of the stress-energy-tensor operator is formally
given as the limit
\begin{equation}\label{VExpVSET_unren}
\langle \psi |T_{\mu \nu}(x) |\psi \rangle = \lim_{x' \to x}
{\cal{T}}_{\mu \nu}(x,x') \left[ -i G^{\mathrm{F}} (x,x')\right]
\end{equation}
where $G^{\mathrm{F}} (x,x')$ is the Feynman propagator
(\ref{FeynmanProp}) which is assumed to possess one of the Hadamard
form displayed in the previous subsection. In
Eq.~(\ref{VExpVSET_unren}), ${\cal{T}}_{\mu \nu}(x,x')$ is a
differential operator which is constructed by point-splitting from
the classical expression (\ref{STclEXP}) of the stress-tensor. It is
a tensor of type (0,2) in $x$ and a scalar in $x'$. It is given by
\begin{eqnarray}\label{OP_tau}
& &  {\cal{T}}_{\mu \nu }= (1-2\xi)g_{\nu}^{\phantom{\nu} \nu'}
\nabla_{\mu}\nabla_{\nu'} + \left(2\xi -\frac{1}{2}\right)g_{\mu
\nu}
g^{\rho \sigma'} \nabla_{\rho}\nabla_{\sigma'}   \nonumber \\
&  & \qquad \qquad -2\xi g_{\mu}^{\phantom{\mu}
\mu'}g_{\nu}^{\phantom{\nu} \nu'}  \nabla_{ \mu'} \nabla_{\nu'} +
2\xi g_{\mu \nu}
\nabla_{\rho} \nabla^{\rho} \nonumber \\
&  & \qquad \qquad +\xi \left(R_{\mu \nu} - \frac{1}{2}g_{\mu \nu}R
\right)
 - \frac{1}{2}g_{\mu \nu}m^2
\end{eqnarray}
where $g_{\mu \nu'}$ denotes the bivector of parallel transport from
$x$ to $x'$ (see Refs.~\cite{DeWittBrehme,DeWitt65}) which is
defined by the partial differential equation
\begin{subequations}
\begin{equation}\label{Geodetic1a} g_{\mu \nu';
\rho}\sigma^{;\rho}=0
\end{equation}
and the boundary condition
\begin{equation}\label{Geodetic1b}
\lim_{x' \to x} g_{\mu \nu'} = g_{\mu \nu}.
\end{equation}
\end{subequations}
Of course, because of the short-distance behavior of the Feynman
propagator, the expression (\ref{VExpVSET_unren}) of the expectation
value of the stress-energy-tensor operator in the Hadamard state
$|\psi \rangle$ is divergent and therefore meaningless. This
pathological behavior comes from the purely geometrical part of the
Hadamard expansion given by (\ref{H2Rep1sing}) for $D=2$ or
(\ref{HevRep1sing}) for $D$ even with $D\not=2$ or by
(\ref{HodRep1sing}) for $D$ odd. More precisely, for $D=2$ the terms
in $\ln \sigma$ and $\sigma \ln \sigma$ which are present in
(\ref{H2Rep1sing}) induce divergences in $1/\sigma$ and $\ln \sigma$
in the expression (\ref{VExpVSET_unren}) of $\langle \psi |T_{\mu
\nu} |\psi \rangle$. For $D$ even with $D\not=2$, the terms in
$1/\sigma^{D/2-1}$, \dots, $1/\sigma $, $\ln \sigma$ and $\sigma \ln
\sigma$ which are present in (\ref{HevRep1sing}) induce divergences
in $1/\sigma^{D/2}$, \dots, $1/\sigma^2$, $1/\sigma$ and $\ln
\sigma$ in the expression (\ref{VExpVSET_unren}) of $\langle \psi
|T_{\mu \nu} |\psi \rangle$ while, for $D$ odd, the terms in
$1/\sigma^{D/2-1}$, \dots, $1/\sigma^{1/2}$ and $\sigma^{1/2} $
which are present in (\ref{HodRep1sing}) induce divergences in
$1/\sigma^{D/2}$, \dots, $1/\sigma^{1/2}$ in this expression.

With Wald \cite{Wald77,Wald78,Wald94} is possible to cure the
pathological behavior of $\langle \psi |T_{\mu \nu} |\psi \rangle$
given by (\ref{VExpVSET_unren}) and to construct from it a
meaningful expression which can act as a source in the semiclassical
Einstein equations (\ref{SCEinsteinEq}) and which can be considered
as the renormalized expectation value with respect to the Hadamard
quantum state $|\psi \rangle$ of the stress-energy tensor operator.
The Hadamard regularization prescription permits us to accomplish
this in the following manner: we first discard in the right-hand
side of (\ref{VExpVSET_unren}) the purely geometrical part
(\ref{H2Rep1sing}) or (\ref{HevRep1sing}) or (\ref{HodRep1sing}) of
$G^F$, i.e. we make the replacement
\begin{eqnarray}\label{VExpVSET_reg1}
& & \lim_{x' \to x} {\cal{T}}_{\mu \nu}(x,x') \, \left[ -i
G^{\mathrm{F}} (x,x')\right] \to \nonumber \\
& & \qquad \qquad \frac{\alpha_D}{2}  \lim_{x' \to x} {\cal{T}}_{\mu
\nu}(x,x') W (x,x').
\end{eqnarray}
We then add to the right-hand side of (\ref{VExpVSET_reg1}) a
state-independent tensor ${\tilde \Theta}_{\mu \nu}$ which only
depends on the parameters $m^2$ and $\xi$ of the theory and on the
local geometry and which ensures the conservation of the resulting
expression. The renormalized expectation value of stress-energy
tensor operator in the Hadamard state $|\psi \rangle$ is therefore
given by
\begin{equation}\label{VExpVSET_reg2}
\langle \psi |T_{\mu \nu}(x) |\psi \rangle_{\mathrm{ren}} =
\frac{\alpha_D}{2}  \lim_{x' \to x} {\cal{T}}_{\mu \nu}(x,x') W
(x,x') + {\tilde \Theta}_{\mu \nu}(x).
\end{equation}
Bearing in mind practical applications, it is also interesting to
reexpress the previous result in terms of the lowest order
coefficients of the covariant Taylor series expansion of the
biscalar $W(x,x')$. By inserting (\ref{CTSExpW1})-(\ref{CTSExpW2})
into (\ref{VExpVSET_reg2}) and by using the expansions (\ref{App2a})
and (see, for example, Refs.~\cite{Christensen1,Christensen2} or
Ref.~\cite{DecaniniFolacci2005a})
\begin{equation}
g_{\nu}^{\phantom{\nu} \nu'} \sigma_{; \mu \nu'} = - g_{\mu \nu} -
(1/6) \, R_{\mu a_1 \nu a_2} \sigma^{;a_1}\sigma^{;a_2}+ O
\left(\sigma^{3/2} \right) \label{App2b}
\end{equation}
as well as the relations (see, for example,
Refs.~\cite{Christensen1,Christensen2})
\begin{subequations}\label{App1}
\begin{eqnarray}
& & g_\mu^{\phantom{\mu} \rho'}g_{\nu \rho'}=g_{\mu \nu} \label{App1debut} \\
& & g_\nu^{\phantom{\nu} \nu'}g_{\mu \nu' ;\rho}=-(1/2) \, R_{\mu
\nu \rho a}\sigma^{;a} + O
\left(\sigma \right) \\
& & g_\nu^{\phantom{\nu} \nu'}g_\rho^{\phantom{\rho} \rho'} g_{\mu
\nu' ; \rho'}=-(1/2) \, R_{\mu \nu \rho a}\sigma^{;a} + O
\left(\sigma \right) \label{App1fin}
\end{eqnarray}
\end{subequations}
we obtain
\begin{eqnarray}\label{VExpVSET_reg3}
& & \langle \psi |T_{\mu \nu} |\psi \rangle_{\mathrm{ren}} =
\frac{\alpha_D}{2}  \left[ -\left( w_{\mu \nu}-\frac{1}{2} g_{\mu
\nu}w^\rho_{\phantom{\rho} \rho}\right)  \right.  \nonumber \\
& & \qquad \left. +\frac{1}{2} (1-2\xi)w_{;\mu \nu} + \frac{1}{2}
\left(2\xi - \frac{1}{2}\right)g_{\mu \nu} \Box w  \right. \nonumber \\
& & \qquad
 \left. + \xi \left(R_{\mu
\nu} - \frac{1}{2}g_{\mu \nu}R \right)w - \frac{1}{2}g_{\mu \nu}m^2w
\right] + {\tilde
\Theta}_{\mu \nu}. \nonumber \\
\end{eqnarray}
Now, by requiring the conservation of $\langle \psi |T_{\mu \nu}
|\psi \rangle_{\mathrm{ren}}$ given by (\ref{VExpVSET_reg3}), we
find that ${\tilde \Theta}_{\mu \nu}$ must satisfy
\begin{equation}\label{EqTTev}
\left[{\tilde \Theta}^{\mu \nu} - (D/4) \alpha_D \, g^{\mu \nu} v_1
\right]_{;\nu}=0
\end{equation}
when $D$ is even and
\begin{equation}\label{EqTTod}
{\tilde \Theta}^{\mu \nu}_{ \phantom{\mu \nu} ;\nu}=0
\end{equation}
when $D$ is odd. Equations (\ref{EqTTev}) and (\ref{EqTTod}) are
derived by using (\ref{HevRep_EQ_W4a}) and (\ref{HevRep_EQ_W5}) for
the former and (\ref{HodRep_EQ_W4a}) and (\ref{HodRep_EQ_W5}) for
the latter.

It is now possible to provide a definitive expression for the
renormalized expectation value of the stress-energy tensor operator
in the Hadamard state $|\psi \rangle$. From (\ref{VExpVSET_reg2})
and by taking into account (\ref{EqTTev}), we have for $D$ even
\begin{eqnarray}\label{VExpVSET_reg2_EV}
& & \langle \psi |T_{\mu \nu}(x) |\psi \rangle_{\mathrm{ren}} =
\frac{\alpha_D}{2} \left[ \lim_{x' \to x} {\cal{T}}_{\mu \nu}(x,x')
W (x,x')  \right. \nonumber \\
& & \qquad \qquad \qquad \quad \qquad \quad \left. +
 \frac{D}{2} \, g_{\mu \nu} v_1\right] + {\Theta}_{\mu \nu}(x).
\end{eqnarray}
This result can be also written in the form
\begin{eqnarray}\label{VExpVSET_regDEFev}
& & \langle \psi |T_{\mu \nu} |\psi \rangle_{\mathrm{ren}} =
\frac{\alpha_D}{2}  \left[ - w_{\mu \nu} +\frac{1}{2}
(1-2\xi)w_{;\mu \nu} \right. \nonumber \\
& & \quad
 \left. + \frac{1}{2} \left(2\xi -
\frac{1}{2}\right)g_{\mu \nu} \Box w  + \xi R_{\mu \nu} w - g_{\mu
\nu}v_1
\right] + {\Theta}_{\mu \nu} \nonumber \\
\end{eqnarray}
which is obtained by inserting (\ref{HevRep_EQ_W4a}) into
(\ref{VExpVSET_reg3}) and by taking into account (\ref{EqTTev}).
From (\ref{VExpVSET_reg2}) and by taking into account
(\ref{EqTTod}), we have for $D$ odd
\begin{equation}\label{VExpVSET_reg2_OD}
\langle \psi |T_{\mu \nu}(x) |\psi \rangle_{\mathrm{ren}} =
\frac{\alpha_D}{2} \, \lim_{x' \to x} {\cal{T}}_{\mu \nu}(x,x') W
(x,x')  + {\Theta}_{\mu \nu}(x).
\end{equation}
This result can be also written in the form
\begin{eqnarray}\label{VExpVSET_regDEFod}
& & \langle \psi |T_{\mu \nu} |\psi \rangle_{\mathrm{ren}} =
\frac{\alpha_D}{2}  \left[ - w_{\mu \nu} +\frac{1}{2}
(1-2\xi)w_{;\mu \nu} \right. \nonumber \\
& & \quad
 \left. + \frac{1}{2} \left(2\xi -
\frac{1}{2}\right)g_{\mu \nu} \Box w  + \xi R_{\mu \nu} w \right] +
{\Theta}_{\mu \nu}
\end{eqnarray}
which is obtained by inserting (\ref{HodRep_EQ_W4a}) into
(\ref{VExpVSET_reg3}) and by taking into account (\ref{EqTTod}). In
Eqs.~(\ref{VExpVSET_reg2_EV})-(\ref{VExpVSET_regDEFod}), the tensor
${\Theta}_{\mu \nu}$ only depends on the parameters $m^2$ and $\xi$
of the theory and on the local geometry and it is now conserved,
i.e. it satisfies
\begin{equation}\label{EqConservTetha}
{\Theta}^{\mu \nu}_{ \phantom{\mu \nu} ;\nu}=0.
\end{equation}
To conclude this subsection, we think it is interesting to recall to
the reader that the two coefficients $w(x)$ and $w_{\mu \nu}(x)$
which appear in the final expressions (\ref{VExpVSET_regDEFev}) and
(\ref{VExpVSET_regDEFod}) and which encode the state-dependence are
obtained as Taylor coefficients of the expansion of the biscalar
$W(x,x')$ but also more directly by the following two formulas
\begin{subequations}\label{w_and wab}
\begin{eqnarray}
& & w(x) = \lim_{x' \to x} W(x,x') \\
& & w_{\mu \nu}(x)= \lim_{x' \to x}  W(x,x')_{;\mu \nu}
\end{eqnarray}
\end{subequations}
which can be derived easily from
(\ref{CTSExpW1})-(\ref{CTSExpW2}) by using (\ref{AppConstrF_2a}) and
(\ref{App2a}). They are useful to treat practical applications.

\subsection{Ambiguities in the renormalized expectation value
of the stress-energy tensor}

As we have previously noted, the renormalized expectation value
$\langle \psi |T_{\mu \nu} |\psi \rangle_{\mathrm{ren}}$ is unique
up to the addition of a local conserved tensor $\Theta_{\mu \nu}$.
This problem plagues the Hadamard renormalization procedure since
its invention (see Sec.~III of Ref.~\cite{Wald78}). It has been
recurrently discussed in the four-dimensional context: we refer to
the monographs of Fulling \cite{Fulling89} and Wald \cite{Wald94}
and to references therein as well as to more recent considerations
developed in
Refs.~\cite{TichyFlanagan98,Salcedo99,MorettiCMP2003,HollandsWald2001,
HollandsWald2002,HollandsWald2003,HollandsWald2005}. In our opinion,
this problem cannot be solved in the lack of a complete quantum
theory of gravity. As a consequence, it induces a serious difficulty
with regard to the study of back reaction effects, the right-hand
side of the semiclassical Einstein equation (\ref{SCEinsteinEq})
being ambiguously defined.

In the present subsection, we shall not consider the ambiguity
problem from a general point of view. We shall only discuss the
standard ambiguity associated with the choice of a mass scale $M $ -
the so-called renormalization mass - introduced in order to make the
argument of the logarithm in Eq.~(\ref{HevRep1}) dimensionless. We
intend to provide a more general (but still incomplete) discussion
in Sec.~IV. The ambiguity associated with the renormalization mass
only exists when the dimension $D$ of spacetime is even. It
corresponds to the replacement of the term $V(x,x') \ln
[\sigma(x,x')+i\epsilon]$ by the term $V(x,x') \ln [M^2
\left(\sigma(x,x')+i\epsilon \right)]$ and therefore to an
indeterminacy in the function $W(x,x')$ previously considered which
corresponds to the replacement
\begin{equation}\label{VExpVSET_amb1}
W(x,x') \to  W(x,x') - V (x,x') \ln M^2
\end{equation}
for which the theory developed in Sec.~II.C remains valid. This
indeterminacy is therefore associated with the term
\begin{equation}\label{VExpVSET_amb2}
\Theta^{M^2}_{\mu \nu}(x)=- \frac{\alpha_D}{2}  \lim_{x' \to x}
{\cal{T}}_{\mu \nu}(x,x') V (x,x') \ln M^2.
\end{equation}
By using Eqs.~(\ref{HevRep2b}), (\ref{CTSExpV1}) and
(\ref{CTSExpV2})), we can see also that the transformation
(\ref{VExpVSET_amb1}) leads to the replacement
\begin{subequations} \label{VExpVSET_amb3}
\begin{eqnarray}
& & w \to w - v_0 \ln M^2 \\
& & w_{\mu \nu} \to w_{\mu \nu} - \left(v_{0 \, \, \mu \nu}+ g_{\mu
\nu}v_1 \right) \ln M^2
\end{eqnarray}
\end{subequations}
into Eq.~(\ref{VExpVSET_regDEFev}) and thus we have
\begin{eqnarray}\label{VExpVSET_amb4}
& & \Theta^{M^2}_{\mu \nu} = -\frac{\alpha_D}{2}  \left[ -
\left(v_{0 \,\, \mu \nu} + g_{\mu \nu} v_1\right) +\frac{1}{2}
(1-2\xi)v_{0 \, ;\mu \nu} \right. \nonumber \\
& & \quad
 \left. + \frac{1}{2} \left(2\xi -
\frac{1}{2}\right)g_{\mu \nu} \Box v_0  + \xi R_{\mu \nu} v_0
\right] \ln M^2.
\end{eqnarray}
As a consequence, the knowledge of the first Taylor coefficients of
the purely geometrical Hadamard coefficients $V_0(x,x')$ and
$V_1(x,x')$ permits us to treat partially the ambiguity problem. It
should be finally recalled that the renormalization mass can be
fixed by imposing additional physical conditions on the renormalized
expectation value of the stress-energy tensor, these conditions
being appropriate to the problem treated.

\subsection{Trace anomaly}

Here, we shall assume that the renormalized expectation value of the
stress-energy tensor $\langle \psi |T_{\mu \nu} |\psi
\rangle_{\mathrm{ren}}$ is given by (\ref{VExpVSET_regDEFev}) for
$D$ even with the geometrical tensor $\Theta_{\mu \nu}$ which
reduces to $\Theta^{M^2}_{\mu \nu}$ given by (\ref{VExpVSET_amb4})
and by (\ref{VExpVSET_regDEFod}) for $D$ odd with the geometrical
tensor $\Theta_{\mu \nu}$ which vanishes. We neglect all the other
possible contributions (see however Sec.~IV.A for a more general
discussion).

By using (\ref{HevRep_EQ_W4a}), we can show that the trace of
$\langle \psi |T_{\mu \nu} |\psi \rangle_{\mathrm{ren}}$ is then
given by
\begin{eqnarray}\label{TR_Anomaly_ev}
& & \langle \psi |T^{\mu}_{\phantom{\mu} \mu} |\psi
\rangle_{\mathrm{ren}} = \frac{\alpha_D}{2}  \left[ -m^2 w +
(D-1)\left(\xi  -\xi_c(D)\right) \Box w \right. \nonumber \\
& & \qquad \qquad \qquad \qquad \qquad
 \left. + 2v_1
\right] + g^{\mu \nu} {\Theta}^{M^2}_{\mu \nu}
\end{eqnarray}
for $D$ even and by using (\ref{HodRep_EQ_W4a}) that it reduces to
\begin{eqnarray}\label{TR_Anomaly_od}
& & \langle \psi | T^{\mu}_{\phantom{\mu} \mu} |\psi
\rangle_{\mathrm{ren}} = \frac{\alpha_D}{2} \left[ -m^2 w +
(D-1)\left(\xi  -\xi_c(D)\right) \Box w \right]  \nonumber \\
\end{eqnarray}
for $D$ odd. Furthermore, we have
\begin{eqnarray}\label{TR_Anomaly_M2}
& & g^{\mu \nu} \Theta^{M^2}_{\mu \nu} = -\frac{\alpha_D}{2} [
-m^2 v_0 \nonumber    \\
& &  \qquad \qquad + (D-1)\left(\xi  -\xi_c(D)\right) \Box v_0 ] \ln
M^2
\end{eqnarray}
which is obtained from (\ref{VExpVSET_amb4}) by using
$v^{\phantom{0} \rho}_{0 \phantom{\rho} \rho}=-D \, v_1+ (m^2 + \xi
R)v_0$, this last relation being easily derived from
(\ref{H2Rep_EQ_V}) or (\ref{HevRep_EQ_V}).

For $m^2=0$ and $\xi= \xi_c(D)$, i.e. when the scalar field theory
is conformally invariant, the trace $g^{\mu \nu} \Theta^{M^2}_{\mu
\nu}$ vanishes and Eq.~(\ref{TR_Anomaly_ev}) yields
\begin{eqnarray}\label{TR_Anomaly_ev_res}
& & \langle \psi |T^{\mu}_{\phantom{\mu} \mu} |\psi
\rangle_{\mathrm{ren}} = \alpha_D \, v_1
\end{eqnarray}
for $D$ even. After renormalization, the expectation value of the
stress-energy tensor has acquired a non-vanishing or ``anomalous"
trace even though the classical stress-energy tensor is traceless
[see Eq.~(\ref{TraceSTclEXP})]. We refer to the monographs of
Birrell and Davies \cite{BirrellDavies}, Fulling \cite{Fulling89}
and Wald \cite{Wald94} as well as to references therein for various
discussions and considerations concerning trace anomalies in quantum
field theory in curved spacetime. For $D$ odd, $m^2=0$ and $\xi=
\xi_c(D)$, Eq.~(\ref{TR_Anomaly_od}) yields
\begin{eqnarray}\label{TR_Anomaly_od_res}
& & \langle \psi | T^{\mu}_{\phantom{\mu} \mu} |\psi
\rangle_{\mathrm{ren}} = 0
\end{eqnarray}
and it appears that the trace anomaly does not exist when the
dimension of spacetime is odd.

\section{Hadamard renormalized stress-energy tensor: Explicit construction}

In this section, we shall mainly discuss the practical aspects of
the Hadamard renormalization of the expectation value of the
stress-energy tensor. This section is written for the reader who
simply wishes to calculate this renormalized expectation value in a
particular case and is not specially interested in the derivation of
all the previous general results.

We assume that we know the explicit expression of the Feynman
propagator $G^{\mathrm{F}}(x,x')$ associated with a given Hadamard
quantum state $|\psi \rangle$. We first obtain the state-dependent
Hadamard biscalar $W(x,x')$ from the relation
\begin{equation}\label{ConstrW}
W(x,x')=\frac{2}{i\alpha_D} \left[ G^{\mathrm{F}}(x,x')-
G^{\mathrm{F}}_{\mathrm{sing}} (x,x') \right]
\end{equation}
where $G^{\mathrm{F}}_{\mathrm{sing}} (x,x')$ is given by
(\ref{H2Rep1sing}) or (\ref{HevRep1sing}) or (\ref{HodRep1sing})
according to the dimension $D$ of spacetime. Of course, we need only
the covariant Taylor series expansion of $W(x,x')$ up to order
$\sigma$ and therefore we do not need to know the terms of the
expansion of $G^{\mathrm{F}}_{\mathrm{sing}} (x,x')$ which vanish
faster than $\sigma(x,x')$ for $x'$ near $x$. For the same reason,
the Feynman propagator $G^{\mathrm{F}}(x,x')$ does not need to be
known exactly: we need only its asymptotic expansion for $x'$ near
$x$ and we do not need to know the terms of this expansion which
vanish faster than $\sigma(x,x')$ for $x'$ near $x$. From the
expansion up to order $\sigma$ of the biscalar $W(x,x')$ we then
obtain the Taylor coefficients $w(x)$ and $w_{\mu \nu}(x)$ either
directly or by using the relations (\ref{w_and wab}). This permits
us to finally construct the renormalized expectation value in the
Hadamard quantum state $|\psi \rangle$ of the stress-energy tensor
by using (\ref{VExpVSET_regDEFev}) and (\ref{VExpVSET_amb4}) or
(\ref{VExpVSET_regDEFod}) according to the parity of $D$. Of course,
for $D$ even, we must in addition construct the geometrical tensor
$\Theta^{M^2}_{\mu \nu}$ from  the Taylor coefficients $v_0$, $v_{0
\,\, \mu \nu}$ and $v_1$ in order to do this last step.

In the subsections below, we shall provide for spacetime dimension
from $D=2$ to $D=6$ the explicit expansion of
$G^{\mathrm{F}}_{\mathrm{sing}} (x,x')$ and for $D=2,4$ and $6$ we
shall in addition give the explicit expression of the geometrical
tensor $\Theta^{M^2}_{\mu \nu}$ as well as of the trace anomaly. We
shall use some of the results we obtained in
Ref.~\cite{DecaniniFolacci2005a}. We have simplified them from the
geometrical identities displayed in our unpublished report
\cite{DecaniniFolacci_arXiv2008}. These geometrical identities are
helpful to expand the Riemann polynomials encountered in our
calculations on the FKWC-bases constructed from group theoretical
considerations in Ref.~\cite{FKWC1992}. They have permitted us to
provide irreducible expressions for all our results. For spacetime
dimension from $7$ to $11$, we shall describe the method permitting
the interested reader to construct explicitly
$G^{\mathrm{F}}_{\mathrm{sing}} (x,x')$ (as well as
$\Theta^{M^2}_{\mu \nu}$ when it is necessary) in a given spacetime
by using the results obtained in Ref.~\cite{DecaniniFolacci2005a}.

\subsection{D=2}

For $D=2$, the expansion of the singular part
\begin{equation}\label{H2Rep1sing_d2}
G^{\mathrm{F}}_{\mathrm{sing}} (x,x') = \frac{i}{4 \pi} \, \left(
V(x,x') \ln [\sigma(x,x')+i\epsilon] \right)
\end{equation}
of the Feynman propagator is obtained, up the required order, for
\begin{equation}
V=V_0 + V_1 \sigma + O\left( \sigma^{3/2} \right)
\label{UetVTot_d2b}
\end{equation}
with
\begin{eqnarray}
&  & V_0 =v_0 - v_{0 \,\, a} \sigma^{;a}+\frac{1}{2!} v_{0 \,\, a b}
\sigma^{;a}\sigma^{;b}  + O \left(\sigma^{3/2} \right)
\label{U0_V0_V1 d2b} \\
& & V_1 =v_1 + O \left(\sigma^{1/2} \right). \label{U0_V0_V1_d2c}
\end{eqnarray}
The Taylor coefficients appearing in Eqs.~(\ref{U0_V0_V1
d2b})-(\ref{U0_V0_V1_d2c}) are given by
\begin{subequations} \label{CoefT_U0_V0_V1_d2 V0}
\begin{eqnarray}
& & v_0= -1 \label{CoefT_U0_V0_V1_d2 V0a}\\
& & v_{0 \,\, a}= 0 \label{CoefT_U0_V0_V1_d2 V0b} \\
& & v_{0 \,\, a b} = -(1/12) Rg_{ab}  \label{CoefT_U0_V0_V1_d2 V0c}
\end{eqnarray}
\end{subequations}
and
\begin{equation}
v_1= -(1/2) \, m^2 - (1/2)  (\xi -1/6)   R .
\label{CoefT_U0_V0_V1_d2 V1}
\end{equation}

The geometrical tensor $\Theta^{M^2}_{\mu \nu}$ which is associated
with the renormalization mass is obtained from (\ref{VExpVSET_amb4})
by using (\ref{CoefT_U0_V0_V1_d2 V0a}), (\ref{CoefT_U0_V0_V1_d2
V0c}) and (\ref{CoefT_U0_V0_V1_d2 V1}) and is given by
\begin{eqnarray}\label{VExpVSET_amb_d2}
& & \Theta^{M^2}_{\mu \nu} = \frac{\ln M^2}{4\,\pi} \left[-(1/2) \,
m^2 \, g_{\mu \nu} \right].
\end{eqnarray}

The trace anomaly (\ref{TR_Anomaly_ev_res}) is obtained by using
$m^2=0$ and $\xi=\xi_c(2)=0$ into (\ref{CoefT_U0_V0_V1_d2 V1}). It
reduces to \cite{BirrellDavies,DeWitt03}
\begin{eqnarray}\label{TR_Anomaly_d2_res}
& & \langle \psi |T^{\mu}_{\phantom{\mu} \mu} |\psi
\rangle_{\mathrm{ren}} = \frac{R}{24 \pi}.
\end{eqnarray}

\subsection{D=3}

For $D=3$, the expansion of the singular part
\begin{equation}\label{HodRep1sing_d3}
G^{\mathrm{F}}_{\mathrm{sing}} (x,x') = \frac{i}{4\sqrt{2} \pi} \,
\left(\frac{U(x,x')}{[\sigma(x,x')+i\epsilon]^{1/2}} \right)
\end{equation}
of the Feynman propagator is obtained, up the required order, for
\begin{equation}\label{UTot_d3}
U=U_0 + U_1 \sigma + O\left( \sigma^{2}  \right)
\end{equation}
with
\begin{eqnarray}
&  & U_0 =u_0 - u_{0 \,\, a} \sigma^{;a}+\frac{1}{2!} u_{0 \,\, a b}
\sigma^{;a}\sigma^{;b}  -\frac{1}{3!} u_{0 \,\, a b c}
\sigma^{;a}\sigma^{;b}\sigma^{;c}  \nonumber \\
& & \qquad +  O \left(\sigma^{2} \right)
\label{U0_U1_d3a} \\
& & U_1 =u_1 - u_{1 \,\, a} \sigma^{;a} + O \left(\sigma \right).
\label{U0_U1_d3b}
\end{eqnarray}
The Taylor coefficients appearing in
Eqs.~(\ref{U0_U1_d3a})-(\ref{U0_U1_d3b}) are given by
\begin{subequations} \label{CoefT_U0_U1_d3_U0}
\begin{eqnarray}
& & u_0=1 \\
& & u_{0 \,\, a}=0 \\
& & u_{0 \,\, a b} = (1/6)   R_{a b} \\
& & u_{0 \,\, a b c} = (1/4)     R_{(a b; c)}
\end{eqnarray}
\end{subequations}
and
\begin{eqnarray}\label{CoefT_U0_U1_d3_U1}
& & u_1= m^2 + (\xi -1/6)   R  \\
& & u_{1 \,\, a}= (1/2)  (\xi -1/6)   R_{;a}.
\end{eqnarray}

\subsection{D=4}

For $D=4$, the expansion of the singular part
\begin{eqnarray}\label{HevRep1sing_d4}
& & G^{\mathrm{F}}_{\mathrm{sing}} (x,x') = \frac{i}{8 \pi^2} \,
\left(\frac{U(x,x')}{\sigma(x,x')+i\epsilon} \right. \nonumber \\
& &  \left. \phantom{\frac{U}{\sigma^{d}}} + V(x,x') \ln
[\sigma(x,x')+i\epsilon] \right)
\end{eqnarray}
of the Feynman propagator is obtained, up the required order, for
\begin{eqnarray}
& & U=U_0 \label{UetVTot_d4a} \\
& & V=V_0 + V_1 \sigma + O\left( \sigma^{3/2} \right)
\label{UetVTot_d4b}
\end{eqnarray}
with
\begin{eqnarray}
&  & U_0 =u_0 - u_{0 \,\, a} \sigma^{;a}+\frac{1}{2!} u_{0 \,\, a b}
\sigma^{;a}\sigma^{;b}  -\frac{1}{3!} u_{0 \,\, a b c}
\sigma^{;a}\sigma^{;b}\sigma^{;c}  \nonumber \\
& & \quad + \frac{1}{4!} u_{0 \,\, a b c d}
\sigma^{;a}\sigma^{;b}\sigma^{;c}\sigma^{;d} + O \left(\sigma^{5/2}
\right)
\label{U0_V0_V1_d4a} \\
&  & V_0 =v_0 - v_{0 \,\, a} \sigma^{;a}+\frac{1}{2!} v_{0 \,\, a b}
\sigma^{;a}\sigma^{;b}  + O \left(\sigma^{3/2} \right)
\label{U0_V0_V1 d4b} \\
& & V_1 =v_1 + O \left(\sigma^{1/2} \right). \label{U0_V0_V1_d4c}
\end{eqnarray}
The Taylor coefficients appearing in
Eqs.~(\ref{U0_V0_V1_d4a})-(\ref{U0_V0_V1_d4c}) are given by
\begin{subequations} \label{CoefT_U0_V0_V1_d4 U0}
\begin{eqnarray}
& & u_0=1 \\
& & u_{0 \,\, a}=0 \\
& & u_{0 \,\, a b} =
(1/6)  \,  R_{a b}  \\
& & u_{0 \,\, a b c} = (1/4)      R_{(a b; c)} \\
& & u_{0 \,\, a b c d} = (3/10)    R_{(a b; c d)} + (1/12)  R_{(a b} R_{c d)} \nonumber \\
& & \qquad \qquad + (1/15)   R_{p (a |q| b} R^{p \phantom{c}
q}_{\phantom{p} c \phantom{q}  d)}
\end{eqnarray}
\end{subequations}
and
\begin{subequations} \label{CoefT_U0_V0_V1_d4 V0}
\begin{eqnarray}
& & v_0= (1/2)   m^2 + (1/2)  (\xi -1/6)   R  \label{CoefT_U0_V0_V1_d4 V0a}\\
& & v_{0 \,\, a}= (1/4)  (\xi -1/6)  R_{;a} \\
& & v_{0 \,\, a b} = (1/12) m^2    R_{a b} +(1/6)  (\xi-3/20)
\, R_{;ab} \nonumber \\
& & \quad  -(1/120)     \Box R_{a b} +(1/12) (\xi -1/6)   R
R_{a b} \nonumber \\
& & \quad  +(1/90)  R^p_{\phantom{p}a} R_{p b} - (1/180)
  R^{pq}R_{p a q b} \nonumber \\
& & \quad  - (1/180)    R^{pqr}_{\phantom{pqr} a }R_{pqr b}
\label{CoefT_U0_V0_V1_d4 V0c}
\end{eqnarray}
\end{subequations}
and
\begin{eqnarray}
& & v_1= (1/8)   m^4 + (1/4)   (\xi -1/6)
m^2 \, R \nonumber \\
& & \quad -(1/24)   (\xi -1/5)   \Box R + (1/8)   (\xi -1/6)^2
\, R^2 \nonumber \\
& & \quad - (1/720)  R_{pq}R^{pq} +  (1/720)   R_{pqrs}R^{pqrs}.
\label{CoefT_U0_V0_V1_d4 V1}
\end{eqnarray}

The geometrical tensor $\Theta^{M^2}_{\mu \nu}$ which is associated
with the renormalization mass is obtained from (\ref{VExpVSET_amb4})
by using (\ref{CoefT_U0_V0_V1_d4 V0a}), (\ref{CoefT_U0_V0_V1_d4
V0c}) and (\ref{CoefT_U0_V0_V1_d4 V1}) and is given by
\begin{eqnarray}\label{VExpVSET_amb_d4}
& & \Theta^{M^2}_{\mu \nu} = \frac{\ln M^2}{2 {(2\pi )}^2}
\left[\phantom{\frac{}{}} -(1/2)(\xi -1/6)  m^2   R_{\mu \nu} \right. \nonumber \\
& & \qquad  \left. +(1/2)
 [\xi^2 -(1/3)  \xi +1/30 ]
R_{;\mu \nu}   -(1/120)   \Box R_{\mu \nu} \right. \nonumber \\
& & \qquad  \left. -(1/2)(\xi -1/6)^2  R R_{\mu \nu}  +(1/90)
R^p_{\phantom{p}\mu} R_{p \nu} \right. \nonumber \\
& & \qquad  \left.- (1/180)  R^{p q}R_{p \mu q \nu} - (1/180)   R^{p
q
r}_{\phantom{p q r} \mu }R_{p q r \nu} \right. \nonumber \\
& & \qquad \qquad \qquad ~~ \left. +g_{\mu \nu}
\left(\phantom{\frac{}{}} (1/8)  m^4   + (1/4)  (\xi -1/6)  m^2  R
 \right.\right. \nonumber \\
& & \qquad  \left.\left. -(1/2)  [\xi^2-(1/3) \xi +1/40]  \Box R  \right.\right. \nonumber \\
& & \qquad  \left.\left. + (1/8)  (\xi -1/6)^2  R^2 - (1/720) R_{p
q}R^{p q} \right.\right. \nonumber \\
& & \qquad  \left.\left.  + (1/720)  R_{p q rs }R^{p q rs}
\phantom{\frac{}{}}\right) \right].
\end{eqnarray}

The trace anomaly (\ref{TR_Anomaly_ev_res}) is obtained by using
$m^2=0$ and $\xi=\xi_c(4)=1/6$ into (\ref{CoefT_U0_V0_V1_d4 V1}). It
reduces to \cite{BirrellDavies,DeWitt03}
\begin{eqnarray}\label{TR_Anomaly_d4_res}
& & \langle \psi |T^{\mu}_{\phantom{\mu} \mu} |\psi
\rangle_{\mathrm{ren}} = \frac{1}{(2\pi)^2} \, \left[(1/720) \Box R
-
(1/720) R_{p q}R^{p q} \right. \nonumber \\
& & \qquad \qquad \left. + (1/720)  R_{p q rs }R^{p q rs} \right].
\end{eqnarray}

\subsection{D=5}

For $D=5$, the expansion of the singular part
\begin{equation}\label{HodRep1sing_d5}
G^{\mathrm{F}}_{\mathrm{sing}} (x,x') =
\frac{i}{16\,{\sqrt{2}}\,{\pi }^2} \,
\left(\frac{U(x,x')}{[\sigma(x,x')+i\epsilon]^{3/2}} \right)
\end{equation}
of the Feynman propagator is obtained, up the required order, for
\begin{equation}\label{UTot_d5}
U=U_0 + U_1 \sigma + U_2 \sigma^2 + O\left( \sigma^{3}  \right)
\end{equation}
with
\begin{eqnarray}
&  & U_0 =u_0 - u_{0 \,\, a} \sigma^{;a}+\frac{1}{2!} u_{0 \,\, a b}
\sigma^{;a}\sigma^{;b}  -\frac{1}{3!} u_{0 \,\, a b c}
\sigma^{;a}\sigma^{;b}\sigma^{;c}  \nonumber \\
& & \quad + \frac{1}{4!} u_{0 \,\, a b c d}
\sigma^{;a}\sigma^{;b}\sigma^{;c}\sigma^{;d}  -\frac{1}{5!} u_{0
\,\, a b c d
e}\sigma^{;a}\sigma^{;b}\sigma^{;c}\sigma^{;d}\sigma^{;e} \nonumber \\
& & \quad + O \left(\sigma^{3} \right)
\label{U0_U1_U2_d5a} \\
& & U_1 =u_1 - u_{1 \,\, a} \sigma^{;a} + \frac{1}{2!} u_{1 \,\, a
b} \sigma^{;a}\sigma^{;b} -\frac{1}{3!} u_{1 \,\, a b c}
\sigma^{;a}\sigma^{;b}\sigma^{;c} \nonumber \\
& & \quad + O \left(\sigma^{2} \right) \label{U0_U1_U2_d5b}  \\
& & U_2 =u_2 - u_{2 \,\, a} \sigma^{;a} + O \left(\sigma \right).
\label{U0_U1_U2_d5c}
\end{eqnarray}
The Taylor coefficients appearing in
Eqs.~(\ref{U0_U1_U2_d5a})-(\ref{U0_U1_U2_d5c}) are given by
\begin{subequations}\label{CoefT_U0_U1_U2_d5 U0}
\begin{eqnarray}
& & u_0=1 \\
& & u_{0 \,\, a}=0 \\
& & u_{0 \,\, a b} =
(1/6)    R_{a b}  \\
& & u_{0 \,\, a b c} =
(1/4)      R_{(a b; c)}  \\
& & u_{0 \,\, a b c d} = (3/10)   R_{(a b; c d)} + (1/12)  R_{(a b} R_{c d)} \nonumber \\
& & \qquad  + (1/15)  R_{p (a |q| b} R^{p \phantom{c}
q}_{\phantom{p} c \phantom{q}  d)} \\
& & u_{0 \,\, a b c d e} =  (1/3)  R_{(a b; c d e)} + (5/12)
R_{(a b} R_{c d; e)} \nonumber \\
& & \qquad + (1/3)  R_{p(a |q| b} R^{p \phantom{c} q}_{\phantom{p} c
\phantom{q} d;e)}
\end{eqnarray}
\end{subequations}
and
\begin{subequations}\label{CoefT_U0_U1_U2_d5 U1}
\begin{eqnarray}
& & u_1= -m^2 -(\xi -1/6)  R  \\
& & u_{1 \,\, a}= -(1/2)  (\xi -1/6)  R_{;a} \\
& & u_{1 \,\, a b} = -(1/6)m^2   R_{a b} -(1/3)  (\xi-3/20)
 R_{;ab}  \nonumber \\
& & \quad +(1/60)    \Box R_{a b} -(1/6)(\xi -1/6)  R R_{a b}
\nonumber \\
& & \quad  -(1/45) R^p_{\phantom{p}a} R_{p b} + (1/90)
 R^{p q}R_{p a q b} \nonumber \\
& & \quad + (1/90)  R^{p qr}_{\phantom{p qr} a }R_{p qr b} \\
& & u_{1 \,\, a b c} =  - (1/4 )  m^2  R_{
(a b; c)} \nonumber \\
& & \quad - (1/4 )  (\xi -2/15)  R _{ ; (a b c)} + (1/40)  (
\Box R_{(a b} )_{;c)}    \nonumber \\
& & \quad      - (1/4 )
(\xi -1/6)  R_{;(a } R_{ bc)} - (1/4 )  (\xi -1/6)  R R_{ (a b; c)}\nonumber \\
& & \quad  - (1/15)  R_{ p (a } R^{p}_{ \phantom{p} b;c) }+ (1/60)
 R_{p
 q ;(a } R^{p \phantom{b} q}_{\phantom{p} b \phantom{q} c)} \nonumber \\
& & \quad
  +
(1/60)  R_{ p  q} R^{p \phantom{(a} q}_{\phantom{p}   (a \phantom{q}
b; c)}
  + (1/30)  R_{pqr (a} R^{pqr}_{\phantom{ p qr} b;  c)} \nonumber \\
\end{eqnarray}
\end{subequations}
and
\begin{subequations}\label{CoefT_U0_U1_U2_d5 U2}
\begin{eqnarray}
& & u_2= -(1/2)  m^4 - (\xi -1/6)  m^2  R  \nonumber \\
& & \quad +(1/6)  (\xi -1/5)  \Box R - (1/2)  (\xi -1/6)^2  R^2 \nonumber \\
& & \quad + (1/180)  R_{p q}R^{p q} -  (1/180)  R_{p qrs}R^{p qrs}
\label{CoefT_U0_U1_U2_d5
U2a} \\
& & u_{2 \,\, a}= -(1/2)  (\xi -1/6)  m^2  R_{;a} \nonumber \\
& & \quad +(1/12)  (\xi -1/5)  (\Box R)_{;a} - (1/2)  (\xi
-1/6)^2  RR_{;a} \nonumber \\
& & \quad  + (1/180)  R_{p q}R^{p q}_{\phantom{p q};a} - (1/180)
 R_{p qrs}R^{p qrs}_{\phantom{p qrs};a}. \nonumber \\
& & \label{CoefT_U0_U1_U2_d5 U2b}
\end{eqnarray}
\end{subequations}

\subsection{D=6}

For $D=6$, the expansion of the singular part
\begin{eqnarray}\label{HevRep1sing_d6}
& & G^{\mathrm{F}}_{\mathrm{sing}} (x,x') = \frac{i}{16\,{\pi }^3}
\,
\left(\frac{U(x,x')}{[\sigma(x,x')+i\epsilon]^2} \right. \nonumber \\
& &  \left. \phantom{\frac{U}{\sigma^{d}}} + V(x,x') \ln
[\sigma(x,x')+i\epsilon] \right)
\end{eqnarray}
of the Feynman propagator is obtained, up the required order, for
\begin{eqnarray}
& & U=U_0 + U_1 \sigma \label{UetVTot_d6a}
\\
& & V=V_0 + V_1 \sigma + O\left( \sigma^{3/2} \right)
\label{UetVTot_d6b}
\end{eqnarray}
with
\begin{eqnarray}
&  & U_0 =u_0 - u_{0 \,\, a} \sigma^{;a}+\frac{1}{2!} u_{0 \,\, a b}
\sigma^{;a}\sigma^{;b}  -\frac{1}{3!} u_{0 \,\, a b c}
\sigma^{;a}\sigma^{;b}\sigma^{;c}  \nonumber \\
& & \quad + \frac{1}{4!} u_{0 \,\, a b c d}
\sigma^{;a}\sigma^{;b}\sigma^{;c}\sigma^{;d}
  -\frac{1}{5!} u_{0 \,\, a b c d
e}\sigma^{;a}\sigma^{;b}\sigma^{;c}\sigma^{;d}\sigma^{;e} \nonumber \\
& & \quad + \frac{1}{6!} u_{0 \,\, a b c d e
f}\sigma^{;a}\sigma^{;b}\sigma^{;c}\sigma^{;d}\sigma^{;e}\sigma^{;f}+
O \left(\sigma^{7/2} \right)
\label{U0_U1_V0_V1_d6a} \\
& & U_1 =u_1 - u_{1 \,\, a} \sigma^{;a} + \frac{1}{2!} u_{1 \,\, a
b} \sigma^{;a}\sigma^{;b} -\frac{1}{3!} u_{1 \,\, a b c}
\sigma^{;a}\sigma^{;b}\sigma^{;c} \nonumber \\
& & \quad  + \frac{1}{4!} u_{1 \,\, a b c d}
\sigma^{;a}\sigma^{;b}\sigma^{;c}\sigma^{;d} +  O \left(\sigma^{5/2}
\right)
\label{U0_U1_V0_V1_d6b}  \\
&  & V_0 =v_0 - v_{0 \,\, a} \sigma^{;a}+\frac{1}{2!} v_{0 \,\, a b}
\sigma^{;a}\sigma^{;b}  + O \left(\sigma^{3/2} \right)
\label{U0_U1_V0_V1_d6c} \\
& & V_1 =v_1 + O \left(\sigma^{1/2} \right). \label{U0_U1_V0_V1_d6d}
\end{eqnarray}
The Taylor coefficients appearing in
Eqs.~(\ref{U0_U1_V0_V1_d6a})-(\ref{U0_U1_V0_V1_d6d}) are given by
\begin{subequations} \label{CoefT_U0_U1_V0_V1_d6 U0}
\begin{eqnarray}
& & u_0=1 \\
& & u_{0 \,\, a}=0 \\
& & u_{0 \,\, a b} = (1/6)    R_{a b}
\end{eqnarray}
and
\begin{eqnarray}& & u_{0 \,\, a b c} =
(1/4)      R_{(a b; c)}  \\
& & u_{0 \,\, a b c d} = (3/10)   R_{(a b; c d)} + (1/12)  R_{(a b} R_{c d)} \nonumber \\
& &  \qquad + (1/15)  R_{p (a |q| b} R^{p \phantom{c}
q}_{\phantom{p} c \phantom{q}  d)} \\
& & u_{0 \,\, a b c d e} =  (1/3)  R_{(a b; c d e)} + (5/12)
R_{(a b} R_{c d; e)} \nonumber \\
& & \qquad + (1/3)  R_{p(a |q| b} R^{p \phantom{c} q}_{\phantom{p}
c \phantom{q} d;e)}\\
& & u_{0 \,\, a b c d e f} = (5/14)  R_{(a b; c d e f)} + (3/4)
R_{(a b }R_{c d; e
f)} \nonumber \\
& & \quad  +(4/7)  R_{p(a |q| b} R^{p \phantom{c} q}_{\phantom{p} c
\phantom{p}  d;e f)} + (5/8)  R_{(a b ; c}R_{ d e ; f)} \nonumber
\\
&  & \quad + (15/28)  R_{p(a |q| b; c} R^{p \phantom{d}
q}_{\phantom{p} d \phantom{q} e; f)} + (5/72)  R_{(a b}R_{c d}R_{e
f)} \nonumber \\
& & \quad + (1/6)  R_{(a b} R^{p \phantom{c} q}_{\phantom{p} c
\phantom{q} d}
R_{|p| e |q |  f)}  \nonumber \\
& & \quad+ (8/63) R^{p}_{\phantom{p}(a |q | b }R^{q}_{\phantom{q} c
|r | d}R^{r}_{\phantom{r} e |p | f)}
\end{eqnarray}
\end{subequations}
and
\begin{widetext}
\begin{subequations} \label{CoefT_U0_U1_V0_V1_d6 U1}
\begin{eqnarray}
& & u_1= -(1/2)  m^2 - (1/2) (\xi -1/6)  R  \\
& & u_{1 \,\, a}= -(1/4) (\xi -1/6)  R_{;a} \\
& & u_{1 \,\, a b} = -(1/12)m^2   R_{a b} -(1/6)
(\xi-3/20)  R_{;ab} +(1/120)    \Box R_{a b}   \nonumber \\
& & \qquad     -(1/12)(\xi -1/6)  R R_{a b} -(1/90)
R^p_{\phantom{p}a} R_{p b} + (1/180)  R^{p q}R_{p a q b}
+ (1/180)   R^{p qr}_{\phantom{p qr} a }R_{p qr b} \\
& & u_{1 \,\, a b c} =  - (1/8 )  m^2  R_{ (a b; c)}  - (1/8 ) (\xi
-2/15)  R _{ ; (a b c)} + (1/80)  (
\Box R_{(a b} )_{;c)}    \nonumber \\
& & \qquad      - (1/8 )  (\xi -1/6)  R_{;(a } R_{ bc)} - (1/8 )
(\xi -1/6)  R R_{ (a b; c)} - (1/30)  R_{ p (a } R^{p}_{
\phantom{p} b;c) }\nonumber \\
& & \qquad + (1/120)  R_{p
 q ;(a } R^{p \phantom{b} q}_{\phantom{p} b \phantom{q} c)}
  +
(1/120)  R_{ p  q} R^{p \phantom{(a} q}_{\phantom{p}   (a
\phantom{q} b; c)}
  + (1/60)  R_{pqr (a} R^{pqr}_{\phantom{ p qr} b;  c)} \\
& & u_{1 \,\, a b c d} = - (3/20)  m^2  R _{
 (a b; c d)} - (1/24)  m^2
 R _{ (a b}R _{ c d)} -(1/30)  m^2
 R_{p (a |q| b} R^{p \phantom{c}
q}_{\phantom{p} c \phantom{q}  d)} \nonumber \\
&  & \qquad  - (1/10)  (\xi-5/42)  R _{ ; (a b c d)} + (1/70)  (
\Box R_{(a b})_{;c d)} -(1/6)
(\xi-3/20)  R_{;(ab}R_{cd)} \nonumber \\
&  & \qquad  - (3/20)  (\xi -1/6)  R R _{
 (a b; c d)} + (1/120)  R _{ (a b}\Box R _{ c d)}
 -(3/70)  R^p_{\phantom{p}  (a } R_{ |p|
b;c d) } + (1/210)  R^{ p  }_{\phantom{ p}  (a }
R_{ b c;  d)p } \nonumber \\
&  & \qquad   + (1/70)  R_{pq ;(a b} R^{p \phantom{c}
q}_{\phantom{p} c \phantom{p} d)}  - (2/105)  R_{p
 (a ;  b |q|} R^{p \phantom{c}
q}_{\phantom{p} c \phantom{p} d)}  + (1/70)  R_{(a b }^{\phantom{(a
b } ; pq  } R_{|p| c |q | d)}
+ (1/105)  R_{pq} R^{p \phantom{a} q}_{\phantom{p} (a \phantom{q} b; c d)}\nonumber \\
&  & \qquad  + (2/105)
 R^{pqr}_{\phantom{pqr} (a} R_{|pqr| b; c d)}
 -(1/4)  (\xi-1/6)  R_{;(a}R_{bc;d)}  - (11/420)  R^{ p  }_{\phantom{ p} (a;b } R_{ |p| c;d)
} \nonumber \\
&  & \qquad - (3/140)  R^{ p }_{\phantom{ p} (a ; b } R_{ c d ); p
}+(17/1680)  R_{ (a b }^{ \phantom{(a b } ; p }R_{ c d ); p }
  + (1/60)  R_{pq;(a } R^{p\phantom{b} q }_{\phantom{p} b \phantom{q} c;
 d)} + (1/210)  R_{p
 (a ; |q |} R^{p\phantom{b} q }_{\phantom{p} b \phantom{q} c;
 d)}
\nonumber \\
&  & \qquad
  + (1/56)
 R^{pqr}_{\phantom{pqr} (a;b} R_{|pqr|  c; d)} + (1/280)
 R^{p \phantom{a} q \phantom{b} ;r}_{\phantom{p} (a \phantom{q} b}R_{|p|c|q|d);r}
  - (1/24)  (\xi -1/6)  R R _{(a b}R _{ c d)} - (1/90)
 R_{(ab }R^{ p
}_{\phantom{ p}  c  } R_{ |p|d)} \nonumber \\
&  & \qquad +  (1/180)  R_{(a b}R^{pq}R_{|p| c
 |q| d)} + (1/90)  R^p_{\phantom{p}(a}R^q_{\phantom{q}b}R_{|p|c|q|d)}
- (1/30)  (\xi-1/6)   RR^{p \phantom{a} q}_{\phantom{p} (a
\phantom{q} b}R_{|p|c|q|d)} \nonumber \\
& & \qquad   +(1/180)  R_{(a b} R^{ pqr}_{\phantom{ pqr} c} R_{|pqr|
d)}  +  (1/315)  R_{p(a}R^{rp s}_{\phantom{rps}
 b}R_{|r|c|s|d)} -(1/315) R_{pq}R^{r\phantom{(a} p}_{\phantom{r} (a \phantom{r} b}
 R^{\phantom{|r| c} q}_{|r| c\phantom{q}d)}  \nonumber \\
 & &  \qquad
  -(2/315) R_{prqs}R^{p\phantom{(a}q}_{\phantom{p}(a \phantom{q}
b}R^{r\phantom{c}s}_{\phantom{r}c \phantom{s} d)} -(1/315) R^{p
\phantom{(a} q}_{\phantom{p}(a \phantom{q}
b}R^{rs}_{\phantom{rs}|p|c}R_{|rs q|d)} +(4/315) R^{p \phantom{(a}
q}_{\phantom{p}(a \phantom{q} b}R^{\phantom{|p|} rs}_{|p|
\phantom{rs} c}R_{|q rs| d)}
\end{eqnarray}
\end{subequations}
and
\begin{subequations}\label{CoefT_U0_U1_V0_V1_d6 V0}
\begin{eqnarray}
& & v_0= -(1/8)  m^4 -
(1/4)  (\xi -1/6)  m^2  R +(1/24)  (\xi -1/5)  \Box R  \nonumber \\
& & \qquad - (1/8)  (\xi -1/6)^2  R^2 + (1/720)  R_{pq}R^{pq} -
(1/720) \, R_{pqrs}R^{pqrs}
\label{CoefT_U0_U1_V0_V1_d6 V0a}\\
& & v_{0 \,\, a}=
-(1/8)  (\xi -1/6)  m^2  R_{;a} +(1/48)  (\xi -1/5)  (\Box R)_{;a}\nonumber \\
& & \qquad  - (1/8)  (\xi -1/6)^2  RR_{;a} + (1/720)
R_{pq}R^{pq}_{\phantom{pq};a} -  (1/720)
R_{pqrs}R^{pqrs}_{\phantom{pqrs};a}
\end{eqnarray}
and
\begin{eqnarray}
& & v_{0 \,\, a b} =  - (1/48)  m^4   R_{a b} -(1/12)  (\xi-3/20)
m^2  R_{;a b} + (1/240)  m^2  \Box R_{a b}
-(1/24) (\xi-1/6)  m^2   R R_{a b} \nonumber \\
& & \quad - (1/180)  m^2 R_{p a} R^p_{\phantom{p}b} + (1/360)  m^2
 R^{pq} R_{p a q b}+ (1/360)  m^2
R^{pqr}_{\phantom{pqr}a} R_{pqr b} \nonumber \\
& & \quad +(1/80) (\xi-4/21) (\Box R)_{;ab}
 -(1/3360) \Box \Box R_{a b} - (1/12)  (\xi -1/6)(\xi-3/20)   R R_{;a b}
   \nonumber \\
& & \quad  +(1/144) (\xi -1/5)   (\Box R) R_{a b} +(1/360)  (\xi
-1/7)  R_{;p (a} R^p_{\phantom{p}b)}
      +(1/240)(\xi -1/6)   R \Box R_{a b} \nonumber \\
& & \quad +(1/1008)  R_{p (a} \Box R^p_{\phantom{p}b)} +(1/1680)
R^{pq} R_{pq;(a b)}  +(1/1260)  R^{pq} R_{p (a ; b)q} -
(1/1680)  R^{pq} R_{ab; pq} \nonumber \\
& & \quad + (1/180)  (\xi -3/14)  R^{;pq}R_{p a q b}  - (1/2520)
(\Box R^{pq})R_{p a q b} +(1/630)  R^{p q;r}_{\phantom{p q;r}
(a} R_{|rqp| b)}  \nonumber \\
& & \quad + (1/420)  R^{p \phantom{(a}; q r}_{\phantom{p } (a}
R_{|pqr | b)}  - (1/1260)  R^{pqrs} R_{pqrs ; (a b) } - (1/16) (\xi
-1/6)^2  R_{;a} R_{;b} \nonumber \\
& & \quad -(1/120)  (\xi -3/14)  R_{;p } R^p_{\phantom{p}(a;b)} +
(1/120)  (\xi -17/84)  R_{;p } R_{ab}^{\phantom{ab}; p}  +
(1/1440)  R^{pq}_{\phantom{pq};a}R_{pq;b}   \nonumber \\
& & \quad   - (1/5040)  R^p_{\phantom{p} a;q} R_{p b}^{\phantom{p
b};q}  + (1/1008)  R^p_{\phantom{p} a;q} R^q_{\phantom{q} b;p } -
(1/2520)  R^{pq;r} R_{rqp (a;b)}  - (1/1680)
R^{pq;r} R_{p a q b;r} \nonumber \\
& & \quad  - (1/1344)  R^{pqrs}_{\phantom{pqrs};a} R_{pqrs ; b } -
(1/1680)  R^{pqr}_{\phantom{pqr}a;s}R_{pqr b}^{\phantom{pqr b};s}   \nonumber \\
& & \quad - (1/48)  (\xi -1/6)^2  R^2 R_{ab}  - (1/180)  (\xi
-1/6)  R R_{p a} R^p_{\phantom{p}b} +(1/4320)  R^{pq}R_{pq}R_{ab} \nonumber \\
& &  \quad - (1/3780)  R^{pq}R_{p a}R_{q b}  + (1/360)  (\xi -1/6)
 R R^{pq}R_{p a q b} +(1/7560)  R^{pr}R^q_{\phantom{q} r}R_{p a
q b} \nonumber \\
& &  \quad + (1/7560)  R^{pq}R^r_{\phantom{r} (a}R_{|rqp| b)}
+(1/360)  (\xi -1/6)  R R^{pqr}_{\phantom{pqr}a }R_{pqr b} -
(1/4320)  R_{ab}R^{pqrs} R_{pqrs } \nonumber \\
& &  \quad  - (1/1890)  R^p_{\phantom{p} (a}R^{qrs}_{\phantom{qrs}
|p } R_{qrs| b)} -(1/3780)  R^{pq}R^{rs}_{\phantom{rs} pa}R_{rsq
b} +(1/1890)  R_{pq}R^{p r q s}R_{r a s b} \nonumber \\
& &  \quad -(1/7560)  R_{pq}R^{p rs}_{\phantom{p
rs}a}R^q_{\phantom{q} rs b} + (1/3780)  R^{pqrs}R_{pq t a
}R_{rs \phantom{t} b}^{\phantom{rs} t}   \nonumber \\
& &  \quad  + (1/378)  R^{p r qs}R^t_{\phantom{t} pq a}R_{t rs b}
-(1/3780)  R^{pqr}_{\phantom{pqr} s } R_{pqr t}R^{s \phantom{a}
t}_{\phantom{s} a \phantom{t} b} \label{CoefT_U0_U1_V0_V1_d6 V0c}
\end{eqnarray}
\end{subequations}
and
\begin{eqnarray} \label{CoefT_U0_U1_V0_V1_d6 V1}
& & v_1= -(1/48)  m^6  - (1/16)  (\xi -1/6)
  m^4  R + (1/48)  (\xi-1/5)  m^2  \Box R \nonumber \\
& & \qquad -(1/16)  (\xi-1/6)^2  m^2  R^2   +(1/1440) m^2  R_{pq}
R^{pq}
  - (1/1440)  m^2  R_{pqrs} R^{pqrs}
   \nonumber \\
& &   \qquad  -(1/480)   (\xi -3/14)  \Box \Box R    + (1/48)
(\xi-1/6) (\xi-1/5)  R\Box R - (1/720)  (\xi -3/14)   R_{;p q}
R^{pq}  \nonumber \\
& &   \qquad  -(1/5040)  R_{pq} \Box R^{pq} + (1/840)  R_{pq ;
rs}R^{prqs}  + (1/96)  [\xi^2- (2/5)  \xi +17/420]  R_{;p}R^{;p}
\nonumber \\
& &   \qquad -(1/20160) R_{pq;r} R^{pq;r}  -(1/10080) R_{pq;r}
R^{pr;q} + (1/4480)  R_{pqrs;t} R^{pqrs;t}
    \nonumber \\
& &    \qquad  - (1/48)  (\xi-1/6)^3  R^3   + (1/1440)  (\xi-1/6)
RR_{pq} R^{pq}   + (1/45360)  R_{pq}
R^{p}_{\phantom{p} r}R^{qr} \nonumber \\
& &   \qquad -(1/15120)  R_{pq}R_{rs}R^{prqs}  -(1/1440) (\xi-1/6)
 RR_{pqrs} R^{pqrs}
  + (1/2160)  R_{pq}R^p_{\phantom{p} rst} R^{qrst }
\nonumber \\
& &   \qquad  -(1/5670)  R_{pqrs}R^{pquv} R^{rs}_{\phantom{rs} uv}
-(11/11340)  R_{prqs} R^{p \phantom{u} q}_{\phantom{p} u \phantom{q}
v} R^{r u s v}.
\end{eqnarray}

The geometrical tensor $\Theta^{M^2}_{\mu \nu}$ which is associated
with the renormalization mass is obtained from (\ref{VExpVSET_amb4})
by using (\ref{CoefT_U0_U1_V0_V1_d6 V0a}),
(\ref{CoefT_U0_U1_V0_V1_d6 V0c}) and (\ref{CoefT_U0_U1_V0_V1_d6 V1})
and is given by
\begin{eqnarray}\label{VExpVSET_amb_d6}
& & \Theta^{M^2}_{\mu \nu} = \frac{\ln M^2}{2\,(2 \pi )^3}
\left[\phantom{\frac{}{}}  (1/8)(\xi-1/6) m^4  R_{\mu \nu} -(1/4)
[\xi^2-(1/3)\xi+1/30]  m^2  R_{;\mu \nu} + (1/240)  m^2  \Box R_{\mu
\nu} \right. \nonumber \\
& & \quad \left. +(1/4) (\xi-1/6)^2  m^2   R R_{\mu \nu} - (1/180)
m^2 R_{p \mu} R^p_{\phantom{p}\nu} + (1/360)  m^2  R^{pq} R_{p \mu q
\nu}+ (1/360)  m^2
R^{pqr}_{\phantom{pqr}\mu} R_{pqr \nu} \right. \nonumber \\
& & \quad \left. +(1/24) [\xi^2-(2/5)\xi+3/70] (\Box R)_{;\mu \nu}
 -(1/3360) \Box \Box R_{\mu \nu} - (1/4) (\xi -1/6)[\xi^2-(1/3)\xi+1/30]  R R_{;\mu \nu}
   \right. \nonumber \\
& & \quad \left.  -(1/24) (\xi -1/6)(\xi -1/5)   (\Box R) R_{\mu
\nu} +(1/360)  (\xi -1/7)  R_{;p (\mu} R^p_{\phantom{p}\nu)}
      +(1/240)(\xi -1/6)   R \Box R_{\mu \nu} \right. \nonumber \\
& & \quad \left. +(1/1008)  R_{p (\mu} \Box R^p_{\phantom{p}\nu)}
+(1/360)(\xi -2/7) R^{pq} R_{pq;(\mu \nu)}  +(1/1260)  R^{pq} R_{p
(\mu ; \nu)q} -
(1/1680)  R^{pq} R_{\mu \nu; pq} \right. \nonumber \\
& & \quad \left. + (1/180)  (\xi -3/14)  R^{;pq}R_{p \mu q \nu}  -
(1/2520)  (\Box R^{pq})R_{p \mu q \nu} +(1/630)  R^{p
q;r}_{\phantom{p q;r}
(\mu} R_{|rqp| \nu)}  \right. \nonumber \\
& & \quad \left. + (1/420)  R^{p \phantom{(\mu}; q r}_{\phantom{p }
(\mu} R_{|pqr | \nu)}  - (1/360)(\xi-3/14) R^{pqrs} R_{pqrs ; (\mu
\nu) } - (1/4) (\xi
-1/6)^2 (\xi-1/4) R_{;\mu} R_{;\nu} \right. \nonumber \\
& & \quad \left. -(1/120)  (\xi -3/14)  R_{;p }
R^p_{\phantom{p}(\mu;\nu)} + (1/120)  (\xi -17/84)  R_{;p } R_{\mu
\nu}^{\phantom{\mu \nu}; p}  +
(1/360) (\xi-1/4) R^{pq}_{\phantom{pq};\mu}R_{pq;\nu}   \right. \nonumber \\
& & \quad \left.  - (1/5040)  R^p_{\phantom{p} \mu;q} R_{p
\nu}^{\phantom{p \nu};q}  + (1/1008)  R^p_{\phantom{p} \mu;q}
R^q_{\phantom{q} \nu;p } - (1/2520)  R^{pq;r} R_{rqp (\mu;\nu)}  -
(1/1680)
R^{pq;r} R_{p \mu q \nu;r} \right. \nonumber \\
& & \quad \left.  - (1/360)(\xi-13/56) R^{pqrs}_{\phantom{pqrs};\mu}
R_{pqrs ; \nu } -
(1/1680)  R^{pqr}_{\phantom{pqr}\mu;s}R_{pqr \nu}^{\phantom{pqr \nu};s}   \right. \nonumber \\
& & \quad \left. + (1/8)  (\xi -1/6)^3  R^2 R_{\mu \nu}  - (1/180)
(\xi
-1/6)  R R_{p \mu} R^p_{\phantom{p}\nu} -(1/720) (\xi -1/6) R^{pq}R_{pq}R_{\mu \nu} \right. \nonumber \\
& & \quad \left. - (1/3780)  R^{pq}R_{p \mu}R_{q \nu}  + (1/360)
(\xi -1/6)  R R^{pq}R_{p \mu q \nu} +(1/7560)
R^{pr}R^q_{\phantom{q} r}R_{p \mu
q \nu} \right. \nonumber \\
& & \quad \left. + (1/7560)  R^{pq}R^r_{\phantom{r} (\mu}R_{|rqp|
\nu)} +(1/360)  (\xi -1/6)  R R^{pqr}_{\phantom{pqr}\mu }R_{pqr
\nu}+
(1/720)(\xi-1/6) R_{\mu \nu}R^{pqrs} R_{pqrs } \right. \nonumber \\
& & \quad \left. - (1/1890)  R^p_{\phantom{p}
(\mu}R^{qrs}_{\phantom{qrs} |p } R_{qrs| \nu)} -(1/3780)
R^{pq}R^{rs}_{\phantom{rs} p\mu}R_{rsq
\nu} +(1/1890)  R_{pq}R^{p r q s}R_{r \mu s \nu} \right. \nonumber \\
& & \quad \left. -(1/7560)  R_{pq}R^{p rs}_{\phantom{p
rs}\mu}R^q_{\phantom{q} rs \nu} + (1/3780)  R^{pqrs}R_{pq t \mu
}R_{rs \phantom{t} \nu}^{\phantom{rs} t}   \right. \nonumber \\
& & \quad \left. + (1/378)  R^{p r qs}R^t_{\phantom{t} pq \mu}R_{t
rs \nu} -(1/3780)  R^{pqr}_{\phantom{pqr} s } R_{pqr t}R^{s
\phantom{\mu} t}_{\phantom{s} \mu \phantom{t} \nu}
 \right. \nonumber \\
& & \qquad ~~   \left. +g_{\mu \nu} \left(\phantom{\frac{}{}}
-(1/48)  m^6 - (1/16)  (\xi -1/6) m^4  R + (1/4)
[\xi^2-(1/3)\xi+1/40]  m^2  \Box R
\right. \right. \nonumber \\
& & \qquad \left.\left.  -(1/16)  (\xi-1/6)^2  m^2  R^2 +(1/1440)
m^2  R_{pq} R^{pq}
  - (1/1440)  m^2  R_{pqrs} R^{pqrs}
   \right. \right. \nonumber \\
& & \qquad \left.\left.  -(1/24)   [\xi^2-(2/5)\xi +11/280]  \Box
\Box R + (1/4)  (\xi-1/6) [\xi^2-(1/3)\xi +1/40]  R\Box R
\right. \right. \nonumber \\
& & \qquad \left.\left. - (1/720)  (\xi -3/14)   R_{;p q} R^{pq}
-(1/360) (\xi-5/28)  R_{pq} \Box R^{pq} + (1/90) (\xi-1/7)
R_{pq ; rs}R^{prqs}  \right. \right. \nonumber \\
& & \qquad \left.\left. + (1/4)  [\xi^3-(13/24)\xi^2+ (17/180) \xi -
53/10080]  R_{;p}R^{;p}
\right. \right. \nonumber \\
& & \qquad \left.\left. -(1/360) (\xi-13/56)  R_{pq;r} R^{pq;r}
-(1/10080) R_{pq;r} R^{pr;q} + (1/360)  (\xi-19/112)  R_{pqrs;t}
R^{pqrs;t}
    \right. \right. \nonumber \\
& & \qquad \left.\left.  - (1/48)  (\xi-1/6)^3  R^3   + (1/1440)
(\xi-1/6)   RR_{pq} R^{pq}   + (1/45360)  R_{pq}
R^{p}_{\phantom{p} r}R^{qr} \right. \right. \nonumber \\
& & \qquad \left.\left. -(1/15120)  R_{pq}R_{rs}R^{prqs} -(1/1440)
 (\xi-1/6)  RR_{pqrs} R^{pqrs}
  + (1/180)  (\xi-1/6)  R_{pq}R^p_{\phantom{p} rst} R^{qrst }
\right. \right. \nonumber \\
& & \qquad \left.\left.  -(1/360)  (\xi-47/252)  R_{pqrs}R^{pquv}
R^{rs}_{\phantom{rs} uv} -(1/90)  (\xi-41/252)  R_{prqs} R^{p
\phantom{u} q}_{\phantom{p} u \phantom{q} v} R^{r u s v}
\phantom{\frac{}{}}\right) \right].
\end{eqnarray}

The trace anomaly (\ref{TR_Anomaly_ev_res}) is obtained by using
$m^2=0$ and $\xi=\xi_c(6)=1/5$ into (\ref{CoefT_U0_U1_V0_V1_d6 V1}).
It reduces to
\begin{eqnarray}\label{TR_Anomaly_d6_res}
& & \langle \psi |T^{\mu}_{\phantom{\mu} \mu} |\psi
\rangle_{\mathrm{ren}} = \frac{1}{(2\pi)^3} \, [(1/33600)  \Box \Box
R     + (1/50400) R_{;p q} R^{pq}   -(1/5040)  R_{pq} \Box R^{pq} +
(1/840)  R_{pq ; rs}R^{prqs} \nonumber \\
& &    \quad  + (1/201600) R_{;p}R^{;p} -(1/20160) R_{pq;r} R^{pq;r}
-(1/10080) R_{pq;r} R^{pr;q} + (1/4480)  R_{pqrs;t} R^{pqrs;t}
\nonumber \\
& &    \quad - (1/1296000)  R^3   + (1/43200) RR_{pq} R^{pq}   +
(1/45360)  R_{pq} R^{p}_{\phantom{p} r}R^{qr} -(1/15120)
R_{pq}R_{rs}R^{prqs} \nonumber \\
& &   \quad  -(1/43200)
 RR_{pqrs} R^{pqrs}
  + (1/2160)  R_{pq}R^p_{\phantom{p} rst} R^{qrst }
 -(1/5670)  R_{pqrs}R^{pquv} R^{rs}_{\phantom{rs} uv}
-(11/11340)  R_{prqs} R^{p \phantom{u} q}_{\phantom{p} u \phantom{q}
v} R^{r u s v}]. \nonumber \\
& &
\end{eqnarray}

\end{widetext}

\subsection{D=7,8,9,10,11}

The complexity of the explicit expressions of
$G^{\mathrm{F}}_{\mathrm{sing}} (x,x')$ and of the geometrical
tensor $\Theta^{M^2}_{\mu \nu}$  greatly increases with the
dimension $D$ of spacetime. That clearly appears in the previous
subsections. For this reason, we cannot write them explicitly for
spacetime dimension from $D=7$ to $D=11$ even though we have at our
disposal all the tools permitting us to carry out all the necessary
calculations. Indeed, in the appendices of
Ref.~\cite{DecaniniFolacci2005a}, we have obtained the covariant
Taylor series expansions of the Van Vleck -Morette determinant
$U_0(x,x')={\Delta ^{1/2}}(x,x')$ up to order $\sigma^{11/2}$ and of
the bitensor $\sigma^{\mu \nu}(x,x')$ up to order $\sigma^{9/2}$. We
have also developed the general theory permitting us to construct
the covariant derivative and the d'Alembertian of an arbitrary
biscalar $F(x,x')$ symmetric in the exchange of $x$ and $x'$. From a
theoretical point of view, all these results could permit us to
solve the recursion relations (\ref{HevRep5A}) and (\ref{HevRep5B})
for $D$ even and the recursion relations (\ref{HodRep3}) for $D$ odd
and therefore to obtain the explicit expressions of
$G^{\mathrm{F}}_{\mathrm{sing}} (x,x')$ up to the required order and
of the geometrical tensor $\Theta^{M^2}_{\mu \nu}$ when necessary.
Of course, this could be realized but at the cost of odious
calculations in a general spacetime.

By contrast, in a given spacetime, i.e. if we know explicitly the
Riemann tensor $R_{\mu \nu \rho \sigma}$ and therefore the Ricci
tensor $R_{\mu \nu }$ and the scalar curvature $R$, interesting
simplifications may occur, the construction of
$G^{\mathrm{F}}_{\mathrm{sing}} (x,x')$ and of $\Theta^{M^2}_{\mu
\nu}$ done explicitly and the renormalization of the expectation
value of the stress-energy tensor ``easily" achieved. For example,
in $D$-dimensional Schwarzschild black hole spacetimes where we have
$R=0$, $R_{\mu \nu }=0$ and more generally in Ricci-flat spacetimes,
considerable simplifications could permit us to obtain explicitly
$G^{\mathrm{F}}_{\mathrm{sing}} (x,x')$ and $\Theta^{M^2}_{\mu \nu}$
even for $D>6$. This certainly also happens in $D$-dimensional
spacetimes such as $\mathrm{AdS}_p \times \mathrm{S}_q$ with $p+q=D$
where the covariant derivative of the Riemann tensor vanishes
($R_{\mu \nu \rho \sigma; \tau}=0$) as well as in $D$-dimensional de
Sitter and Anti-de Sitter spacetimes, i.e. in maximally symmetric
spacetimes, where $R_{\mu \nu \rho \sigma}=[R/D(D-1)](g_{\mu
\rho}g_{\nu \sigma}-g_{\mu \sigma}g_{\nu \rho})$ with $R=
\mathrm{Cte}$.

\section{Important remarks and complements}

In this section, we shall complete our study by discussing some
aspects of the Hadamard renormalization of the stress-energy tensor
which are more or less directly related to the explicit calculations
described in Secs.~II and III. They are helpful in order to simplify
some of the results displayed above. Furthermore, they permit us to
discuss more generally the ambiguity problem and the trace anomaly
as well as to clarify the links existing between the Hadamard
formalism and the more popular method based on regularization and
renormalization in the effective action.

\subsection{Ambiguities and trace anomaly}

As already noted in Sec.~II, the renormalized expectation value
$\langle \psi |T_{\mu \nu} |\psi \rangle_{\mathrm{ren}}$ is unique
up to the addition of a local conserved tensor $\Theta_{\mu \nu}$.
For $D$ even, we have been able to construct the standard ambiguity
associated with the choice of the renormalization mass $M$ [see
Eq.~(\ref{VExpVSET_amb4})] and, in Sec.~III, we have explicitly
obtained its expression for $D=2,4$ and $6$ [see
Eqs.~(\ref{VExpVSET_amb_d2}),(\ref{VExpVSET_amb_d4}) and
(\ref{VExpVSET_amb_d6})]. In the present subsection, following
Wald's arguments of Ref.~\cite{Wald78}, we shall push further our
discussion and provide for $D=2,3,4,5$ and $6$, the bases (i.e., all
the independent conserved local tensors) permitting us to
constructed the most general expression for the tensor $\Theta_{\mu
\nu}$. Here, we adhere to a conventional point of view \cite{Wald78}
by discarding ambiguities diverging as $m^2 \to 0$. It should be
however noted that a less conventional point of view has been
considered by Tichy and Flanagan in Ref.~\cite{TichyFlanagan98}.

In order to extend Wald's arguments, it is important to keep in mind
that $\Theta_{\mu \nu}$ is a local conserved tensor of dimension
$(\mathrm{mass})^D$ and that it can be obtained by functional
derivation with respect to the metric tensor from a geometrical
Lagrangian of dimension $(\mathrm{mass})^D$. We note also that
$g_{\mu \nu}$ is dimensionless while $R$, $R_{\mu \nu}$ and $R_{\mu
\nu \rho \sigma}$ have dimension $(\mathrm{mass})^2$.

\subsubsection{$D=2$}

For $D=2$, there are only two ``independent" geometrical Lagrangians
of dimension $(\mathrm{mass})^2$ which remain finite in the massless
limit: ${\cal L}=m^2$ and ${\cal L}= R$. However, by functional
derivation, the latter does not provide any contribution to
$\Theta_{\mu \nu}$ because, in two dimensions, the Euler number
\begin{equation} \label{Euler_D2}
\int_{\cal M} d^2 x \sqrt{-g}R
\end{equation}
is a topological invariant. $\Theta_{\mu \nu}$ is then necessarily
proportional to the functional derivative of ${\cal L}=m^2$ and
therefore of the form
\begin{equation}\label{Theta_munu_amb2}
\Theta_{\mu \nu}=A \, m^2 g_{\mu \nu}
\end{equation}
where $A$ is a dimensionless constant.

It is interesting to note that $\Theta_{\mu \nu}$ given by
(\ref{Theta_munu_amb2}) vanishes for $m^2=0$ and therefore does not
modify the trace anomaly (\ref{TR_Anomaly_d2_res}).

\subsubsection{$D=3$}

For $D=3$, there are only two independent geometrical Lagrangians of
dimension $(\mathrm{mass})^3$ which remain finite in the massless
limit: ${\cal L}=m^3$ and ${\cal L}=m R$. So, it is natural to
consider that $\Theta_{\mu \nu}$ is necessarily a linear combination
of their functional derivatives $m^3 g_{\mu \nu}/2$ and $m
[(1/2)Rg_{\mu \nu}-R_{\mu \nu}]$, i.e. that
\begin{equation}\label{Theta_munu_amb3}
\Theta_{\mu \nu}=A \, m^3 g_{\mu \nu} + + B\, m^2[R_{\mu
\nu}-(1/2)Rg_{\mu \nu}]
\end{equation}
where $A$ and $B$ are dimensionless constants.

It should be noted that $\Theta_{\mu \nu}$ given by
(\ref{Theta_munu_amb3}) vanishes for $m=0$. Thus, it cannot be used
in order to modified (\ref{TR_Anomaly_od_res}). In other words, the
trace anomaly does not exist for $D=3$ even if we take into account
the possible ambiguities of the Hadamard renormalization process.

\subsubsection{$D=4$}

For $D=4$, there are five ``independent" geometrical Lagrangians of
dimension $(\mathrm{mass})^4$ which remain finite in the massless
limit: ${\cal L}=m^4$, ${\cal L}=m^2 R$, ${\cal L}= R^2$, ${\cal L}=
R_{pq}R^{pq}$ and ${\cal L}= R_{pqrs}R^{pqrs}$. By functional
derivation with respect to the metric tensor, they define the
conserved tensors $m^4 g_{\mu \nu}/2$, $m^2[(1/2)Rg_{\mu \nu}-R_{\mu
\nu}]$ as well as the three conserved tensors of rank 2 and order 4
\begin{subequations} \label{FD_04_1}
\begin{eqnarray}
  H_{\mu \nu}^{(4,2)(1)} &\equiv& \frac{1}{ \sqrt{-g}}
\frac{\delta}{\delta g^{\mu \nu}} \int_{\cal M} d^D
x \sqrt{-g}~R^2   \label{FD_04_1a}\\
&=& 2 \, R_{;\mu \nu}-2\, RR_{\mu \nu} \nonumber \\
&&   \quad + g_{\mu \nu}[-2 \, \Box R + (1/2) \, R^2],
\label{FD_04_1b}
\end{eqnarray}
\end{subequations}
\begin{subequations}\label{FD_04_2}
\begin{eqnarray}
  H_{\mu \nu}^{(4,2)(2)}  &\equiv& \frac{1}{ \sqrt{-g}}
\frac{\delta}{\delta g^{\mu \nu}} \int_{\cal M} d^D
x \sqrt{-g}~R_{pq}R^{pq}  \label{FD_04_2a}\\
&=& R_{;\mu \nu} - \Box R_{\mu \nu} -2 \, R^{pq}R_{p\mu q\nu } \nonumber \\
&&    \quad + g_{\mu \nu} [ -(1/2) \, \Box R +(1/2) \, R_{pq}R^{pq}], \nonumber \\
&& \label{FD_04_2b}
\end{eqnarray}
\end{subequations}
\begin{subequations}\label{FD_04_3}
\begin{eqnarray}
 H_{\mu \nu}^{(4,2)(3)} &\equiv& \frac{1}{ \sqrt{-g}}
\frac{\delta}{\delta g^{\mu \nu}} \int_{\cal M} d^D
x \sqrt{-g}~R_{pqrs}R^{pqrs}   \label{FD_04_3a}\\
&=& 2 \, R_{;\mu \nu} - 4\, \Box R_{\mu \nu} +4R^p_{\phantom{p}
 \mu}R_{p\nu}-4R^{pq}R_{p\mu q \nu} \nonumber  \\
 &&  \quad -2R^{pqr}_{\phantom{pqr}\mu}R_{pqr\nu}
 + g_{\mu \nu} [ (1/2) \, R_{pqrs}R^{pqrs}]. \nonumber \\
&& \label{FD_04_3b}
\end{eqnarray}
\end{subequations}
$\Theta_{\mu \nu}$ is therefore necessarily of the form
\begin{eqnarray}\label{GenFormAmb_4}
&& \Theta_{\mu \nu}= A\, m^4 g_{\mu \nu} + B\, m^2[R_{\mu
\nu}-(1/2)Rg_{\mu
\nu}]  \nonumber \\
&& \qquad + C_1 \, H_{\mu \nu}^{(4,2)(1)} + C_2 \,
H_{\mu \nu}^{(4,2)(2)}+ C_3 \, H_{\mu \nu}^{(4,2)(3)} \nonumber \\
&&
\end{eqnarray}
where $A$, $B$, $C_1$, $C_2$ and $C_3$ are dimensionless constants.
Here, it should be also noted that it is possible to simplify the
previous expression because, in a four-dimensional background, the
Euler number
\begin{equation} \label{Euler_D4}
\int_{\cal M} d^4 x \sqrt{-g}~{\cal L}_{(2)},
\end{equation}
where ${\cal L}_{(2)}$ is the quadratic Gauss-Bonnet Lagrangian
given by
\begin{equation}\label{actionLovelock_2}
{\cal L}_{(2)}=R^2-4R_{pq}R^{pq}+R_{pqrs}R^{pqrs},
\end{equation}
is a topological invariant. By functional derivation of
(\ref{Euler_D4}) we obtain
\begin{equation}\label{GB_dim4}
H_{\mu \nu}^{(4,2)(1)}-4 \, H_{\mu \nu}^{(4,2)(2)} + H_{\mu
\nu}^{(4,2)(3)}=0
\end{equation}
which could be helpful in order to eliminate one of the three
conserved tensors of rank 2 and order 4 into (\ref{GenFormAmb_4}).
In other words, without loss of generality it is possible to use
$C_1=0$ or $C_2=0$ or $C_3=0$ into (\ref{GenFormAmb_4}).

It should be noted that the ``basis" exhibited above which has
permitted us to provide the general form for the tensor $\Theta_{\mu
\nu}$ can be used to simplify considerably the expression
(\ref{VExpVSET_amb_d4}) obtained for $\Theta^{M^2}_{\mu \nu}$.
Indeed, from Eqs.~(\ref{FD_04_1})-(\ref{FD_04_3}), we can write
\begin{eqnarray}\label{VExpVSET_amb_d4_simp2}
& & \Theta^{M^2}_{\mu \nu}  = \frac{\ln M^2}{(4 \pi )^2} \times
\left(  +(1/2)(\xi-1/6)^2 \, H_{\mu \nu}^{(4,2)(1)} \right. \nonumber \\
& &\quad  \left. -(1/180) \,
H_{\mu \nu}^{(4,2)(2)} + (1/180)  \, H_{\mu \nu}^{(4,2)(3)} \right. \nonumber \\
& &\quad  \left.  - (\xi-1/6)m^2 [R_{\mu \nu} -(1/2)R g_{\mu \nu}] \right. \nonumber \\
& &\quad  \left. +(1/4)m^4 g_{\mu \nu}  \phantom{\frac{}{}} \right).
\end{eqnarray}

Finally, it is interesting to note that $\Theta_{\mu \nu}$ given by
(\ref{GenFormAmb_4}) can be used in order to modify the trace
anomaly (\ref{TR_Anomaly_d4_res}). Indeed, for $m^2=0$ and by using
Eqs.~(\ref{Tr_H_421})-(\ref{Tr_H_423}) with $D=4$, we obtain
\begin{eqnarray}\label{GenFormAmb_4_trace}
&& g^{\mu \nu} \Theta_{\mu \nu}= [-6 C_1 -2 C_2 -2 C_3] \Box R.
\end{eqnarray}
For example, by taking $C_1=1/4320(2\pi)^2$ and $C_2=C_3=0$, we can
remove the $\Box R$ term from (\ref{TR_Anomaly_d4_res}). This
elimination can be achieved by adding a finite $R^2$ term to the
gravitational Lagrangian [see Eq.~(\ref{FD_04_1})] and is in
accordance with the discussion we shall develop in Sec.~IV.B. On the
contrary, the $R_{pq}R^{pq}$ term and the $R_{pqrs}R^{pqrs}$ term
cannot be modified. We refer to Sec.~6.3 of
Ref.~\cite{BirrellDavies} for various physical comments concerning
the possible modifications of the trace anomaly in a four
dimensional gravitational background.

\subsubsection{$D=5$}

For $D=5$, there are five independent geometrical Lagrangians of
dimension $(\mathrm{mass})^5$ which remain finite in the massless
limit: ${\cal L}=m^5$, ${\cal L}=m^3 R$, ${\cal L}= m R^2$, ${\cal
L}= m R_{pq}R^{pq}$ and ${\cal L}= m R_{pqrs}R^{pqrs}$. By
functional derivation, they define the conserved tensors $m^5 g_{\mu
\nu}/2$, $m^3[(1/2)Rg_{\mu \nu}-R_{\mu \nu}]$ as well as the three
conserved tensors of rank 2 and order 4 $m H_{\mu \nu}^{(4,2)(1)}$,
$m H_{\mu \nu}^{(4,2)(2)}$ and $m H_{\mu \nu}^{(4,2)(3)}$.
$\Theta_{\mu \nu}$ is therefore necessarily of the form
\begin{eqnarray}\label{GenFormAmb_5}
&& \Theta_{\mu \nu}= A\, m^5 g_{\mu \nu} + B\, m^3[R_{\mu
\nu}-(1/2)Rg_{\mu
\nu}]  \nonumber \\
&& \qquad + C_1 \, m H_{\mu \nu}^{(4,2)(1)} + C_2 \,
m H_{\mu \nu}^{(4,2)(2)}+ C_3 \, m H_{\mu \nu}^{(4,2)(3)} \nonumber \\
&&
\end{eqnarray}
where $A$, $B$, $C_1$, $C_2$ and $C_3$ are dimensionless constants.
Here, the conserved tensors $H_{\mu \nu}^{(4,2)(1)}$, $H_{\mu
\nu}^{(4,2)(2)}$ and $H_{\mu \nu}^{(4,2)(3)}$ are still respectively
defined by Eqs.~(\ref{FD_04_1a}), (\ref{FD_04_2a}) and
(\ref{FD_04_3a}) but now, in these equations, $D=4$ must be replaced
by $D=5$. Their explicit expressions (\ref{FD_04_1b}),
(\ref{FD_04_2b}) and (\ref{FD_04_3b}) remain unchanged. For $D=5$,
it is not possible to simplify Eq.~(\ref{GenFormAmb_5}) by using
(\ref{GB_dim4}). Indeed, for $D>4$ this topological constraint is
not valid because the Euler number (\ref{Euler_D4}) does not remain
a topological invariant.

It should be noted that $\Theta_{\mu \nu}$ given by
(\ref{GenFormAmb_5}) vanishes for $m=0$. Thus, it cannot be used in
order to modified (\ref{TR_Anomaly_od_res}). In other words, the
trace anomaly does not exist for $D=5$ even if we take into account
the possible ambiguities of the Hadamard renormalization process.

\subsubsection{$D=6$}

For $D=6$, there are fifteen ``independent" geometrical Lagrangians
of dimension $(\mathrm{mass})^6$ which remain finite in the massless
limit: ${\cal L}=m^6$, ${\cal L}=m^4 R$ and the three Riemann
polynomials of rank 0 and order 4 ${\cal L}= m^2 R^2$, ${\cal L}=
m^2 R_{pq}R^{pq}$, ${\cal L}= m^2 R_{pqrs}R^{pqrs}$ as well as the
ten Riemann monomials of rank 0 and order 6 (see
Refs.~\cite{FKWC1992,DecaniniFolacci_arXiv2008}) ${\cal L}=R\Box R$,
${\cal L}=R_{pq} \Box R^{pq}$, ${\cal L}=R^3$, ${\cal L}=RR_{pq}
R^{pq}$, ${\cal L}=R_{pq} R^{p}_{\phantom{p} r}R^{qr}$, ${\cal
L}=R_{pq}R_{rs}R^{prqs}$, ${\cal L}=RR_{pqrs} R^{pqrs}$, ${\cal
L}=R_{pq}R^p_{\phantom{p} rst} R^{qrst }$, ${\cal
L}=R_{pqrs}R^{pquv} R^{rs}_{\phantom{rs} uv}$, ${\cal L}=R_{prqs}
R^{p \phantom{u} q}_{\phantom{p} u \phantom{q} v} R^{r u s v}$. By
functional derivation, they define the conserved tensors $m^6 g_{\mu
\nu}/2$, $m^4[(1/2)Rg_{\mu \nu}-R_{\mu \nu}]$ and the three
conserved tensors of rank 2 and order 4 $m^2 H_{\mu
\nu}^{(4,2)(1)}$, $m^2 H_{\mu \nu}^{(4,2)(2)}$ and $m^2 H_{\mu
\nu}^{(4,2)(3)}$ as well as the ten conserved tensors of rank 2 and
order 6
\begin{eqnarray}
& & H_{\mu \nu}^{\lbrace{2,0\rbrace}(1)}  \equiv \frac{1}{
\sqrt{-g}}\frac{\delta}{\delta g^{\mu \nu}} \int_{\cal M} d^D x
\sqrt{-g}~R\Box R  \label{FD_06_1} \\
&& H_{\mu \nu}^{\lbrace{2,0\rbrace}(3)}  \equiv  \frac{1}{
\sqrt{-g}}\frac{\delta}{\delta g^{\mu \nu}} \int_{\cal M} d^D x
\sqrt{-g}~R_{pq} \Box R^{pq}  \label{FD_06_2} \\
& &
  H_{\mu \nu}^{(6,3)(1)} \equiv \frac{1}{
\sqrt{-g}}\frac{\delta}{\delta g^{\mu \nu}} \int_{\cal M} d^D x
\sqrt{-g}~R^3  \label{FD_06_3} \\
& & H_{\mu \nu}^{(6,3)(2)} \equiv \frac{1}{
\sqrt{-g}}\frac{\delta}{\delta g^{\mu \nu}} \int_{\cal M} d^D x
\sqrt{-g}~RR_{pq} R^{pq} \label{FD_06_4} \\
& & H_{\mu \nu}^{(6,3)(3)}  \equiv \frac{1}{
\sqrt{-g}}\frac{\delta}{\delta g^{\mu \nu}} \int_{\cal M} d^D x
\sqrt{-g}~R_{pq} R^{p}_{\phantom{p} r}R^{qr}  \label{FD_06_5} \\
& & H_{\mu \nu}^{(6,3)(4)}  \equiv \frac{1}{
\sqrt{-g}}\frac{\delta}{\delta g^{\mu \nu}} \int_{\cal M} d^D x
\sqrt{-g}~R_{pq}R_{rs}R^{prqs}
\label{FD_06_6} \\
& & H_{\mu \nu}^{(6,3)(5)} \equiv \frac{1}{
\sqrt{-g}}\frac{\delta}{\delta g^{\mu \nu}} \int_{\cal M} d^D x
\sqrt{-g}~RR_{pqrs} R^{pqrs} \label{FD_06_7} \\
& & H_{\mu \nu}^{(6,3)(6)} \equiv \frac{1}{
\sqrt{-g}}\frac{\delta}{\delta g^{\mu \nu}} \int_{\cal M} d^D x
\sqrt{-g}~R_{pq}R^p_{\phantom{p} rst} R^{qrst } \nonumber \\
&&\label{FD_06_8} \\
& &H_{\mu \nu}^{(6,3)(7)} \equiv  \frac{1}{
\sqrt{-g}}\frac{\delta}{\delta g^{\mu \nu}} \int_{\cal M} d^D x
\sqrt{-g}~R_{pqrs}R^{pquv} R^{rs}_{\phantom{rs} uv} \nonumber \\
&&\label{FD_06_9} \\
& & H_{\mu \nu}^{(6,3)(8)} \equiv \frac{1}{
\sqrt{-g}}\frac{\delta}{\delta g^{\mu \nu}} \int_{\cal M} d^D x
\sqrt{-g} ~R_{prqs} R^{p \phantom{u} q}_{\phantom{p} u
\phantom{q} v} R^{r u s v}. \nonumber \\
&&\label{FD_06_10}
\end{eqnarray}
Here we do not provide the explicit expressions of these ten
tensors. They are very complicated ones and can be found in
Ref.~\cite{DecaniniFolacci2007} [see Eqs.~(2.22)-(2.31) of this
article]. As far as the conserved tensors $H_{\mu \nu}^{(4,2)(1)}$,
$H_{\mu \nu}^{(4,2)(2)}$ and $H_{\mu \nu}^{(4,2)(3)}$ are concerned,
they are still respectively defined by Eqs.~(\ref{FD_04_1a}),
(\ref{FD_04_2a}) and (\ref{FD_04_3a}) but now, in these equations,
$D=4$ must be replaced by $D=6$. Their explicit expressions
(\ref{FD_04_1b}), (\ref{FD_04_2b}) and (\ref{FD_04_3b}) remain
unchanged. $\Theta_{\mu \nu}$ is therefore necessarily of the form
\begin{eqnarray}\label{GenFormAmb_6}
&& \Theta_{\mu \nu}= A\, m^6 g_{\mu \nu} + B\, m^4[(1/2)Rg_{\mu
\nu}-R_{\mu \nu}]  \nonumber \\
&& \quad + C_1 \, m^2 H_{\mu \nu}^{(4,2)(1)} + C_2 \,
m^2 H_{\mu \nu}^{(4,2)(2)}+ C_3 \, m^2 H_{\mu \nu}^{(4,2)(3)} \nonumber \\
&& \quad + D_1 \, H_{\mu \nu}^{\lbrace{2,0\rbrace}(1)} + D_2 \,
H_{\mu \nu}^{\lbrace{2,0\rbrace}(3)} + D_3 \, H_{\mu \nu}^{(6,3)(1)} \nonumber \\
&& \quad + D_4 \, H_{\mu \nu}^{(6,3)(2)} + D_5 \, H_{\mu
\nu}^{(6,3)(3)} + D_6 \, H_{\mu \nu}^{(6,3)(4)} \nonumber \\
&& \quad + D_7 \, H_{\mu \nu}^{(6,3)(5)} + D_8 \, H_{\mu
\nu}^{(6,3)(6)} + D_9 \, H_{\mu \nu}^{(6,3)(7)} \nonumber \\
&& \quad + D_{10} \, H_{\mu \nu}^{(6,3)(8)}
\end{eqnarray}
where $A$, $B$, $C_1$, $C_2$ and $C_3$ as well as $D_1$, ... $D_9$
and $D_{10}$ are dimensionless constants. Finally, it should be
noted that it is possible to simplify the previous expression for
$\Theta_{\mu \nu}$ because, in a six-dimensional background, the
Euler number
\begin{equation} \label{Euler_D6}
\int_{\cal M} d^6 x \sqrt{-g}~{\cal L}_{(3)},
\end{equation}
where ${\cal L}_{(3)}$ is the cubic Lovelock Lagrangian explicitly
given by
\begin{eqnarray}\label{actionLovelock_3}
&& {\cal L}_{(3)}= R^3 -12 \,  RR_{pq} R^{pq} + 16 \, R_{pq}
R^{p}_{\phantom{p} r}R^{qr}  \nonumber \\
&&    + 24 \, R_{pq}R_{rs}R^{prqs} + 3 \,RR_{pqrs} R^{pqrs}
  -24\, R_{pq}R^p_{\phantom{p} rst} R^{qrst } \nonumber \\
&&    + 4 \, R_{pqrs}R^{pquv} R^{rs}_{\phantom{rs} uv}  -8\,
R_{prqs} R^{p \phantom{u} q}_{\phantom{p} u \phantom{q} v} R^{r u s
v},
\end{eqnarray}
is a topological invariant. By functional derivation of
(\ref{Euler_D6}) we obtain the relation
\begin{eqnarray}\label{GB_d6} &&H_{\mu \nu}^{(6,3)(1)}
-12 \, H_{\mu \nu}^{(6,3)(2)} +16\, H_{\mu \nu}^{(6,3)(3)}
\nonumber\\
&& \quad  +24\, H_{\mu \nu}^{(6,3)(4)} +3\, H_{\mu \nu}^{(6,3)(5)}
-24\,
H_{\mu \nu}^{(6,3)(6)} \nonumber\\
&& \quad +4\, H_{\mu \nu}^{(6,3)(7)} -8\, H_{\mu \nu}^{(6,3)(8)} =0.
\end{eqnarray}
Equations (\ref{GB_d6}) could be helpful in order to eliminate into
(\ref{GenFormAmb_6}) one of the conserved tensors of rank 2 and
order 6.

Of course, the ``basis" exhibited above and which has permitted us
to provide the general form for the tensor $\Theta_{\mu \nu}$ can be
used to simplify considerably the expression (\ref{VExpVSET_amb_d6})
obtained for $\Theta^{M^2}_{\mu \nu}$. By using Eqs.~(2.22)-(2.31)
of Ref.~\cite{DecaniniFolacci2007} and after a tedious calculation,
we obtain the compact expression
\begin{eqnarray}\label{VExpVSET_amb_d6_simp2} & &
\Theta^{M^2}_{\mu \nu}  = \frac{\ln M^2}{(4 \pi )^3} \times
\nonumber \\
&& \left( [(1/12)\,\xi^2- (1/30)\, \xi+1/336] \,
H_{\mu \nu}^{\lbrace{2,0\rbrace}(1)} \right. \nonumber\\
& &\quad  \left. + (1/840) \, H_{\mu \nu}^{\lbrace{2,0\rbrace}(3)} -
(1/6)(\xi-1/6)^3 \, H_{\mu \nu}^{(6,3)(1)} \right. \nonumber\\
& &\quad  \left. + (1/180) \, (\xi-1/6) \, H_{\mu \nu}^{(6,3)(2)}  -
(4/2835) \, H_{\mu \nu}^{(6,3)(3)} \right. \nonumber\\
& &\quad  \left. +(1/945) \,
H_{\mu \nu}^{(6,3)(4)} -(1/180) \, (\xi-1/6) \, H_{\mu \nu}^{(6,3)(5)} \right. \nonumber \\
& &\quad  \left. + (1/7560) \, H_{\mu \nu}^{(6,3)(6)}  +(17/45360)
\,
H_{\mu \nu}^{(6,3)(7)} \right. \nonumber\\
& &\quad  \left. -(1/1620) \, H_{\mu \nu}^{(6,3)(8)}
-(1/2)(\xi-1/6)^2m^2 \, H_{\mu \nu}^{(4,2)(1)} \right. \nonumber \\
& &\quad  \left. +(1/180)m^2 \,
H_{\mu \nu}^{(4,2)(2)} - (1/180)m^2 \, H_{\mu \nu}^{(4,2)(3)} \right. \nonumber \\
& &\quad  \left.  +(1/2)(\xi-1/6)m^4 [R_{\mu \nu} -(1/2)R g_{\mu
\nu}]\right. \nonumber \\
& &\quad  \left.  -(1/12)m^6 g_{\mu \nu}  \phantom{\frac{}{}}
\right).
\end{eqnarray}

Finally, it is interesting to note that $\Theta_{\mu \nu}$ given by
(\ref{GenFormAmb_6}) could permit us to modify the trace anomaly
(\ref{TR_Anomaly_d6_res}). Indeed, for $m^2=0$ and by using
Eqs.~(\ref{Tr_H_20b1})-(\ref{Tr_H_638}) with $D=6$, we obtain

\begin{widetext}
\begin{eqnarray}\label{GenFormAmb_6_trace}
&& g^{\mu \nu} \Theta_{\mu \nu}=  [-10 D_1 - 3 D_2]  \Box \Box R +
[-2D_1 - 30D_3 - 4D_4 - D_6/2 - 2D_7]  R\Box R \nonumber \\
& & \qquad + [ - 8D_2 - 4D_4 - 6D_5 + 2D_6 - 4D_7 - D_8 +3D_{10}/2]
R_{;p q} R^{pq}  \nonumber \\
& &   \qquad  +[2 D_2 - 10 D_4 - 3 D_5 - 6 D_6 - 2 D_8 -3
D_{10}] R_{pq} \Box R^{pq} \nonumber \\
& &   \qquad + [8 D_2 - 4 D_6 - 40 D_7 - 14 D_8 - 24 D_9 +3 D_{10}]
R_{pq ; rs}R^{prqs} \nonumber \\
& &   \qquad  + [-2 D_1 - 3D_2/2 - 30 D_3 -
    6 D_4 - 3D_5/2 - 3D_6/4 - 4\ D7 - D_8/2]
R_{;p}R^{;p}
\nonumber \\
& &   \qquad + [10 D_2 - 10 D_4 - 3 D_5 - 10 D_6 - 8 D_8 - 12 D_9-3
D_{10}] R_{pq;r} R^{pq;r} \nonumber \\
& &   \qquad +[- 18 D_2 - 6 D_5 + 9 D_6 + 6 D_8 + 12 D_9 +3 D_{10}]
R_{pq;r} R^{pr;q} + [- 10 D_7 - 2 D_8 - 3 D_9 +3 D_{10}/4] \,
R_{pqrs;t} R^{pqrs;t}
    \nonumber \\
& &    \qquad     + [- 10 D_2 - 6 D_5 + 5 D_6+3 D_{10}] \,(  R_{pq}
R^{p}_{\phantom{p} r}R^{qr} - R_{pq}R_{rs}R^{prqs} )
\nonumber \\
& &    \qquad
   +[- 10 D_7 - 2 D_8 - 3 D_9 +3 D_{10}/4] \,
   (  2 R_{pq}R^p_{\phantom{p} rst} R^{qrst } -R_{pqrs}R^{pquv}
R^{rs}_{\phantom{rs} uv} -4 R_{prqs} R^{p \phantom{u}
q}_{\phantom{p} u \phantom{q} v} R^{r u s v}).
\end{eqnarray}
\end{widetext}
It should be noted that three of the ten scalar Riemann monomials of
order 6, namely $R^3$, $RR_{pq}R^{pq}$ and $RR_{pqrs}R^{pqrs}$, do
not appear in (\ref{GenFormAmb_6_trace}). As a consequence, it is
impossible to remove such terms from the trace anomaly
(\ref{TR_Anomaly_d6_res}). On the contrary, by choosing correctly
the coefficients $D_i$, it is possible to remove any other term from
(\ref{TR_Anomaly_d6_res}).

\subsection{Infinities and gravitational actions}

In the previous sections, we have constructed the renormalized
expectation value of the stress-energy operator for a massive scalar
field in a general spacetime of arbitrary dimension by assuming that
the Wald's axiomatic approach (see
Refs.~\cite{Wald77,Wald78,Wald94}) remains valid for all dimensions.
In the Wald's axiomatic approach, the treatment of the divergences
present in the formal expression (\ref{VExpVSET_unren}) does not
necessitate a particular study i.e. absorbtion into renormalized
gravitational parameters. These divergences are simply discarded and
the cosmological constant $\Lambda$ and the Newton's gravitational
constant $G$ (as well as the other coupling constants associated
with higher-order gravitational terms if we need to consider such
terms) appearing in the semiclassical Einstein equations
(\ref{SCEinsteinEq}) are directly the physical gravitational
parameters while the expectation value $\langle \psi |T_{\mu \nu}
|\psi \rangle_\mathrm{ren}$ constructed from the Hadamard biscalar
$W(x,x')$ is automatically the physically meaningful source.

In the present subsection, we shall depart from the path marked out
by Wald. We shall briefly describe one way to deal with the
divergent part of (\ref{VExpVSET_unren}) by extending the approach
developed by Christensen in Refs.~\cite{Christensen1,Christensen2}
(see also Adler and coworkers in Refs.~\cite{AdlerETAL1,AdlerETAL2}
for a related but slightly different approach). We intend to discuss
at more length this very technical aspect of our work in a paper in
preparation \cite{DecaniniFolacci2008}. However, in order to be as
completed as possible, we shall here provide partial results related
to the present work.

For a given spacetime dimension, we can formally evaluate the
divergent part of (\ref{VExpVSET_unren}) and express the result of
our calculation as a power series in $\sigma^{;a} (x,x')$. By
consistently averaging this power series over all the angular
directions joining $x'$ and $x$ and by adding to it, for $D$ even,
the opposite of (\ref{VExpVSET_amb4}) as well as
$-(D/4)\alpha_Dg_{\mu \nu}v_1$, we find a final divergent expression
constructed from ``simple" conserved geometrical tensors which can
be absorbed into a bare gravitational Lagrangian. It is important to
note that the averaging process of the direction-dependant terms
adopted in
Refs.~\cite{AdlerETAL1,AdlerETAL2,Christensen1,Christensen2}) must
be modified in order to take into account spacetime dimension.

For $D=2$, we obtain for the averaged divergent part of
(\ref{VExpVSET_unren}) an expression of the form
\begin{eqnarray}\label{TMN_sing_d2}
& & \langle \psi |T_{\mu \nu} |\psi \rangle_{\mathrm{sing}} \sim A
\frac{g_{\mu \nu}}{\sigma}+[B_1 m^2 g_{\mu \nu}+ B_2 Rg_{\mu
\nu}]  \ln (M^2\sigma) \nonumber \\
& & + \, \mathrm{finite~terms~in} ~~ m^2 g_{\mu \nu}  \,
\mathrm{and}  \, Rg_{\mu \nu}.
\end{eqnarray}
Here $A$, $B_1$ and $B_2$ are dimensionless constants. This singular
tensor cannot be absorbed into a bare gravitational Lagrangian of
Einstein-Hilbert type because, in two dimensions, the Euler number
(\ref{Euler_D2}) being a topological invariant, the tensor $R_{\mu
\nu}-(1/2)Rg_{\mu \nu}$ vanishes identically. However, this tensor
can be absorbed into the Polyakov non-local bare gravitational
action
\begin{equation}\label{BareAction_D2}
S_\mathrm{grav}= \int _{\cal M} d^2 x\sqrt{-g}\left(a_B
R\frac{1}{\Box} R - 2\Lambda_B \right).
\end{equation}
It should be noted that the non-local Lagrangian ${\cal
L}=R\frac{1}{\Box} R$ provides, by functional derivation with
respect to the metric tensor, a contribution in $2 Rg_{\mu \nu}$ but
also a non-local contribution proportional to
\begin{eqnarray}\label{Nonlocal_Tmunu_d2}
& & -2\left(\frac{1}{\Box} R\right)_{;\mu \nu} +
\left(\frac{1}{\Box} R\right)_{;\mu }\left(\frac{1}{\Box}
R\right)_{; \nu} \nonumber \\
& & \qquad -\frac{1}{2}g_{\mu \nu} \left(\frac{1}{\Box}
R\right)_{;p}\left(\frac{1}{\Box} R\right)^{;p}.
\end{eqnarray}
We think that these two contributions must be added to
(\ref{Theta_munu_amb2}). In the particular case of a two-dimensional
background, it is not natural to follow Wald's prescription and to
construct the conserved tensor $\Theta_{\mu \nu}$ from a purely
local Lagrangian.

For $D=3$, we obtain for the averaged divergent part of
(\ref{VExpVSET_unren}) an expression of the form
\begin{eqnarray}\label{TMN_sing_d3}
& & \langle \psi |T_{\mu \nu} |\psi \rangle_{\mathrm{sing}} \sim A
\frac{g_{\mu \nu}}{\sigma^{3/2}}+B\frac{[R_{\mu \nu}-(1/2)Rg_{\mu
\nu}]}{\sigma^{1/2}} \nonumber \\
& & + \, \mathrm{finite~terms~in} ~~ m^3 g_{\mu \nu}  \,
\mathrm{and}  \, m[R_{\mu \nu}-(1/2)Rg_{\mu \nu}]. \nonumber \\
& &
\end{eqnarray}
Here $A$ and $B$ are dimensionless constants. This singular tensor
can be absorbed into a bare gravitational action given by
\begin{eqnarray}\label{BareAction_D3}
& & S_\mathrm{grav}= -\frac{1}{16\pi G_B}\int _{\cal M} d^3
x\sqrt{-g}\left(R - 2\Lambda_B \right).
\end{eqnarray}

For $D=4$, we obtain for the averaged divergent part of
(\ref{VExpVSET_unren}) an expression of the form
\begin{eqnarray}\label{TMN_sing_d4}
& & \langle \psi |T_{\mu \nu} |\psi \rangle_{\mathrm{sing}} \sim A
\frac{g_{\mu \nu}}{\sigma^{2}}+B\frac{[R_{\mu \nu}-(1/2)Rg_{\mu
\nu}]}{\sigma} \nonumber \\
& & + \left(C_1H_{\mu \nu}^{(4,2)(1)}+C_2H_{\mu
\nu}^{(4,2)(2)}+C_3H_{\mu \nu}^{(4,2)(3)}\right) \ln (M^2\sigma) \nonumber \\
& & + \, \mathrm{finite~terms~in} ~~ m^4 g_{\mu \nu}, m^2[R_{\mu
\nu}-(1/2)Rg_{\mu \nu}], \nonumber \\
& & \qquad  H_{\mu \nu}^{(4,2)(1)}, H_{\mu \nu}^{(4,2)(2)} \,
\mathrm{and}  \, H_{\mu \nu}^{(4,2)(3)}.
\end{eqnarray}
Here $A$, $B$, $C_1$, $C_2$ and $C_3$ are dimensionless constants.
This singular tensor can be absorbed into a bare gravitational
action given by
\begin{eqnarray}\label{BareAction_D4}
& & S_\mathrm{grav}= -\frac{1}{16\pi G_B}\int _{\cal M} d^4
x\sqrt{-g}\left(R - 2\Lambda_B \phantom{\alpha^{(3)}_B}\right. \nonumber \\
& & \qquad  \left. + \alpha^{(1)}_B R^2 + \alpha^{(2)}_B
R_{pq}R^{pq} + \alpha^{(3)}_B R_{pqrs}R^{pqrs} \right). \nonumber \\
& &
\end{eqnarray}
In this bare gravitational action, the term in $\alpha^{(3)}_B
R_{pqrs}R^{pqrs}$ could be removed because, as we have already
noted, the Euler number (\ref{Euler_D4}) is a topological invariant
in four dimensions. Similarly, it would have been possible to remove
the $H_{\mu \nu}^{(4,2)(3)}$ term from (\ref{TMN_sing_d4}).

For $D=5$, we obtain for the averaged divergent part of
(\ref{VExpVSET_unren}) an expression of the form
\begin{eqnarray}\label{TMN_sing_d5}
& & \langle \psi |T_{\mu \nu} |\psi \rangle_{\mathrm{sing}} \sim A
\frac{g_{\mu \nu}}{\sigma^{5/2}}+B\frac{[R_{\mu \nu}-(1/2)Rg_{\mu
\nu}]}{\sigma^{3/2}} \nonumber \\
& & + \frac{\left(C_1H_{\mu \nu}^{(4,2)(1)}+C_2H_{\mu
\nu}^{(4,2)(2)}+C_3H_{\mu \nu}^{(4,2)(3)}\right)}{\sigma^{1/2}}  \nonumber \\
& & + \, \mathrm{finite~terms~in} ~~ m^5 g_{\mu \nu}, m^3[R_{\mu
\nu}-(1/2)Rg_{\mu \nu}], \nonumber \\
& & \qquad  mH_{\mu \nu}^{(4,2)(1)}, mH_{\mu \nu}^{(4,2)(2)} \,
\mathrm{and}  \, mH_{\mu \nu}^{(4,2)(3)}.
\end{eqnarray}
Here $A$, $B$, $C_1$, $C_2$ and $C_3$ are dimensionless constants.
This singular tensor can be absorbed into a bare gravitational
action given by
\begin{eqnarray}\label{BareAction_D5}
& & S_\mathrm{grav}= -\frac{1}{16\pi G_B}\int _{\cal M} d^5
x\sqrt{-g}\left(R - 2\Lambda_B \phantom{\alpha^{(3)}_B} \right. \nonumber \\
& & \qquad  \left. + \alpha^{(1)}_B R^2 + \alpha^{(2)}_B
R_{pq}R^{pq} + \alpha^{(3)}_B
R_{pqrs}R^{pqrs} \right). \nonumber \\
\end{eqnarray}
Of course, this bare gravitational action cannot be simplified
because, for $D>4$, the Euler number (\ref{Euler_D4}) does not
remain a topological invariant. Similarly, the $H_{\mu
\nu}^{(4,2)(3)}$ term cannot be removed from (\ref{TMN_sing_d5}).

Finally, for $D=6$, we obtain for the averaged divergent part of
(\ref{VExpVSET_unren}) an expression of the form
\begin{eqnarray}\label{TMN_sing_d6}
& & \langle \psi |T_{\mu \nu} |\psi \rangle_{\mathrm{sing}} \sim A
\frac{g_{\mu \nu}}{\sigma^{3}}+B\frac{[R_{\mu \nu}-(1/2)Rg_{\mu
\nu}]}{\sigma^2} \nonumber \\
& & + \frac{\left(C_1H_{\mu \nu}^{(4,2)(1)}+C_2H_{\mu
\nu}^{(4,2)(2)}+C_3H_{\mu \nu}^{(4,2)(3)}\right)}{\sigma} \nonumber \\
& & + \left(D_1H_{\mu \nu}^{\lbrace{2,0\rbrace}(1)}+D_2H_{\mu
\nu}^{\lbrace{2,0\rbrace}(3)}+D_3H_{\mu \nu}^{(6,3)(1)} \right. \nonumber \\
& & \left. + D_4 H_{\mu \nu}^{(6,3)(2)} +D_5 H_{\mu \nu}^{(6,3)(3)}+
D_6 H_{\mu \nu}^{(6,3)(4)} \right. \nonumber \\
& & \left. +D_7 H_{\mu \nu}^{(6,3)(5)} +D_8 H_{\mu \nu}^{(6,3)(6)} +
D_9 H_{\mu \nu}^{(6,3)(7)} \right. \nonumber \\
& & \left. + D_{10} H_{\mu \nu}^{(6,3)(8)}
\right)\ln (M^2\sigma) \nonumber \\
& & + \, \mathrm{finite~terms~in} ~~ m^6 g_{\mu \nu}, m^4[R_{\mu
\nu}-(1/2)Rg_{\mu \nu}], \nonumber \\
& & \qquad  m^2H_{\mu \nu}^{(4,2)(1)}, m^2H_{\mu \nu}^{(4,2)(2)},
m^2 H_{\mu \nu}^{(4,2)(3)}, \nonumber \\
& & \qquad H_{\mu \nu}^{\lbrace{2,0\rbrace}(1)}, H_{\mu
\nu}^{\lbrace{2,0\rbrace}(3)}, H_{\mu \nu}^{(6,3)(1)}, H_{\mu
\nu}^{(6,3)(2)}, H_{\mu \nu}^{(6,3)(3)}, \nonumber \\
& & \qquad H_{\mu \nu}^{(6,3)(4)}, H_{\mu \nu}^{(6,3)(5)}, H_{\mu
\nu}^{(6,3)(6)}, H_{\mu \nu}^{(6,3)(7)} \, \nonumber \\
& & \qquad \mathrm{and}  \, H_{\mu \nu}^{(6,3)(8)}.
\end{eqnarray}
Here $A$, $B$, $C_1$, $C_2$, $C_3$, $D_1$, \dots, $D_9$ and $D_{10}$
are dimensionless constants. This singular tensor can be absorbed
into a bare gravitational action given by
\begin{eqnarray}\label{BareAction_D6}
& & S_\mathrm{grav}= -\frac{1}{16\pi G_B}\int _{\cal M} d^6
x\sqrt{-g}\left(R - 2\Lambda_B \phantom{\beta^{(8)}_B}\right. \nonumber \\
& &  \left. + \alpha^{(1)}_B R^2 + \alpha^{(2)}_B R_{pq}R^{pq} +
\alpha^{(3)}_B
R_{pqrs}R^{pqrs} \right. \nonumber \\
& &  \left. + \beta^{(1)}_B R^3 + \beta^{(2)}_B  RR_{pq} R^{pq} +
\beta^{(3)}_B R_{pq}
R^{p}_{\phantom{p} r}R^{qr}  \right. \nonumber \\
& &  \left.   + \beta^{(4)}_B R_{pq}R_{rs}R^{prqs} + \beta^{(5)}_B
RR_{pqrs} R^{pqrs}
  \right. \nonumber \\
& &  \left. +\beta^{(6)}_B R_{pq}R^p_{\phantom{p} rst} R^{qrst }  +
\beta^{(7)}_B R_{pqrs}R^{pquv} R^{rs}_{\phantom{rs} uv} \right. \nonumber \\
& &  \left. +\beta^{(8)}_B R_{prqs} R^{p \phantom{u} q}_{\phantom{p}
u \phantom{q} v} R^{r u s v} \right).
\end{eqnarray}
This bare gravitational action could be simplified by using the fact
that the Euler number (\ref{Euler_D6}) is a topological invariant.
This result could be used to remove from the bare gravitational
action a term such as $\beta^{(8)}_B R_{prqs} R^{p \phantom{u}
q}_{\phantom{p} u \phantom{q} v} R^{r u s v} $. Similarly, it would
have been possible to remove the $H_{\mu \nu}^{(6,2)(8)}$ term from
(\ref{TMN_sing_d6}).

To conclude this subsection, it is important to note that the
previous results must be taken with a grain of salt. Indeed,
Eqs.~(\ref{TMN_sing_d2}), (\ref{TMN_sing_d3}), (\ref{TMN_sing_d4}),
(\ref{TMN_sing_d5}) and (\ref{TMN_sing_d6}) are formal relations:
they display on a very condensed form the true behavior of the
averaged divergent part of (\ref{VExpVSET_unren}) in the limit of
small $\sigma$ (for more details, we refer to
Ref.~\cite{DecaniniFolacci2008}).

\subsection{Hadamard renormalization versus renormalization in the effective action}

Field quantization in curved spacetime can be addressed very
efficiently by using the effective action
\cite{DeWitt65,AvramidiLNP2000,DeWitt03}. This basic object
contains, in principle, all the information about a given quantum
field theory but, unfortunately, it is not usually possible to
express it explicitly. Even in the very simple case of the scalar
field theory considered in the present article, we have only an
approximation for the associated effective action, the so-called
DeWitt-Schwinger approximation
\cite{DeWitt65,BirrellDavies,AvramidiNPB1991,Avramidi_PhD,AvramidiLNP2000,DeWitt03},
which may be represented by the asymptotic series \cite{DeWitt03}
\begin{eqnarray}\label{DS_EffAct}
&& W^\mathrm{DS}=\int_{\cal M} d^D x \sqrt{-g(x)} \times \nonumber \\
&& \quad  \left[\frac{1}{2 (4\pi)^{D/2}} \int_0^{+\infty}
\frac{d(is)}{(is)^{D/2+1}}e^{-m^2 is} \Lambda(x;s) \right]
\end{eqnarray}
where $\Lambda(x;s)$ is a purely geometrical object (see
Ref.~\cite{DeWitt03})  which satisfies
\begin{equation}\label{DS_EffAct_Lambda_inf}
\lim_{s \to + \infty} \, e^{-m^2 is} \Lambda (x;s) =0
\end{equation}
and which can be formally written for $s\to 0$ on the form
\begin{equation}\label{DS_EffAct_Lambda}
\Lambda (x;s) \approx \sum_{k=0}^{+\infty} a_k(x) (is)^k .
\end{equation}
Here, $a_k(x)$ are the diagonal DeWitt coefficients. The four first
ones can be found in Refs.~\cite{DeWitt03,Gilkey75,Gilkey84} and,
because we have previously assumed that spacetime has no boundary,
we have for their global (or integrated) expressions
\begin{eqnarray}
& & \int_{\cal M} d^D x\sqrt{-g} ~a_0=\int_{\cal M} d^D x \sqrt{-g},
\label{IntDeWittCoeff_0} \\
& &  \int_{\cal M} d^D x\sqrt{-g} ~a_1= \int_{\cal M} d^D x
\sqrt{-g} \left[ -(\xi -1/6) \, R \right], \label{IntDeWittCoeff_1}
\\
 & & \int_{\cal M} d^D x\sqrt{-g} ~a_2= \int_{\cal M} d^D x
\sqrt{-g}
\left[(1/2) \, (\xi -1/6)^2 \, R^2 \right. \nonumber \\
& & \qquad \left. - (1/180) \, R_{pq}R^{pq} + (1/180) \,
R_{pqrs}R^{pqrs} \right] \label{IntDeWittCoeff_2}
\end{eqnarray}
and
\begin{widetext}
\begin{eqnarray}\label{IntDeWittCoeff_3}
&& \int_{\cal M} d^D x\sqrt{-g} ~a_3= \int_{\cal M} d^D x \sqrt{-g}
\left([(1/12)\,\xi^2- (1/30)\, \xi+1/336]
  \, R\Box R +(1/840)\,  R_{pq} \Box R^{pq} \right. \nonumber \\
&&    \qquad  \left. - (1/6) \, (\xi-1/6)^3 \, R^3   + (1/180) \,
(\xi-1/6) \,  RR_{pq} R^{pq}  - (4/2835) \, R_{pq}
R^{p}_{\phantom{p} r}R^{qr} +(1/945) \, R_{pq}R_{rs}R^{prqs}
\right. \nonumber \\
&&  \qquad \left. -(1/180) \, (\xi-1/6) \, RR_{pqrs} R^{pqrs}
  + (1/7560) \, R_{pq}R^p_{\phantom{p} rst} R^{qrst }  +(17/45360) \, R_{pqrs}R^{pquv}
R^{rs}_{\phantom{rs} uv} \right. \nonumber \\
&&  \qquad \left. -(1/1620) \, R_{prqs} R^{p \phantom{u}
q}_{\phantom{p} u \phantom{q} v} R^{r u s v} \right).
\end{eqnarray}
\end{widetext}
The DeWitt-Schwinger representation (\ref{DS_EffAct}) of the
effective action is a purely local geometrical object which contains
all the information on the ultraviolet behavior of the quantum
theory of the scalar field obeying the wave equation (\ref{WEQ}) but
which does not take into account its state-dependence. By functional
derivation of (\ref{DS_EffAct}) with respect to the metric tensor,
we can construct the formal stress-energy tensor
\begin{equation}\label{ST_EffAct 1}  \langle  ~T^\mathrm{DS}_{\mu
\nu } ~ \rangle =\frac{2}{ \sqrt{-g}} \frac{\delta W^\mathrm{DS}}
{\delta g^{\mu \nu }}.
\end{equation}
Of course, it is also a purely local geometrical object which is
furthermore state-independent. However, in spite of this last
drawback, it has been extensively used, in the four-dimensional
context, in order i) to understand the regularization and
renormalization of the true (i.e., state-dependent) stress-energy
tensor (see, for example, Ref.~\cite{BirrellDavies}) or ii) to
provide approximations valid in the large mass limit for this true
stress-energy tensor (see, for example,
Refs.~\cite{AvramidiLNP2000,Matyjasek2000,PiedraDeOca2007a,DecaniniFolacci2007}).
In the present subsection, we shall briefly discuss some aspects of
the renormalization of the formal stress-energy tensor
(\ref{ST_EffAct 1}) directly linked to our propose in order to shed
light, from a different point of view, on the results obtained above
but also to advocate, with in mind practical applications, the use
of the Hadamard method we have developed.

First, it is important to note that the effective action
$W^\mathrm{DS}$ is divergent at the lower limit of the integral over
$s$ for all the positive values of the dimension $D$. For $D=2$ and
$D=3$, this divergent behavior is associated with the integrated
DeWitt coefficients (\ref{IntDeWittCoeff_0}) and
(\ref{IntDeWittCoeff_1}); for $D=4$ and $D=5$, it is associated with
the integrated DeWitt coefficients (\ref{IntDeWittCoeff_0}),
(\ref{IntDeWittCoeff_1}) and (\ref{IntDeWittCoeff_2}); and for
$D=6$, it is associated with the integrated DeWitt coefficients
(\ref{IntDeWittCoeff_0}), (\ref{IntDeWittCoeff_1}),
(\ref{IntDeWittCoeff_2}) and (\ref{IntDeWittCoeff_3}). As a
consequence, from (\ref{ST_EffAct 1}) and by using
(\ref{FD_04_1})-(\ref{FD_04_3}) and
(\ref{FD_06_1})-(\ref{FD_06_10}), we can very easily obtain results
analogous to those described in Sec.~IV.B concerning the formal
expression of the divergent part of the stress-energy tensor.

The treatment of the divergent behavior of (\ref{ST_EffAct 1}) can
be achieved by first regularizing the effective action
(\ref{DS_EffAct}), then by absorbing its divergent part into a bare
gravitational action and finally by functionally deriving the
renormalized effective action so obtained. By considering the
dimensionality $D$ of spacetime as a complex number, the effective
action $W^\mathrm{DS}$ can be regularized by analytic continuation
and its divergent part can be extracted coherently and naturally
absorbed into a bare gravitational action
\cite{BirrellDavies,DeWitt03}. The resulting renormalized effective
action can be written in the form
\begin{widetext}
\begin{equation}\label{DS_EffAct_ren_Dpair}
 W_\mathrm{ren}^\mathrm{DS}=\int_{\cal M} d^D x \sqrt{-g(x)}
\times  \left[-\frac{1}{2 [(D/2)!] (4\pi)^{D/2}}  \int_0^{+\infty}
d(is) \ln{(4\pi M^2 is)}\left(\frac{\partial }{\partial \, (is)}
\right)^{D/2+1} \left[e^{-m^2 is} \Lambda(x;s) \right] \right]
\end{equation}
for $D$ even and in the form
\begin{equation}\label{DS_EffAct_ren_Dimpair}
 W_\mathrm{ren}^\mathrm{DS}=\int_{\cal M} d^D x \sqrt{-g(x)} \times
  \left[\frac{1}{2 [\Gamma(D/2)/\sqrt{\pi}] (4\pi)^{D/2}} \int_0^{+\infty}
\frac{d(is)}{(is)^{1/2}}\left(\frac{\partial }{\partial \, (is)}
\right)^{D/2+1/2}\left[e^{-m^2 is} \Lambda(x;s) \right] \right]
\end{equation}
\end{widetext}
for $D$ odd. Formulas (\ref{DS_EffAct_ren_Dpair}) and
(\ref{DS_EffAct_ren_Dimpair}) generalize results displayed in
Ref.~\cite{DeWitt03} for $D=2,3,4$. In
Eq.~(\ref{DS_EffAct_ren_Dpair}), $M$ is an arbitrary mass scale
parameter (the renormalization mass) which is necessary in
dimensional regularization because only dimensionless quantities can
be analytically continued. $M$ remains in the renormalized effective
action for $D$ even. Now, by inserting (\ref{DS_EffAct_ren_Dpair})
or (\ref{DS_EffAct_ren_Dimpair}) into (\ref{ST_EffAct 1}), we can
obtain a renormalized stress-energy tensor for $D$ even or $D$ odd.
Of course, the object calculated in that way is only a
state-independent approximation of the true expectation value of the
stress-energy operator. Furthermore, because in order to obtain
(\ref{DS_EffAct_ren_Dpair}) and (\ref{DS_EffAct_ren_Dimpair}) we
have discarded not only infinite terms involving the integrated
DeWitt coefficients but also finite ones which have been absorbed by
finite renormalization, this object is also ambiguously defined. The
corresponding ambiguities are obtained by functional derivation of
the integrated DeWitt coefficients and are those displayed in
Sec.~IV.A.

Let us now consider the part of the renormalized effective action
(\ref{DS_EffAct_ren_Dpair}) associated with the renormalization mass
$M$. By using (\ref{DS_EffAct_Lambda_inf}) we obtain for its
expression
\begin{eqnarray}\label{Amb-DWS_EffAct}
& &  W_{M^2}^\mathrm{DS}=\int_{\cal M} d^D x \sqrt{-g(x)} \left[
\frac{\ln M^2}{2 [(D/2)!] (4\pi)^{D/2} } \right.
\nonumber \\
& & \qquad \left.  \left(\frac{\partial }{\partial \, (is)}
\right)^{D/2}\left[e^{-m^2 is} \Lambda(x;s) \right] \right]_{s=0}
\end{eqnarray}
and, from (\ref{ST_EffAct 1}) and (\ref{DS_EffAct_Lambda}), it
provides a geometrical ambiguity associated with the stress-energy
tensor given by
\begin{eqnarray}\label{VExpVSET_amb_Dgen}
& & \Theta^{M^2}_{\mu \nu} = \frac{\ln M^2}{2 (4\, \pi)^{D/2} }
\times \frac{2}{ \sqrt{-g}} \frac{\delta } {\delta g^{\mu \nu }}
\left(\int_{\cal M} d^D x \sqrt{-g(x)}  \times   \right. \nonumber \\
& & \qquad \qquad \left. \sum_{k=0}^{D/2} \frac{(-1)^k}{k!}(m^2)^k
a_{D/2-k}(x) \right).
\end{eqnarray}
This ambiguity is in fact equivalent to that obtained from the
Hadamard formalism in Sec.~II [see Eq.~(\ref{VExpVSET_amb4})].
Indeed, for $D=2$ it reduces to
\begin{eqnarray}\label{VExpVSET_amb_d2_simp1}
& & \Theta^{M^2}_{\mu \nu} = \frac{\ln M^2}{2 (4\, \pi) } \times \nonumber \\
& & \quad  \frac{2}{ \sqrt{-g}} \frac{\delta } {\delta g^{\mu \nu }}
\left(\int_{\cal M} d^2 x \sqrt{-g(x)} \left[ a_1(x) - m^2 a_0(x)
\right] \right) \nonumber \\
& &
\end{eqnarray}
which permits us to recover (\ref{VExpVSET_amb_d2}) from
(\ref{IntDeWittCoeff_0}) and (\ref{IntDeWittCoeff_1}). For $D=4$, it
reduces to
\begin{eqnarray}\label{VExpVSET_amb_d4_simp1}
& & \Theta^{M^2}_{\mu \nu} = \frac{\ln M^2}{2\,({4 \pi })^2} \times
\frac{2}{ \sqrt{-g}} \frac{\delta } {\delta g^{\mu \nu }}
\left(\int_{\cal M} d^4 x \sqrt{-g(x)} [a_2(x) \right.
\nonumber \\
& & \left.  \phantom{\int_{\cal M}} - m^2 a_1(x) + (m^4/2) a_0(x) ]
\right)
\end{eqnarray}
which permits us to recover (\ref{VExpVSET_amb_d4_simp2}) from
(\ref{IntDeWittCoeff_0})-(\ref{IntDeWittCoeff_2}) by using
(\ref{FD_04_1})-(\ref{FD_04_3}). For $D=6$, it reduces to
\begin{eqnarray}\label{VExpVSET_amb_d6_simp1}
& & \Theta^{M^2}_{\mu \nu} = \frac{\ln M^2}{2 \,({4 \pi })^3} \times
\frac{2}{ \sqrt{-g}} \frac{\delta } {\delta g^{\mu \nu }}
\left(\int_{\cal M} d^6 x \sqrt{-g(x)} [a_3(x) \right.
\nonumber \\
& & \left.  \phantom{\int_{\cal M}} -m^2 a_2(x)+ (m^4/2) a_1(x) -
(m^6/6) a_0(x) ] \right)
\end{eqnarray}
which permits us to recover (\ref{VExpVSET_amb_d6_simp2}) from
(\ref{IntDeWittCoeff_0})-(\ref{IntDeWittCoeff_3}) by using
(\ref{FD_04_1})-(\ref{FD_04_3}) and
(\ref{FD_06_1})-(\ref{FD_06_10}).

We shall now conclude the present subsection by comparing the
respective merits of the Hadamard formalism developed in this
article and of the approach based on renormalization in the
effective action. Renormalization in the effective action is a
powerful tool which permits us to understand the structure of the
ultraviolet divergences contained in the unrenormalized expression
of the stress-energy tensor and to discuss the ambiguity problem.
Because it uses functional derivation with respect to the metric
instead of the point-splitting method, it permits us to obtain very
easily the results mentioned above with a formalism which is rather
independent of the dimension of spacetime. Hadamard formalism, if we
depart from the axiomatic point of view advocated by Wald, does not
seem so interesting. Unfortunately, calculations based on
renormalization in the effective action cannot permit us to take
into account the state-dependence of the considered quantum theory
and therefore to obtain, in a general framework, the full
renormalized expectation value of the stress-energy operator. In
fact, bearing in mind practical calculations, Hadamard formalism is
much more efficient than the method based on renormalization in the
effective action even if, at first sight and because of its use of
the point-splitting method, it seems rather heavier. It is also
important to note that, in the present article, we have achieved the
major part of the boring job. The reader who simply wishes to
calculate the renormalized expectation value of the stress-energy
tensor in a particular case must only extract from the available
Feynman propagator the first two coefficients of the biscalar
$W(x,x')$ by using the formulas displayed in Sec.~III. If he wants
furthermore to discuss the ambiguities of the renormalized
stress-energy tensor obtained, he can used the expressions displayed
in Sec.~IV.A. He has nothing else to do!

\section{Conclusion and perspectives}

In this article, we have developed the Hadamard renormalization of
the stress-energy tensor for a massive scalar field theory defined
on a general spacetime of arbitrary dimension. For spacetime
dimension up to 6, we have explicitly described the renormalization
procedure while for spacetime dimension from 7 to 11, we have
provided the framework permitting the interested reader to perform
this procedure explicitly in a given spacetime.

Our formalism is very general: we do not assume any symmetry for
spacetime and we do not limit our study to the massless or the
conformally invariant scalar fields. As a consequence, we have
provided a powerful formalism which could permit us to deal with
some particular aspects of the quantum physics of extra spatial
dimensions in a rather general way or, more precisely, in a more
general way than usual (see references in Sec.~I). We think that
this formalism could be immediately used to discuss, from a more
general point of view, the consequence of the presence of extra
spatial dimensions upon:

\quad - The stabilization of Randall-Sundrum braneworld models of
cosmological interest (in connection with the inflationary scenario
and the dark energy problem).

\quad - The quantum violations of the classical energy conditions
(in connection with the singularity theorems of Hawking and Penrose)
as well as of the averaged null energy condition (in connection with
the existence of traversable wormholes and time-machines).

\quad - The fluctuations of the stress-energy tensor (in connection
with the validity of semiclassical gravity and again with the
singularity theorems of Hawking and Penrose).

Furthermore, we think it would be very interesting to revisit
holographic renormalization from the point of view of the Hadamard
formalism and, above all, to use the Hadamard renormalization
procedure developed in this article to perform calculations of
stress-energy tensors for higher-dimensional black holes. Indeed,
even though such a subject has been a central topic of
four-dimensional semiclassical gravity, very little has been
realized in the higher-dimensional framework. This is rather
incomprehensible since string theory (or more precisely the
so-called TeV-scale quantum gravity \cite{ADD98,AADD98,ADD99})
predicts the possibility of production of such black holes at CERN's
Large Hadron Collider
\cite{ArgyresDimopoulosMarch98,BanksFischler99,
EmparanHorowitzMyers2000} with a production rate around
$1~\mathrm{H}z$ \cite{DimopoulosLandsberg2001,GiddingsThomas2002}.
In this context, the semiclassical Einstein equations
(\ref{SCEinsteinEq}) could permit us to partially describe the black
hole evaporation and to test TeV-scale quantum gravity.

Of course, with the various applications previously mentioned in
mind, it is necessary to extend our present work to more general
field theories and more particularly to the graviton field. In order
to perform such a generalization, it is first of all necessary to
carry out the program described at the end of the conclusion of
Ref.~\cite{DecaniniFolacci2005a}, i.e. to construct the covariant
Taylor series expansions for the off-diagonal Hadamard coefficients
for these field theories by going beyond existing results.

\begin{acknowledgments}
We would like to thank Valter Moretti for valuable comments and for
bringing the Hollands and Wald references
\cite{HollandsWald2001,HollandsWald2002,HollandsWald2003,HollandsWald2005}
to our attention. We are also grateful to Bruce Jensen for various
discussions some years ago and to Rosalind Fiamma for help with the
English.
\end{acknowledgments}

\appendix

\section{Traces for the anomalous trace of the stress tensor}

In this appendix, we provide the expressions for the traces of the
three conserved tensors of rank 2 and order 4 $H_{\mu
\nu}^{(4,2)(1)}$, $H_{\mu \nu}^{(4,2)(2)}$ and $H_{\mu
\nu}^{(4,2)(3)}$ and of the ten conserved tensors of rank 2 and
order 6 $H_{\mu \nu}^{\lbrace{2,0\rbrace}(1)}$, $H_{\mu
\nu}^{\lbrace{2,0\rbrace}(3)}$, $H_{\mu \nu}^{(6,2)(1)}$, $H_{\mu
\nu}^{(6,2)(2)}$, $H_{\mu \nu}^{(6,2)(3)}$, $H_{\mu
\nu}^{(6,2)(4)}$, $H_{\mu \nu}^{(6,2)(5)}$, $H_{\mu
\nu}^{(6,2)(6)}$, $H_{\mu \nu}^{(6,2)(7)}$ and $H_{\mu
\nu}^{(6,2)(8)}$.

From Eqs.~(\ref{FD_04_1a}), (\ref{FD_04_2a}) and (\ref{FD_04_3a}),
we easily obtain
\begin{eqnarray}
& & g^{\mu \nu} H_{\mu \nu}^{(4,2)(1)} = -(2D-2) \, \Box R +(D/2-2)
\, R^2, \label{Tr_H_421} \\
& & g^{\mu \nu} H_{\mu \nu}^{(4,2)(2)} = -(D/2) \, \Box R +(D/2-2)
\, R_{pq}R^{pq}, \nonumber \\
& &\label{Tr_H_422}\\
& & g^{\mu \nu} H_{\mu \nu}^{(4,2)(3)} = -2 \, \Box R +(D/2-2) \,
R_{pqrs}R^{pqrs}. \label{Tr_H_423}
\end{eqnarray}

From Eqs.~(2.22)-(2.31) of Ref.~\cite{DecaniniFolacci2007}, we
obtain after tedious calculations using results and geometrical
identities of Ref.~\cite{DecaniniFolacci_arXiv2008}

\begin{widetext}
\begin{eqnarray}
& & g^{\mu \nu} H_{\mu \nu}^{\lbrace{2,0\rbrace}(1)} = -(2D-2) \,
\Box\Box R -2 \, R \Box R -(D/2-1) \, R_{;p}R^{;p} \label{Tr_H_20b1}
\end{eqnarray}
\begin{eqnarray}\label{Tr_H_20b3}
& & g^{\mu \nu} H_{\mu \nu}^{\lbrace{2,0\rbrace}(3)} =  -(D/2) \,
\Box \Box R +  (4-2D) \, R_{;p q} R^{pq} + (D-4)\, R_{pq} \Box
R^{pq} + (2D-4) \, R_{pq ; rs}R^{prqs}   \nonumber \\
& & \qquad \qquad  - (D/2-3/2) \, R_{;p}R^{;p}
  + (5D/2-5) \, R_{pq;r} R^{pq;r} - (4D-6)\, R_{pq;r}
R^{pr;q} -(2D-2)\, R_{pq} R^{p}_{\phantom{p} r}R^{qr}  \nonumber \\
& & \qquad \qquad  +(2D-2) \, R_{pq}R_{rs}R^{prqs}
\end{eqnarray}
\begin{eqnarray}\label{Tr_H_631}
& & g^{\mu \nu} H_{\mu \nu}^{(6,3)(1)} =  -(6D-6) \, R\Box R -(6D-6)
\, R_{;p}R^{;p} +(D/2-3) \, R^3
\end{eqnarray}
\begin{eqnarray}\label{Tr_H_632}
& & g^{\mu \nu} H_{\mu \nu}^{(6,3)(2)} =  -(D/2+1) \, R\Box R- (D-2)
\, R_{;p q} R^{pq} -(2D-2) \,  R_{pq} \Box R^{pq} -D \, R_{;p}R^{;p}
 \nonumber \\
& & \qquad\qquad  -(2D-2) \,  R_{pq;r} R^{pq;r} + (D/2-3) \, RR_{pq}
R^{pq}
\end{eqnarray}
\begin{eqnarray}\label{Tr_H_633}
& & g^{\mu \nu} H_{\mu \nu}^{(6,3)(3)} = - (3D/2-3) \,   R_{;p q}
R^{pq} -3 \,  R_{pq} \Box R^{pq} - (3D/8-3/4) \,R_{;p}R^{;p} -3 \,
  R_{pq;r} R^{pq;r} \nonumber \\
& & \qquad \qquad -(3D/2-3) \, R_{pq;r} R^{pr;q}
      - D \,  R_{pq} R^{p}_{\phantom{p} r}R^{qr}   +(3D/2-3)
\,  R_{pq}R_{rs}R^{prqs}
\end{eqnarray}
\begin{eqnarray}\label{Tr_H_634}
& & g^{\mu \nu} H_{\mu \nu}^{(6,3)(4)} = -(1/2) \,  R\Box R
  + (D/2-1) \,  R_{;p q} R^{pq} - D \, R_{pq} \Box R^{pq} - (D-2) \,
R_{pq ; rs}R^{prqs}   -(3/4) \, R_{;p}R^{;p}
 \nonumber \\
& & \qquad \qquad  -(2D-2) \,  R_{pq;r} R^{pq;r} + (2D-3) \,
R_{pq;r} R^{pr;q} + (D-1) \,  R_{pq} R^{p}_{\phantom{p} r}R^{qr}
-(D/2+2) \, R_{pq}R_{rs}R^{prqs}
\end{eqnarray}
\begin{eqnarray}\label{Tr_H_635}
& & g^{\mu \nu} H_{\mu \nu}^{(6,3)(5)} =  -2 \, R\Box R
   - 4 \,   R_{;p q} R^{pq} - (8D-8) \,
R_{pq ; rs}R^{prqs} -4 \, R_{;p}R^{;p}
   -(2D-2) \,  R_{pqrs;t} R^{pqrs;t}
     \nonumber \\
& & \qquad \qquad   + (D/2-3) \,  RR_{pqrs} R^{pqrs}
  -(4D-4) \,   R_{pq}R^p_{\phantom{p} rst} R^{qrst }+ (2D-2) \,  R_{pqrs}R^{pquv}
R^{rs}_{\phantom{rs} uv}  \nonumber \\
& & \qquad \qquad + (8D-8) \,  R_{prqs} R^{p \phantom{u}
q}_{\phantom{p} u \phantom{q} v} R^{r u s v}
\end{eqnarray}
\begin{eqnarray}\label{Tr_H_636}
& & g^{\mu \nu} H_{\mu \nu}^{(6,3)(6)} =   -  R_{;p q} R^{pq}-2\,
R_{pq} \Box R^{pq} - (2D+2) \, R_{pq ; rs}R^{prqs}   -(1/2) \,
R_{;p}R^{;p}
 -(D+2) \,   R_{pq;r} R^{pq;r}  \nonumber \\
& & \qquad \qquad  +D \, R_{pq;r} R^{pr;q} -(D/4+1/2) \, R_{pqrs;t}
R^{pqrs;t}
       - 4 \,  R_{pq}R^p_{\phantom{p} rst} R^{qrst }
  +(D/4+1/2) \,  R_{pqrs}R^{pquv} R^{rs}_{\phantom{rs}
uv} \nonumber \\
& & \qquad \qquad +(D+2) \,  R_{prqs} R^{p \phantom{u}
q}_{\phantom{p} u \phantom{q} v} R^{r u s v}
\end{eqnarray}
\begin{eqnarray}\label{Tr_H_637}
& & g^{\mu \nu} H_{\mu \nu}^{(6,3)(7)} =  -24 \, R_{pq ; rs}R^{prqs}
-12 \, R_{pq;r} R^{pq;r} +12 \, R_{pq;r} R^{pr;q} -3 \, R_{pqrs;t}
R^{pqrs;t}
     \nonumber \\
& & \qquad \qquad  - 6 \,  R_{pq}R^p_{\phantom{p} rst} R^{qrst }
 + (D/2) \,   R_{pqrs}R^{pquv}
R^{rs}_{\phantom{rs} uv} +12 \,  R_{prqs} R^{p \phantom{u}
q}_{\phantom{p} u \phantom{q} v} R^{r u s v}
\end{eqnarray}
\begin{eqnarray}\label{Tr_H_638}
& & g^{\mu \nu} H_{\mu \nu}^{(6,3)(8)} =  (3/2) \, R_{;p q} R^{pq} -
3 \, R_{pq} \Box R^{pq} + 3 \, R_{pq ; rs}R^{prqs}    - 3 \,
R_{pq;r} R^{pq;r} +3 \, R_{pq;r} R^{pr;q} \nonumber \\
& & \qquad\qquad  + (3/4) \,R_{pqrs;t} R^{pqrs;t}
      +3 \,  R_{pq}
R^{p}_{\phantom{p} r}R^{qr}  -3 \, R_{pq}R_{rs}R^{prqs} + (3/2) \,
R_{pq}R^p_{\phantom{p} rst} R^{qrst }
 \nonumber \\
& & \qquad \qquad   -(3/4) \,   R_{pqrs}R^{pquv}
R^{rs}_{\phantom{rs} uv} + (D/2-6) \, R_{prqs} R^{p \phantom{u}
q}_{\phantom{p} u \phantom{q} v} R^{r u s v}.
\end{eqnarray}

\end{widetext}


\bibliography{S-E-Tensor-RegH}

\end{document}